\documentclass[a4paper,11pt]{article}
\pdfoutput=1
\usepackage[top=1in, bottom=1in, left=1in, right=1in]{geometry}
\usepackage{jcappub}

\usepackage{tikz,xcolor,hyperref}
\usepackage{amsmath, amssymb, amsthm, graphicx, epsfig, fancyhdr, slashed}
\DeclareUnicodeCharacter{2212}{\textminus{}}
\usepackage[normalem]{ulem}
\usepackage{tikzsymbols}
\usepackage{natbib}
\usepackage{float}
\usepackage[compatibility=false]{caption}
\usepackage{bm}
\usepackage{adjustbox}
\usepackage[compat=1.1.0]{tikz-feynman}
\usepackage{tikzsymbols}
\usepackage{tikz}
\usepackage{orcidlink}
\usepackage{latexsym}
\usepackage{booktabs}
\usepackage{subfig}
\usepackage{amsfonts}
\usepackage{bigints}
\usepackage{xcolor}
\usepackage{braket}
\usepackage{indentfirst}
\usepackage{slashed}
\usepackage{multirow,array,tabularx}
\usepackage{blindtext}
\usepackage{titlesec}
\useunder{\uline}{\ul}{}
\usepackage{environ}
\usepackage{mathtools}
\usepackage{upgreek}
\newcommand{\prd}{Phys.\ Rev.\ D}
\usepackage{placeins}

\usepackage{relsize}
\usepackage{tcolorbox}

\newcommand{\gs}{g_\star}
\newcommand{\gss}{g_{\star s}}
\newcommand{\Trh}{T_\text{rh}}
\newcommand{\mPhi}{m_{\Phi}}

\newcommand{\ODPhi}{\mathcal{O}_{D\Phi}}
\newcommand{\OfPhi}{\mathcal{O}_{f\Phi}}
\newcommand{\Hrh}{\mathcal{H}_\text{rh}}
\newcommand{\arh}{a_\text{rh}}
\newcommand{\Tmax}{T_\text{max}}

\newcommand{\rR}{\rho_R}
\newcommand{\rp}{\rho_\phi}

\newcommand{\DNeff}{\Delta N_\text{eff}}
\newcommand{\Gp}{\Gamma_\phi}

\newcommand{\mueff}{\mu_\text{eff}}

\newcommand{\ndm}{n_{\rm dm}}
\newcommand{\Ndm}{N_{\rm dm}}
\newcommand{\mdm}{m_{\rm dm}}
\newcommand{\adm}{a_{\rm dm}}

\newcommand{\lNP}{\Lambda_{\rm NP}}
\newcommand{\fmax}{f_\text{max}}

\newcommand{\sv}{\langle\sigma v\rangle}
\newcommand{\afo}{a_\text{fo}}
\newcommand{\Tfo}{T_\text{fo}}

\usepackage[title]{appendix}
\allowdisplaybreaks[4]

\newcommand{\GeV}{{\rm GeV}}

\renewcommand{\arraystretch}{1.25}
\newcolumntype{Y}{>{\centering\arraybackslash}X}

\title{From WIMP to FIMP during reheating: collider vs non-collider probes for p-wave annihilation}
\author[a,b]{Dipankar Pradhan\orcidlink{0000-0002-2450-6677}, }
\emailAdd{dipankar.pradhan@iopb.res.in}
\author[c]{Niloy Mondal\orcidlink{0009-0006-5837-9772},}
\emailAdd{niloy18@iitg.ac.in}
\author[c]{Abhik Sarkar\orcidlink{0000-0003-1449-2934},}
\emailAdd{sarkar.abhik@iitg.ac.in}
\author[c]{Anupam Ghosh\orcidlink{0000-0003-4163-4491},}
\emailAdd{anupamg@rnd.iitg.ac.in}
\author[d]{Shashwat Sharma\orcidlink{0009-0002-2266-0467},}
\emailAdd{ph23resch11016@iith.ac.in}
\author[f]{Mathew Thomas Arun\orcidlink{0000-0003-3264-3628},}
\emailAdd{mathewthomas@iisertvm.ac.in}
\author[g]{and Basabendu Barman\orcidlink{0000-0003-0374-7655}}
\emailAdd{basabendu.b@srmap.edu.in}
\affiliation[a]{\,\,Institute of Physics, Sachivalaya Marg, Bhubaneswar, Odisha 751005, India.}
\affiliation[b]{\,\,Homi Bhabha National Institute, BARC Training School Complex, Anushakti Nagar, Mumbai 400094, India.}
\affiliation[c]{\,\,Department of Physics, Indian Institute of Technology Guwahati, North Guwahati, 781039, India.}
\affiliation[d]{\,\,Department of Physics, Indian Institute of Technology Hyderabad, Kandi, Sangareddy, Telangana-502285, India.}
\affiliation[f]{\,\,School of Physics, Indian Institute of Science Education and Research, Thiruvananthapuram 695551,
Kerala, India.}
\affiliation[g]{\,\,Department of Physics, School of Engineering and Sciences, SRM University-AP, Amaravati 522240, India.}
\abstract{By examining the transition from freeze-out to freeze-in dark matter (DM) production within the framework of perturbative reheating, where DM interacts with the visible sector through effective operators of dimension six, we have investigated how a broad range of new physics probes can reveal the nature of the pre-BBN Universe. Incorporating constraints from direct and indirect DM searches, invisible decay measurements, collider experiments, and gravitational wave observations, our analysis demonstrates that both current and forthcoming experimental sensitivities can serve as powerful tools for probing as well as constraining the post-inflationary era, together with new physics beyond the SM. Our analysis demonstrates that collider experiments at both the intensity and energy frontiers can impose strong bounds on derivative operators whose interactions are typically {\it p-wave suppressed}, and therefore only weakly constrained by astrophysical observations. In particular, these complementary searches can significantly restrict the allowed reheating temperature, DM mass and effective interaction scale required to reproduce the observed DM abundance for DM produced during the epoch of reheating.}
\begin{document}
\begin{flushright}
IOP/BBSR/2026-06
\end{flushright}
\maketitle
\flushbottom
\section{Introduction}
\label{sec:intro}
In recent decades, modern cosmology has evolved into a precision science, profoundly reshaping our understanding of the Universe’s earliest epochs. High-accuracy observations of cosmic microwave background (CMB) temperature fluctuations reveal that, on sufficiently large scales, the Universe is strikingly uniform and isotropic. This strongly supports the existence of an inflationary epoch, during which the Universe underwent a phase of accelerated expansion. At much later times, the observed primordial abundances of light nuclei show extraordinary agreement with the theoretical framework of big-bang nucleosynthesis (BBN), confirming that the Universe once existed as a hot, dense plasma governed by well-established nuclear physics and near-thermal equilibrium. Yet, despite these major successes, the vast cosmological interval separating inflation from BBN remains one of the least understood chapters in cosmic evolution. Connecting these two pillars of early-Universe physics requires understanding an enormous span of energy—and correspondingly time—over which direct observational guidance remains limited. This intermediate era, commonly identified with {\it (re)heating}, is therefore of central importance. It is during this epoch that the immense vacuum-like energy driving inflation must be converted into the thermal bath of Standard Model (SM) particles that ultimately seeds the hot big bang. Reheating thus establishes the initial conditions for the radiation-dominated Universe and plays a decisive role in shaping the thermal history that follows. Beyond simply repopulating the visible sector, reheating may also govern the genesis of new relic species beyond the SM, e.g., dark matter (DM). In this sense, it provides a unique bridge between cosmology and particle physics, linking the Universe’s earliest thermalization processes to present-day observational mysteries. 

A growing body of cosmological, astrophysical, and collider data has placed increasingly severe pressure on the canonical weakly interacting massive particle (WIMP) framework (see, e.g., Refs.~\cite{Roszkowski:2017nbc, Arcadi:2017kky, Chang:2019xva, Darme:2019wpd,Arcadi:2024ukq}), motivating serious consideration of DM genesis beyond the WIMP paradigm. Among the most compelling alternatives is the feebly interacting massive particle (FIMP) paradigm, in which the DM never attains thermal equilibrium with the visible sector. Instead, the DM population is gradually built up from the SM bath. Once the bath temperature falls below the characteristic scale governing these interactions (typically set by the heavier of the mediator or DM masses), the production rate becomes exponentially suppressed, causing the comoving DM abundance to asymptote to a fixed value through the freeze-in mechanism~\cite{McDonald:2001vt, Hall:2009bx, Bernal:2017kxu}. Realizing this scenario requires extremely weak portal interactions between visible and dark sectors so that DM remains out of equilibrium throughout cosmic history. Such suppression may arise either from minuscule renormalizable couplings, corresponding to infrared (IR) freeze-in, or from higher-dimensional effective operators suppressed by a large new physics scale, giving rise to Ultra-Violet (UV) freeze-in~\cite{Hall:2009bx,Elahi:2014fsa,Barman:2020plp}. The UV realization is especially compelling because the final dark matter abundance becomes directly sensitive to the maximum temperature $\Tmax$ attained by the SM plasma~\cite{Giudice:2000ex,Bernal:2019mhf,Bernal:2020qyu,Garcia:2020eof,Garcia:2020wiy,Ahmed:2021try,Mambrini:2021zpp,Kaneta:2021pyx,Ahmed:2021fvt,Barman:2022tzk,Haque:2023yra,Becker:2023tvd,Bernal:2024ykj,Barman:2024ujh}. Since $\Tmax$ is governed by the details of reheating and inflaton dynamics, UV freeze-in provides a particularly powerful connection between dark sector phenomenology and the thermal history of the very early Universe.

Motivated by this broader framework, the present study investigates whether the reheating epoch can be probed through a combined interplay of collider observables and complementary new-physics searches, with particular emphasis on post-inflationary DM production. To pursue this as economically as possible, we employ an effective field theory (EFT) framework to characterize the DM–SM interaction\footnote{Within the standard WIMP paradigm in a radiation dominated Universe, such analyses have been extensively explored; see, for example, Refs.~\cite{Beltran:2008xg,Cao:2009uw,Goodman:2010ku,Buckley:2011kk,Abdallah:2015ter,Berlin:2014tja,Kahlhoefer:2017dnp,Bertuzzo:2017lwt,Belyaev:2018pqr,Bhattacharya:2021edh,Barman:2021hhg}.}. We assume that the entirety of the observed DM abundance is generated from the thermal plasma during reheating\footnote{See also, for example, Refs.~\cite{Garcia:2020eof,Garcia:2020wiy,Barman:2021ugy,Barman:2022qgt,Bernal:2022wck,Haque:2023yra,Becker:2023tvd,Barman:2024ujh} for related studies.}, while this plasma itself originates from the perturbative decay or scattering of the inflaton condensate into pairs of Higgs-like bosons. Because freeze-out and freeze-in dynamics are controlled by the same underlying masses and effective coupling, we systematically delineate the full parameter space across which DM transitions between the WIMP and FIMP paradigms as the interaction strength varies. To our knowledge, this constitutes the first investigation of the WIMP--FIMP transition in such a generic cosmological background using effective DM-SM interactions (see, e.g.,~\cite{Silva-Malpartida:2023yks,Silva-Malpartida:2024emu} for related analyses in radiation dominated and early matter-dominated cosmologies with a Higgs-portal DM model). Within this setup,  any observation of the DM signal—whether in direct detection, indirect probes, invisible decay of hadrons, or collider experiments—can be mapped directly onto two fundamental aspects of reheating: (a) the maximum temperature achieved by the Universe, and (b) background equation of state of the pre-BBN epoch (via the shape of the inflaton potential during reheating). 

This paper is organized as follows. In Sec.~\ref{sec:framework}, we present the underlying particle physics framework. Sec.~\ref{sec:reheat} discusses the post-inflationary inflaton dynamics and the associated dark matter production mechanisms during reheating. Constraints from non-collider probes, including DM--nucleon/electron scattering experiments, indirect detection searches, and inflationary gravitational wave observations, are analyzed in Sec.~\ref{sec:noncollider}. In Sec.~\ref{sec:collider}, we study the phenomenology at both the intensity and energy frontiers, including bounds from invisible meson decays, current constraints from the Large Hadron Collider (LHC), and future search prospects. Finally, our conclusions are presented in Sec.~\ref{sec:concl}.
\section{Effective field theory framework}
\label{sec:framework}
Although abundant cosmological and astrophysical evidence supports the existence of dark DM, it has so far escaped detection in all terrestrial experiments and remains elusive within the present framework of particle physics. Therefore, without committing to a specific origin of dark sector interactions, we adopt an EFT framework to investigate DM phenomenology. To include the dark sector to the beyond SM phenomenology, we extend the SM gauge symmetry ($\mathcal{G}_{\rm SM}$) to include the dark gauge group $\mathcal{G}_{\rm DM}$. Within this approach, we parametrize the effective DM--SM interactions as,
\begin{equation}
\mathcal{L}_{\rm int}
=
\frac{c}{\lNP^{d+d'-4}}\,
\mathcal{O}_{\rm SM}^{(d)}\,
\mathcal{O}_{\rm DM}^{(d')} \ ,
\end{equation}
where, $\lNP$ denotes the new physics scale, which is unknown and to be fixed by the relevant phenomenology. The operator $\mathcal{O}_{\rm SM}^{(d)}$, with mass dimension $d$, is constructed from SM fields and is invariant under $\mathcal{G}_{\rm SM}$, while $\mathcal{O}_{\rm DM}^{(d')}$, with mass dimension $d'$, consists of DM fields and is an invariant of the dark symmetry group $\mathcal{G}_{\rm DM}$, ensuring DM stability over cosmological timescales. The coefficient $c$ is a dimensionless Wilson coefficient, which we consider to be unity throughout our analysis for simplicity. The overall mass dimension of the operator is $\widetilde{d}=d+d'$. We further assume that \(\mathcal{O}_{\rm SM}\) (\(\mathcal{O}_{\rm DM}\)) is a singlet under \(\mathcal{G}_{\rm SM}\) (\(\mathcal{G}_{\rm DM}\)). Since all dark sector fields transform nontrivially under $\mathcal{G}_{\rm DM}$, the operator $\mathcal{O}_{\rm DM}$ must contain at least two dark fields. For definiteness, we take $\mathcal{G}_{\rm DM}=\mathbb{Z}_2$. This interaction governs DM freeze-out in the WIMP scenario and freeze-in in the FIMP scenario, and determines the relevant signals in direct, indirect, and collider searches. 

The main motivation of the present work, as discussed earlier, is to understand how p-wave suppressed DM--SM interactions can be tested across different experimental frontiers. To illustrate this, we consider two representative operators that lead to p-wave suppressed DM annihilation at leading order:
\begin{align}
& \mathcal{O}_{f\Phi}=(\bar{f}\gamma^\mu f)\,(\Phi^\star i \overleftrightarrow{\partial_\mu}\Phi),
\label{eq:Ofphi}
\\
& \mathcal{O}_{D\Phi}=(H^\dagger \overleftrightarrow{D^\mu} H)(\Phi^\star \overleftrightarrow{\partial_\mu}\Phi)
\label{eq:Odphi}
\,.
\end{align}
Here, $\Phi$ denotes a complex scalar DM field\footnote{Similar p-wave suppressed operators can also be constructed for fermionic DM, see, e.g.,~\cite{Buckley:2011kk,DeSimone:2013gj,Li:2018orw}. In this work, however, we restrict ourselves to the spin-0 DM case for simplicity.}, $f$ represents a SM fermion\footnote{A leptophilic version of $\mathcal{O}_{f\Phi}$ has been considered in~\cite{Borah:2025ema}. We, however, include all the SM quarks and leptons for the present analysis.}, and $H$ is the SM Higgs doublet. We stress that our objective is not to perform a comprehensive study of all possible operators that produce p-wave suppressed DM annihilation. Rather, our aim is to highlight a generic and important feature of such scenarios. In conventional freeze-out scenarios, such p-wave suppression generally weakens present-day indirect detection sensitivity, often rendering these interactions only mildly constrained (or in some cases effectively invisible) to standard astrophysical searches. This apparent freedom, however, is deceptive. As we will demonstrate, these operators are instead subject to stringent and often dominant constraints from complementary probes of new physics, including invisible meson decay searches and collider bounds, particularly from the LHC. Therefore, rather than analyzing an exhaustive operator basis, our goal is to demonstrate how p-wave suppressed DM--SM interactions, if present, can still be probed effectively through a combination of different observations, including those originated from cosmology of the early Universe (as discussed in the next section). Finally, for the analysis, we will consider each operator to be present one at a time.

Before moving on, let us briefly discuss possible UV completion of the operators introduced in  Eqs.~\eqref{eq:Ofphi} and~\eqref{eq:Odphi}. The effective operators introduced can naturally arise from UV complete theories containing a heavy neutral vector mediator. A simple realization is obtained by extending the SM gauge symmetry with an additional broken $U(1)_X$ (see, e.g., Refs.~\cite{Leike:1998wr,Langacker:2009im,Das:2016zue,Basso:2008iv}) gauge group under which the DM field $\Phi$ is charged. The corresponding gauge boson, denoted by $Z^\prime_\mu$, couples to the conserved scalar current $
\Phi^\star\, i\overleftrightarrow{\partial^\mu}\Phi$. If the SM fermions are also charged under $U(1)_X$, the renormalizable interactions take the form,
\begin{align}
\mathcal{L}\supset
g_\Phi Z^\prime_\mu\left(
\Phi^\star i\overleftrightarrow{\partial^\mu}\Phi\right)+g_f Z^\prime_\mu\bar f\gamma^\mu f,
\end{align}
where $\mathfrak{g}_\Phi$ and $g_f$ denote the corresponding gauge couplings. In the limit where the mediator mass satisfies $M_{Z^\prime}\gg \sqrt{s}$, $\sqrt{s}$ being the center of mass energy of a $Z'$-mediated 2-to-2 scattering process, integrating out the heavy $Z^\prime$ field generates the effective operator
\begin{align}
\mathcal{L}_{\rm eff}\supset
-\frac{g_\Phi g_f}{M_{Z^\prime}^2}
(\bar f\gamma^\mu f)
(\Phi^\star i\overleftrightarrow{\partial_\mu}\Phi)\,,
\end{align}
which reproduces $\mathcal{O}_{f\Phi}$. Similarly, if the SM Higgs doublet carries a non-zero $U(1)_X$ charge, one obtains the interaction
\begin{align}
\mathcal{L}
\supset
g_H Z^\prime_\mu
(H^\dagger i\overleftrightarrow{D^\mu}H),
\end{align}
which, after integrating out the heavy mediator, generates
\begin{align}
\mathcal{L}_{\rm eff}
\supset
-\frac{g_\Phi g_H}{M_{Z^\prime}^2}
(H^\dagger \overleftrightarrow{D^\mu}H)
(\Phi^\star \overleftrightarrow{\partial_\mu}\Phi),
\end{align}
corresponding to the operator $\mathcal{O}_{D\Phi}$. Therefore, both effective interactions considered in this work can emerge naturally in well-motivated UV-complete scenarios with heavy vector mediators. Although not explicitly mentioned here, it is also possible to realize $Z'$-mediated flavour violation in this set-up (see, e.g., Ref.~\cite{Langacker:2000ju,Bobeth:2016llm}). In our case, such flavour-violating interactions play a crucial role in constraining the relevant parameter space, as we will illustrate in Sec.~\ref{sec:collider}.
\section{Cosmology during perturbative reheating}
\label{sec:reheat}
We consider the post-inflationary oscillation of the inflaton $\phi$ at the bottom of a monomial potential $V(\phi)$ of the form
\begin{equation}\label{eq:inf-pot}
V(\phi) = \lambda\, \frac{\phi^n}{\Lambda^{n - 4}}\,,
\end{equation}
where $\lambda$ is a dimensionless coupling and $\Lambda$ is the scale of inflation. The potential in Eq.~\eqref{eq:inf-pot} can naturally arise in a number of inflationary scenarios, for example, the $\alpha$-attractor T- or E-models~\cite{Kallosh:2013hoa, Kallosh:2013yoa, Kallosh:2013maa, Kallosh:2015lwa}, or the Starobinsky model~\cite{Starobinsky:1980te, Starobinsky:1981vz, Starobinsky:1983zz, Kofman:1985aw}. Now, given a particular inflationary model, for example, in $\alpha$-attractor T-model~\cite{Kallosh:2013hoa, Kallosh:2013yoa}
\begin{equation}\label{V_model}
V(\phi ) =\lambda\ M_P^4 \left[\tanh \left(\frac{\phi}{\sqrt{6\, \alpha}\, M_P}\right)\right]^n \simeq \lambda\ M_P^4\times
\begin{dcases}
1 & \; \text{for}\; \phi \gg M_P,\\[10pt]
\left(\frac{\phi}{\sqrt{6\,\alpha}\,M_P}\right)^n & \; \text{for}\; \phi\ll M_P\,,
\end{dcases}
\end{equation}
where $M_P\simeq 2.4\times 10^{18}$ GeV is the reduced Planck mass. Now, the equation of motion for the oscillating inflaton condensate reads~\cite{Turner:1983he}
\begin{equation} \label{eq:eom0}
\ddot\phi + (3\, \mathcal{H} + \Gp)\, \dot\phi + V'(\phi) = 0\,,
\end{equation} 
where
\begin{align}
\mathcal{H} = \sqrt{\frac{\rp+\rR}{3\,M_P^2}}\,,    
\end{align}
denotes the Hubble expansion rate, $\Gp$ the inflaton dissipation rate, dots $(\dot {\phantom .})$ derivatives with respect to time, and primes ($'$) derivatives with respect to the field. During reheating, it is legitimate to approximate $\mathcal{H}\simeq\sqrt{\rp/(3M_P^2)}$. Defining the energy density and pressure of $\phi$ as $\rp \equiv \frac12\, \dot\phi^2+ V(\phi)$ and $p_\phi \equiv \frac12\, \dot\phi^2 - V(\phi)$, together with the EoS parameter $w \equiv p_\phi/\rp = (n - 2) / (n + 2)$~\cite{Turner:1983he}, one can write the evolution of the inflaton energy density as
\begin{equation} \label{eq:drhodt}
\frac{d\rp}{dt} + \frac{6\, n}{2 + n}\, \mathcal{H}\, \rp = - \frac{2\, n}{2 + n}\, \Gp\, \rp\,.
\end{equation}
During reheating $a_I \ll a \ll \arh$, where $a$ is the scale factor, the term associated with expansion, that is, $\mathcal{H}\, \rp$ typically dominates over the interaction term $\Gp\, \rp$. Then it is possible to solve Eq.~\eqref{eq:drhodt} analytically, leading to
\begin{equation} \label{eq:rpsol}
\rp(a) \simeq \rp (\arh) \left(\frac{\arh}{a}\right)^\frac{6\, n}{2 + n}\,.
\end{equation}
Here, $a_I$ and $\arh$ correspond to the scale factor at the {\it end of inflation} and at the {\it end of reheating}, respectively. Since the Hubble rate during reheating is dominated by the inflaton energy density, it follows that
\begin{equation} \label{eq:Hubble}
\mathcal{H}(a) \simeq \mathcal{H}(\arh) \times
\begin{dcases}
\left(\frac{\arh}{a}\right)^\frac{3\, n}{n + 2} &\text{ for } a \leq \arh\,,\\[10pt]
\left(\frac{\arh}{a}\right)^2 &\text{ for } \arh \leq a\,.
\end{dcases}
\end{equation}
At the end of the reheating ($a = \arh$), the energy densities of the inflaton and radiation are equal, $\rR(\arh) = \rp(\arh) = 3\, M_P^2\, \mathcal{H}(\arh)^2$. Note that, to avoid affecting the success of BBN, the reheating temperature $\Trh$ must satisfy $\Trh > T_\text{BBN} \simeq 4$~MeV~\cite{Sarkar:1995dd, Kawasaki:2000en, Hannestad:2004px, DeBernardis:2008zz, deSalas:2015glj,Hasegawa:2019jsa}.
The evolution of the SM radiation energy density $\rR$, on the other hand, is governed by the Boltzmann equation of the form~\cite{Garcia:2020wiy}
\begin{equation} \label{eq:rR}
\frac{d\rR}{dt} + 4\, \mathcal{H}\, \rR = + \frac{2\, n}{2 + n}\, \Gp\, \rp\,.
\end{equation}
Using Eq.~\eqref{eq:rpsol}, one can solve Eq.~\eqref{eq:rR} and further obtain 
\begin{equation} \label{eq:rR_int}
\rR(a) \simeq \frac{2\, \sqrt{3}\, n}{2 + n}\, \frac{M_P}{a^4} \int_{a_I}^a \Gp(a')\, \sqrt{\rp(a')}\, a'^3\, da'\,.
\end{equation}
Note that here a general scale factor dependence of $\Gp$ has been assumed, which may arise, for example, from the field-dependent inflaton mass. In the present setup, the effective mass $m_\phi(a)$ for the inflaton can be obtained from the second derivative of Eq.~\eqref{eq:inf-pot}, which reads
\begin{equation}\label{eq:inf-mass1}
m_\phi(a)^2 \equiv \frac{d^2V}{d\phi^2} = n\, (n - 1)\, \lambda\, \frac{\phi^{n - 2}}{\Lambda^{n - 4}}
\simeq n\, (n-1)\, \lambda^\frac{2}{n}\, \Lambda^\frac{2\, (4 - n)}{n} \rp(a)^{\frac{n-2}{n}}\,,
\end{equation}
or equivalently,
\begin{equation}\label{eq:inf-mass}
m_\phi(a) \simeq m_I \left(\frac{a_I}{a}\right)^\frac{3 (n-2)}{n+2}\,,
\end{equation}
where $m_I\equiv m_\phi(a_I)$. Note, for $n \neq 2$, $m_\phi$ has a field dependence that, in turn, would lead to a time-dependent inflaton dissipation rate.  
\subsection{Reheating dynamics}
We consider reheating happens through the perturbative decay\footnote{For $n>4$, the inflaton energy density redshifts faster than radiation, causing the equation of state to evolve from a stiff phase ($w>1/3$) to radiation domination ($w=1/3$). In this regime, the coherent inflaton condensate fragments into relativistic quanta with energy density scaling as $a^{-4}$~\cite{Lozanov:2016hid,Lozanov:2017hjm}. This fragmentation, driven by inflaton self-interactions, involves non-perturbative effects such as parametric and tachyonic resonance~\cite{Kofman:1997yn,Dufaux:2006ee} and typically requires lattice simulations for a complete treatment (see, e.g.,~\cite{Greene:1997fu,Green:1999yh,Amin:2011hj,Lozanov:2016hid,Figueroa:2016wxr,Lozanov:2017hjm,Garcia:2023eol,Garcia:2023dyf,Barman:2025lvk}).} of the inflaton condensate into a pair of bosons via a SM gauge invariant interaction of the form $\mu\,\phi\,|H|^2$. In this case, the decay rate reads
\begin{equation} \label{eq:bos_gamma}
\Gp(a) = \frac{\mueff^2}{8\pi\, m_\phi(a)}\,,
\end{equation}
where again the effective coupling $\mueff \ne \mu$ (if $n\neq2$) can be obtained after averaging over oscillations. Proceeding in a similar fashion as before,
\begin{equation} \label{eq:rR_bos}
\rR(a) \simeq \frac{3\, n}{1 + 2\, n}\, M_P^2\, \Gp(\arh)\, \mathcal{H}(\arh) \left(\frac{\arh}{a}\right)^\frac{6}{2 + n} \left[1 - \left(\frac{a_I}{a}\right)^\frac{2\, (1 + 2 n)}{2 + n}\right]\,.
\end{equation}
The temperature of the SM bath evolves as
\begin{equation} \label{eq:Tevol}
T(a) \simeq \Trh \left(\frac{\arh}{a}\right)^\xi\,,
\end{equation}
with which the SM temperature and the Hubble rate evolve similar as Eqs.~\eqref{eq:Tevol} and~\eqref{eq:Hevol}, respectively, with
\begin{equation} \label{eq:TBos}
\xi = \frac32\, \frac{1}{n + 2}\,.
\end{equation}
\begin{figure}
\centering
\includegraphics[scale=0.375]{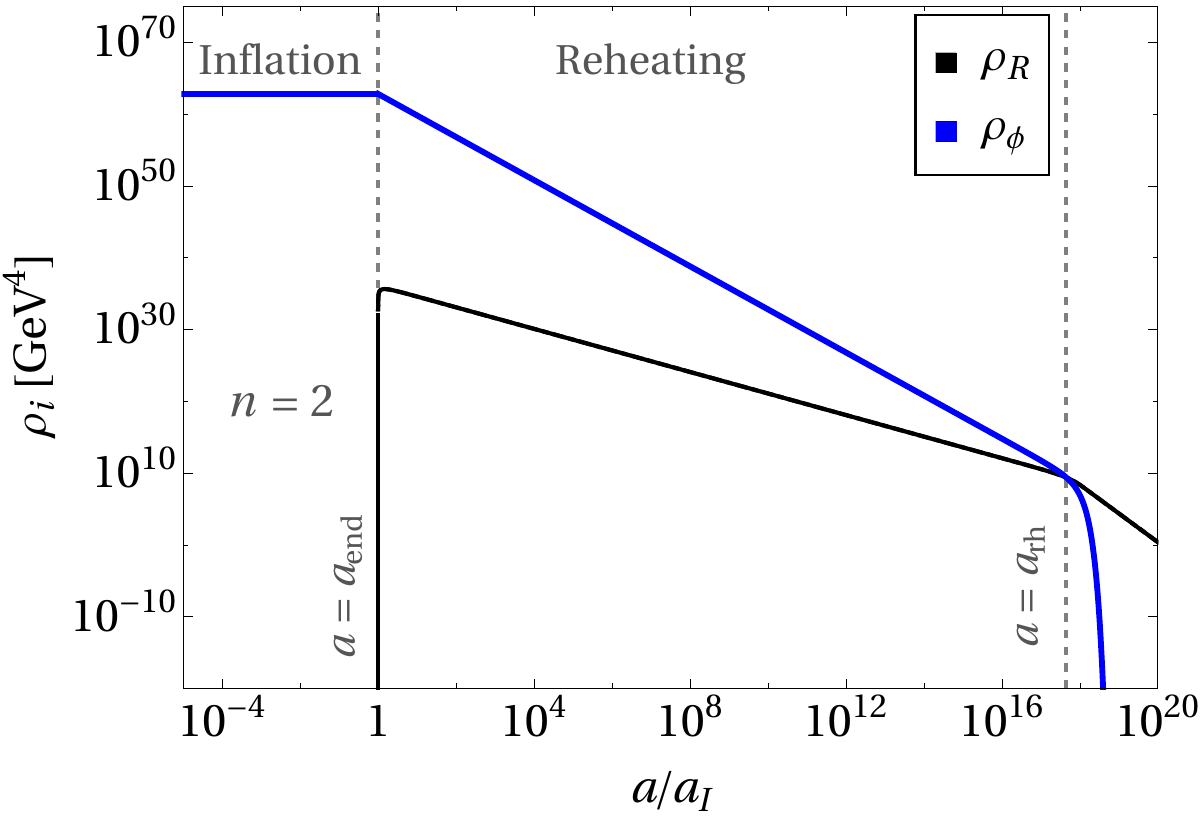}~\includegraphics[scale=0.375]{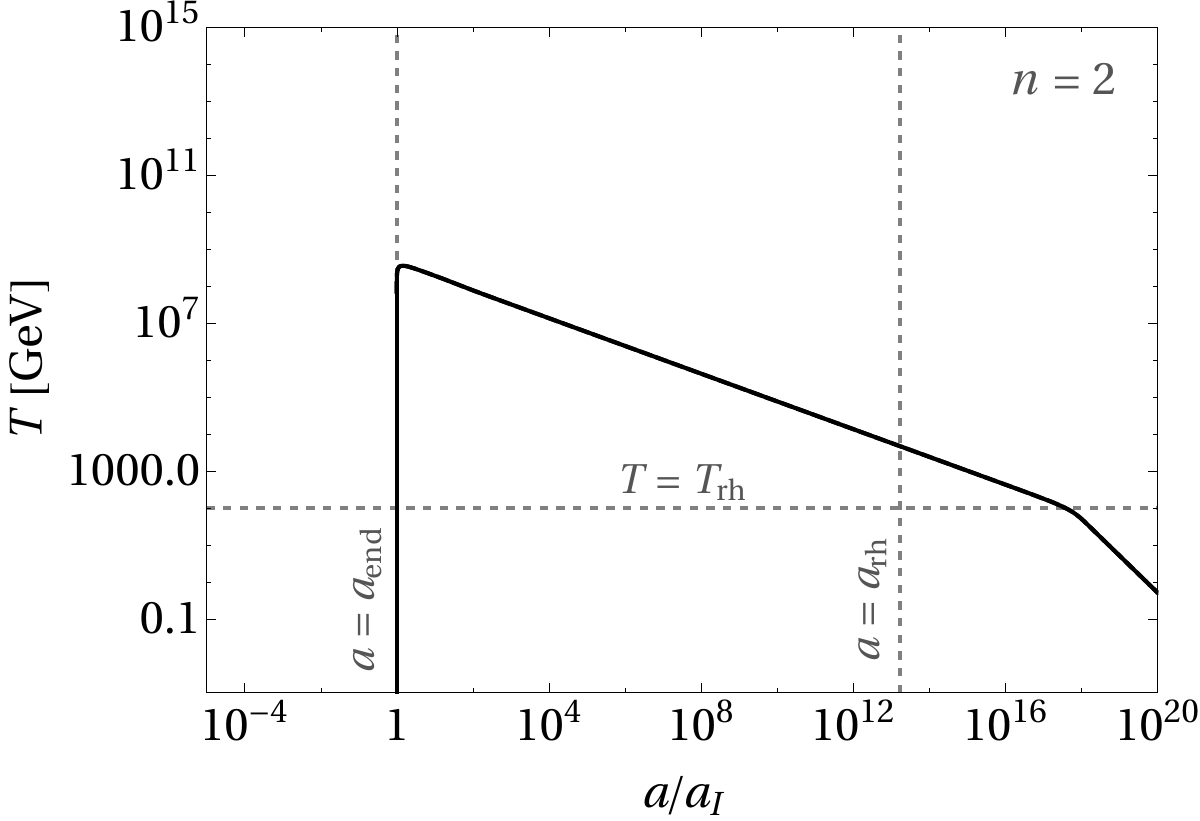}\\[10pt]
\includegraphics[scale=0.375]{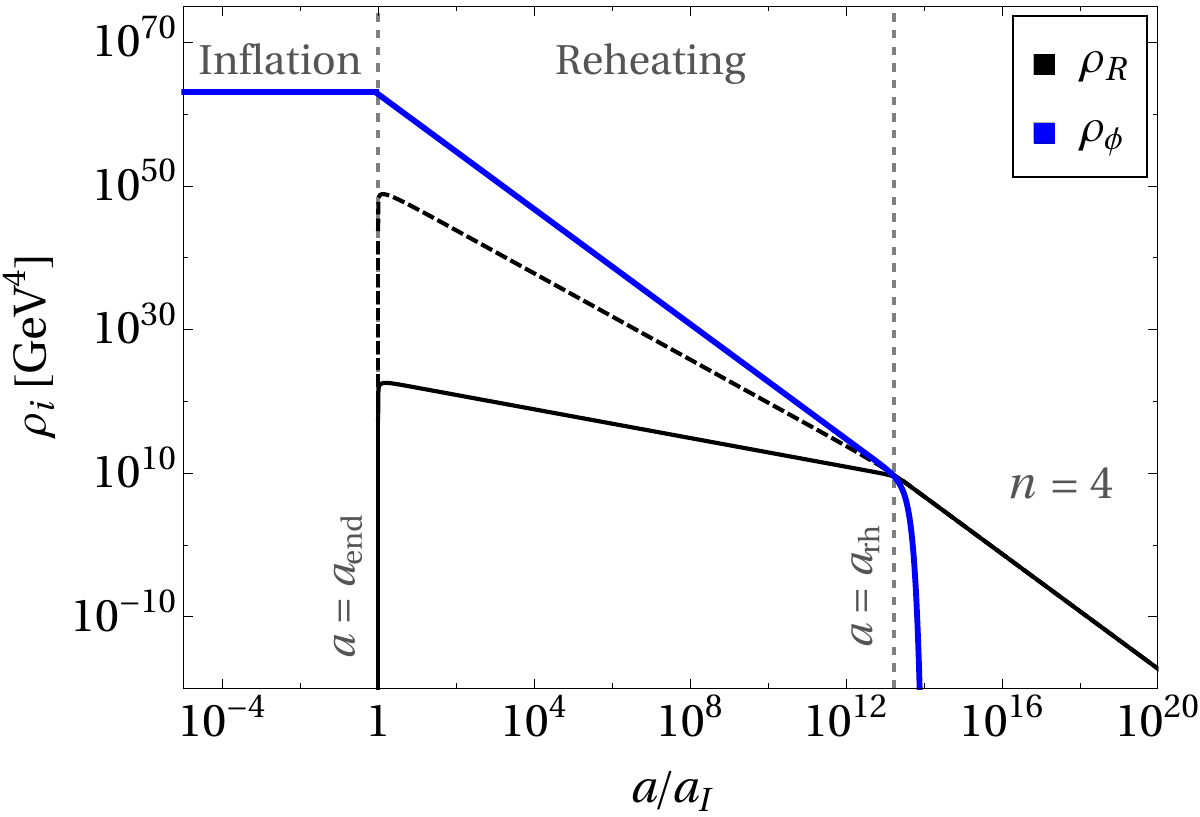}~\includegraphics[scale=0.375]{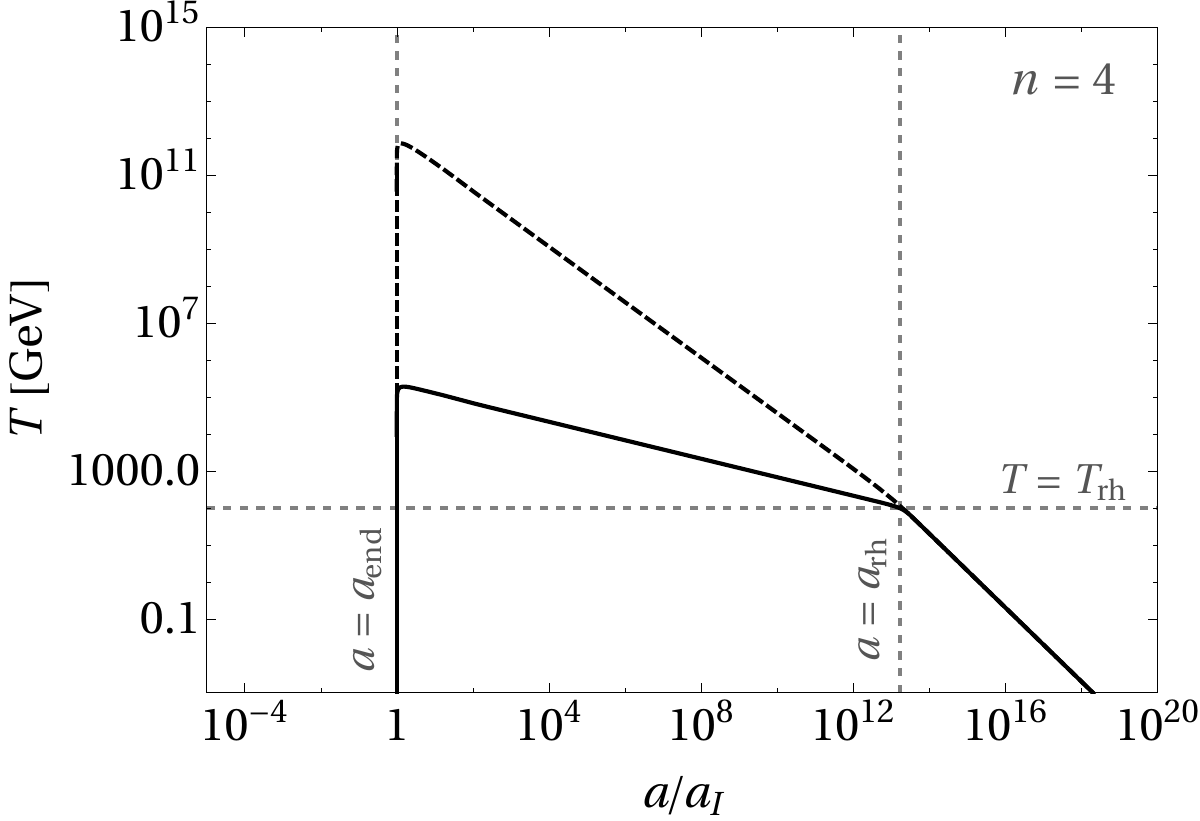}\\[10pt]
\includegraphics[scale=0.375]{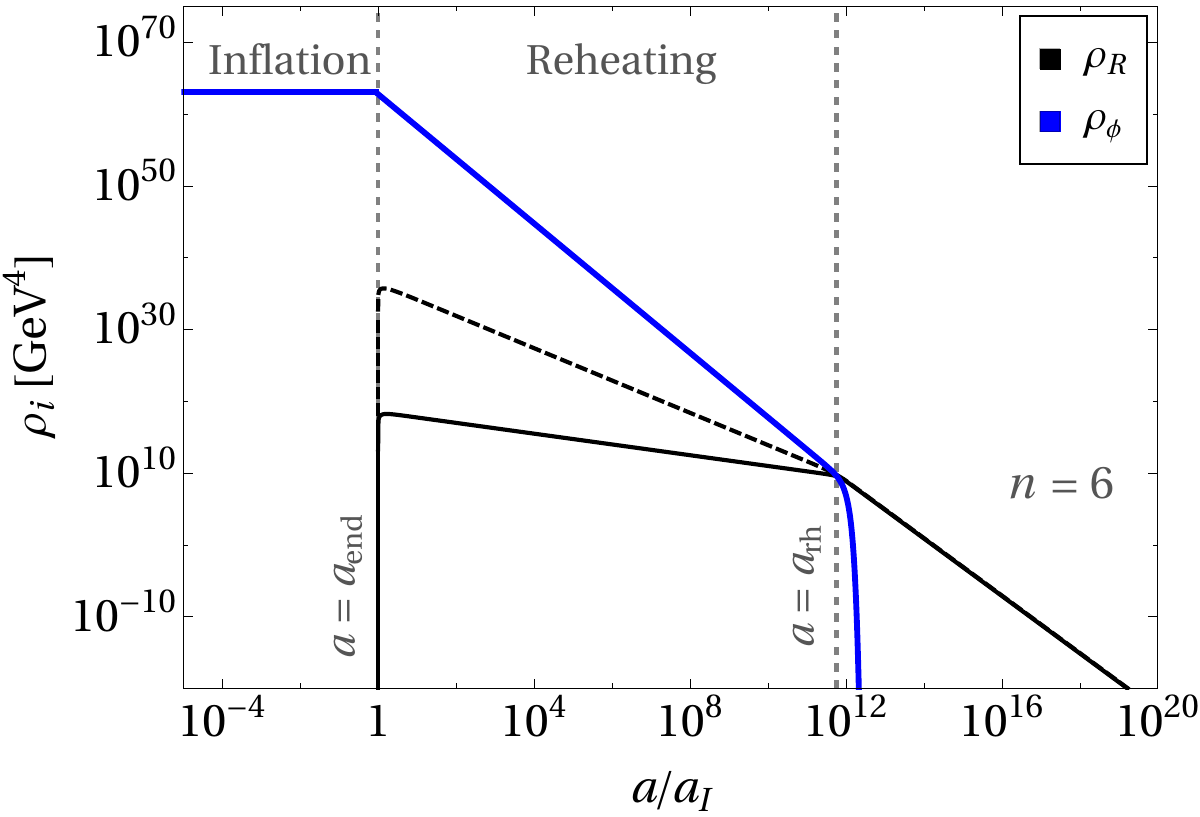}~\includegraphics[scale=0.375]{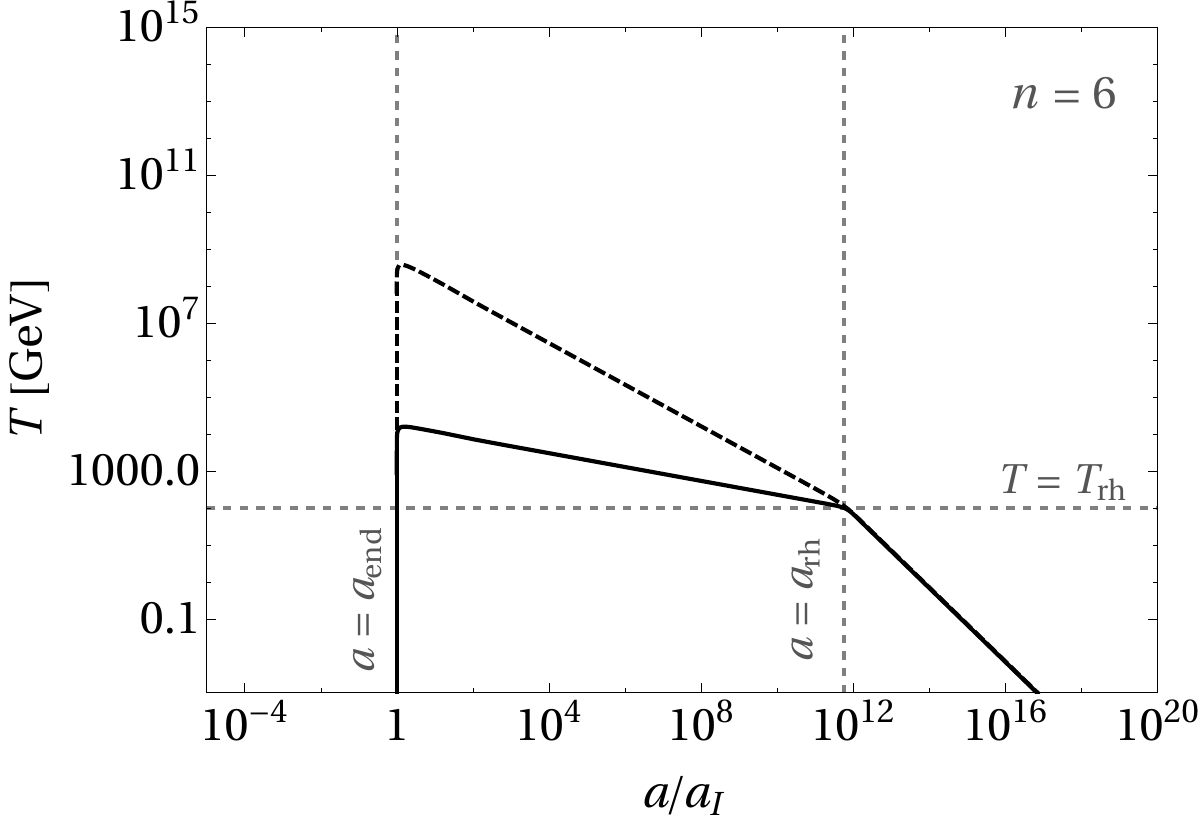}
\caption{Evolution of inflaton (in blue) and radiation (in black) energy densities with the scale factor $a$, for different choices of $n$ (see text for details). In each case, the black solid (dashed) curves correspond to reheating via inflaton decay (scattering). In all cases we have fixed $\Trh=100$ GeV.}
\label{fig:rho-T}
\end{figure}

The bosonic reheating scenario can also be realized via 2-to-2 scattering of the inflaton condensate into a pair of bosonic final states via an interaction $\sigma\,\phi^2\,\left|H\right|^2$, where $\sigma$ is the corresponding interaction strength. In this case the inflaton dissipation rate reads~\cite{Garcia:2020wiy,Bernal:2023wus},
\begin{align}
& \Gp(a)=\frac{\sigma_{\rm eff}^2\,\rp}{8\pi\,m_\phi(a)^3}\,,   
\end{align}
where $\sigma_{\rm eff}$, as before is the effective coupling, obtained after averaging over several oscillations. In this case, the radiation energy density evolves as,
\begin{align}
& \rR(a)\simeq\frac{3n}{2n-5}\,M_P^2\,\Gp(\arh)\,\mathcal{H}(\arh)\,\left(\frac{\arh}{a}\right)^\frac{18}{n+2}\,\left[1-\left(\frac{a_I}{a}\right)^\frac{4n-10}{n+2}\right]\,,   
\end{align}
for $n>5/2$, while for $n=5/2$ we have,
\begin{align}
& \rR(a)\simeq\frac{4}{3}\,M_P^2\,\mathcal{H}(\arh)\,\Gp(\arh)\,\log\left[\frac{a}{a_I}\right]\,.   
\end{align}
In this case, we note, the bath temperature evolves following Eq.~\eqref{eq:Tevol}, with
\begin{align}\label{eq:Tbos22}
&\xi = \frac{9}{2}\,\frac{1}{n+2}\,.   
\end{align}
Note that, both the bosonic decay and bosonic scattering are the inevitable gauge-invariant interactions channels that can be realized in the presence of the inflaton. For $n=2$ (and more generally, for $n<5/2$), the radiation energy density redshifts faster than that of nonrelativistic matter (i.e., the inflaton), preventing it from overtaking the inflaton energy density; consequently, the Universe never becomes radiation dominated. Trading the scale factor with temperature, the Hubble expansion rate during reheating (cf. Eq.~\eqref{eq:Hubble}) can be rewritten as
\begin{equation} \label{eq:Hevol}
\mathcal{H}(T) \simeq \mathcal{H}(\Trh) \left(\frac{T}{\Trh}\right)^{\frac{3\, n}{2 + n}\, \frac{1}{\xi}}\,.
\end{equation}
\begin{table}[ht]
\centering
\renewcommand{\arraystretch}{1.5}{
\begin{tabular}{c|c c c c}
\hline
Reheating channel & Generic & $n=2$ & $n=4$ & $n=6$ \\ 
\hline
Bosonic decay 
& $T \propto a^{-\frac{3}{2n+4}}$ 
& $T \propto a^{-3/8}$ 
& $T \propto a^{-1/4}$ 
& $T \propto a^{-3/16}$ \\
Bosonic scattering 
& $T \propto a^{-\frac{9}{2n+4}}$ 
& $T \propto a^{-9/8}$ 
& $T \propto a^{-3/4}$ 
& $T \propto a^{-9/16}$ \\
\hline
\end{tabular}}
\caption{Scaling of the radiation temperature $T$ with the scale factor $a$ for different production channels.}
\label{tab:T}
\end{table}

In Fig.~\ref{fig:rho-T}, we illustrate the evolution of the inflaton and radiation energy densities as functions of the scale factor, considering reheating via either perturbative bosonic inflaton decay or 2-to-2 scattering of the inflaton condensate. In each scenario, we also display the corresponding evolution of the thermal bath temperature in the right column. As discussed above, for $n=2$, reheating proceeds solely through inflaton decay, as illustrated in the top panel. Noteworthy, for $n>2$, the SM bath attains a higher temperature when reheating occurs via scattering rather than decay. This follows from the fact that the effective rate for bosonic scattering scales as $\rho_\phi/m_\phi^3$, whereas for perturbative decay it scales as $1/m_\phi$, rendering scattering significantly more efficient. The scaling of the bath temperature with the scale factor for different values of $n$ is summarized in Tab.~\ref{tab:T}, and is consistent with the behavior shown in Fig.~\ref{fig:rho-T}. Notably, for $n=2$, the temperature scales as $T\propto a^{-9/8}$, closely resembling the standard radiation-dominated case.
\subsection{Dark matter production during reheating}
We investigate both thermal and non-thermal pathways for DM production. A central observation is that, irrespective of whether DM freezes out (WIMP) or freezes in (FIMP), the evolution of its number density $\ndm$ is governed by the same Boltzmann equation (BEQ),
\begin{equation} \label{eq:boltzdm}
\frac{d\ndm}{dt} + 3\,\mathcal{H}\,\ndm
= - \sv \left(\ndm^2 - n_{\rm eq}^2\right),
\end{equation}
where $\sv$ denotes the thermally averaged 2-to-2 annihilation cross section, into SM states in the freeze-out regime and into DM pairs in the freeze-in regime, and $n_{\rm eq}$ is the equilibrium DM density in the non-relativistic limit. Since we focus on DM production during reheating, when the SM entropy is not conserved, we rewrite Eq.~\eqref{eq:boltzdm} in terms of the comoving DM number $\Ndm\equiv\ndm\,a^3$ as,
\begin{equation}\label{eq:BE-Ndm}
\frac{d\Ndm}{da}
= - \frac{\sv}{a^4\,\mathcal{H}}
\left(\Ndm^2 - N_{\rm eq}^2\right)\,.
\end{equation}
The correct relic abundance is obtained when
\begin{equation}
\mdm\,Y_0
= \frac{\Omega_{\rm DM} h^2\,\rho_c}{s_0\,h^2}
\simeq 4.3 \times 10^{-10}~\text{GeV},
\end{equation}
with $Y_0=\ndm(T_0)/s_0$ is the present DM yield, $s_0\simeq2.69\times10^3~\text{cm}^{-3}$ the present entropy density~\cite{ParticleDataGroup:2020ssz}, $\rho_c\simeq1.05\times10^{-5}h^2~\text{GeV}/\text{cm}^3$ the critical energy density, and $\Omega_{\rm DM}h^2\simeq0.12$ the observed DM relic abundance~\cite{Planck:2018vyg}. For analytic insight, we parametrize the thermally averaged cross section as~\cite{Elahi:2014fsa,Bernal:2019mhf,Kaneta:2019zgw,Barman:2023ktz},
\begin{align}
\langle\sigma v\rangle \equiv \frac{T^k}{\lNP^{k+2}}\,,
\end{align}
with $k=2\,(d-5)$ for an effective operator of dimension $d$, assuming $\mdm\ll T$. In practice, we solve the full set of Boltzmann equations numerically using exact expressions for $\langle\sigma v\rangle$. It is important to note that for the effective description to remain valid throughout reheating, one must ensure $\Tmax\lesssim\lNP$~\cite{Giudice:2000ex}, since DM production takes place entirely from the SM bath\footnote{This limit, $\Tmax\lesssim\lNP$, is necessary only for the UV freeze-in scenario. For the freeze-out of cold dark matter, it is sufficient but not necessary, since the dark-matter freeze-out number density is determined by the dynamics around the freeze-out temperature rather than by the thermal history at earlier times. However, in this case the condition $\Tfo < \lNP$ must be satisfied.}. With these preliminaries in place, we now derive analytical solutions of Eq.~\eqref{eq:BE-Ndm} considering two DM production mechanisms.
\subsubsection{Freeze-out during reheating}
We begin with the WIMP framework, assuming that DM tracks its thermal equilibrium abundance prior to freeze-out. When freeze-out occurs out of equilibrium during reheating, Eq.~\eqref{eq:BE-Ndm} yields,
\begin{align}\label{eq:Ndm-FO}
& \frac{\Ndm(\arh)}{\arh^3}\simeq
\frac{3}{2n+4}\,\left(\frac{\lNP}{\Trh}\right)^k\,\Hrh\,\lNP^2
\begin{dcases}
(k+4)\,\left(\frac{\afo}{\arh}\right)^{\frac{3\,(k+4)}{2n+4}}\,, & \text{for bosonic decay}\,,\\[10pt]
(3k+4)\,\left(\frac{\afo}{\arh}\right)^{\frac{3\,(3k+4)}{2n+4}}\,, & \text{for bosonic scattering}\,,
\end{dcases}
\end{align}
where $\afo=\arh\,\left(\Trh/\Tfo\right)^\xi$ denotes the scale factor at DM freeze-out and $\Hrh\equiv \mathcal{H}(\Trh)=(\pi/3)\,\sqrt{\gs(\Trh)/10}\,(\Trh^2/M_P)$. Since entropy is conserved after reheating, the final DM yield directly follows as $Y_0=\Ndm(\arh)/(s(\arh)\,\arh^3)$. The freeze-out temperature can be determined using the freeze-out condition: $n_{\rm eq}^{\rm DM}\cdot\langle\sigma v\rangle=\mathcal{H}$ evaluated at $T=T_{\rm fo}$, giving
{\small
\begin{align}\label{eq:xfo}
& x_{\rm fo}=\frac{2k+3}{2}
\begin{dcases}
\mathcal{W}_{-1}\!\left[\frac{2}{2k+3}\,\Bigg\{\left(\frac{\Trh}{\mdm}\right)^{2n}\,\left(\frac{\mdm}{\lNP}\right)^k\,\left(\frac{g_{\rm dm}\,\mdm^3}{(2\pi)^{3/2}\,\lNP^2\,\Hrh}\right)\Bigg\}^\frac{2}{2k+3}\right] & \text{for bosonic decay}\,,\\[10pt]
\mathcal{W}_{-1}\!\left[\frac{2}{2k+3}\,\Bigg\{\left(\frac{\Trh}{\mdm}\right)^\frac{2n}{3}\,\left(\frac{\mdm}{\lNP}\right)^k\,\left(\frac{g_{\rm dm}\,\mdm^3}{(2\pi)^{3/2}\,\lNP^2\,\Hrh}\right)\Bigg\}^\frac{2}{2k+3}\right] & \text{for bosonic scattering}\,,
\end{dcases}
\end{align}}
with $\mathcal{W}_{-1}$ representing the $-1$ branch of the Lambert $\mathcal{W}$-function and
\begin{align}
n_{\rm eq}^{\rm DM}\simeq g_{\rm dm}\,\left(\frac{\mdm\,T}{2\pi}\right)^{3/2}\,e^{-\mdm/T}\,,    
\end{align}
is the DM equilibrium number density.
\begin{figure}[htb!]
\centering
\includegraphics[scale=0.375]{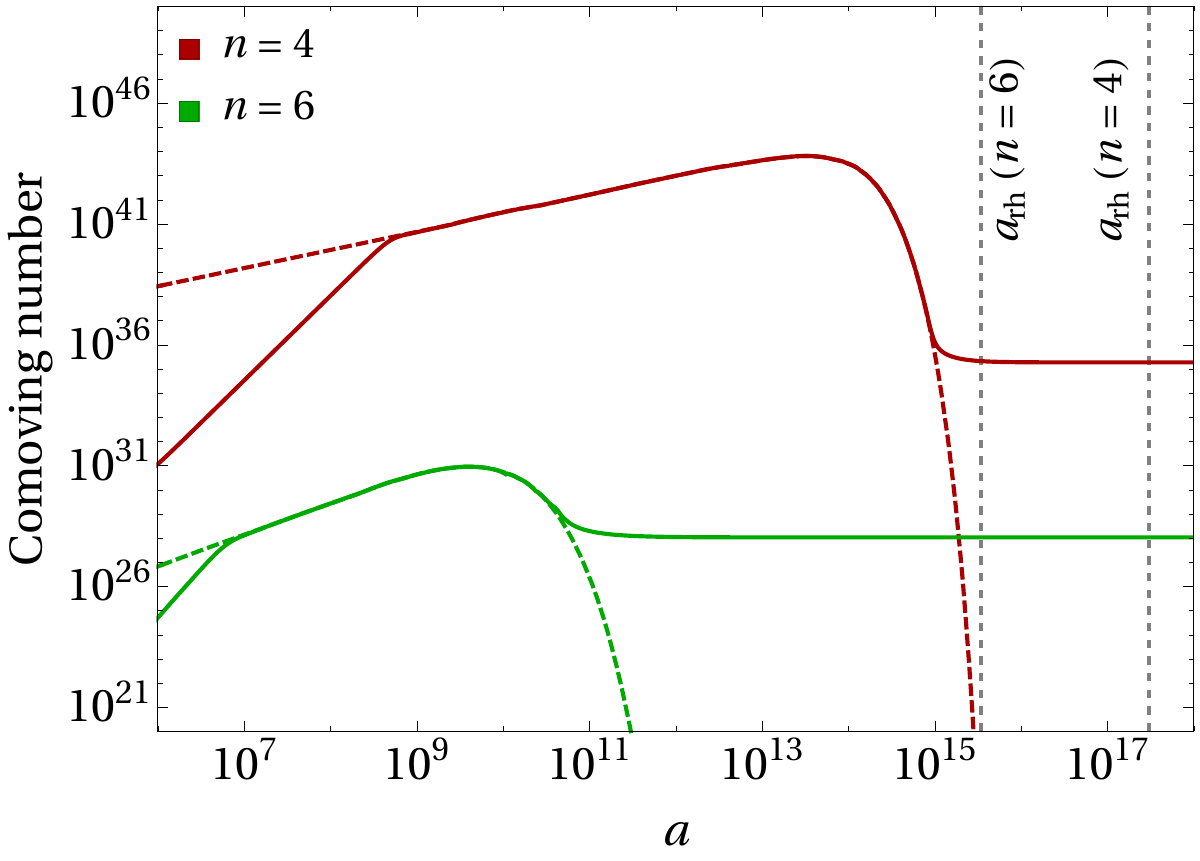}~\includegraphics[scale=0.375]{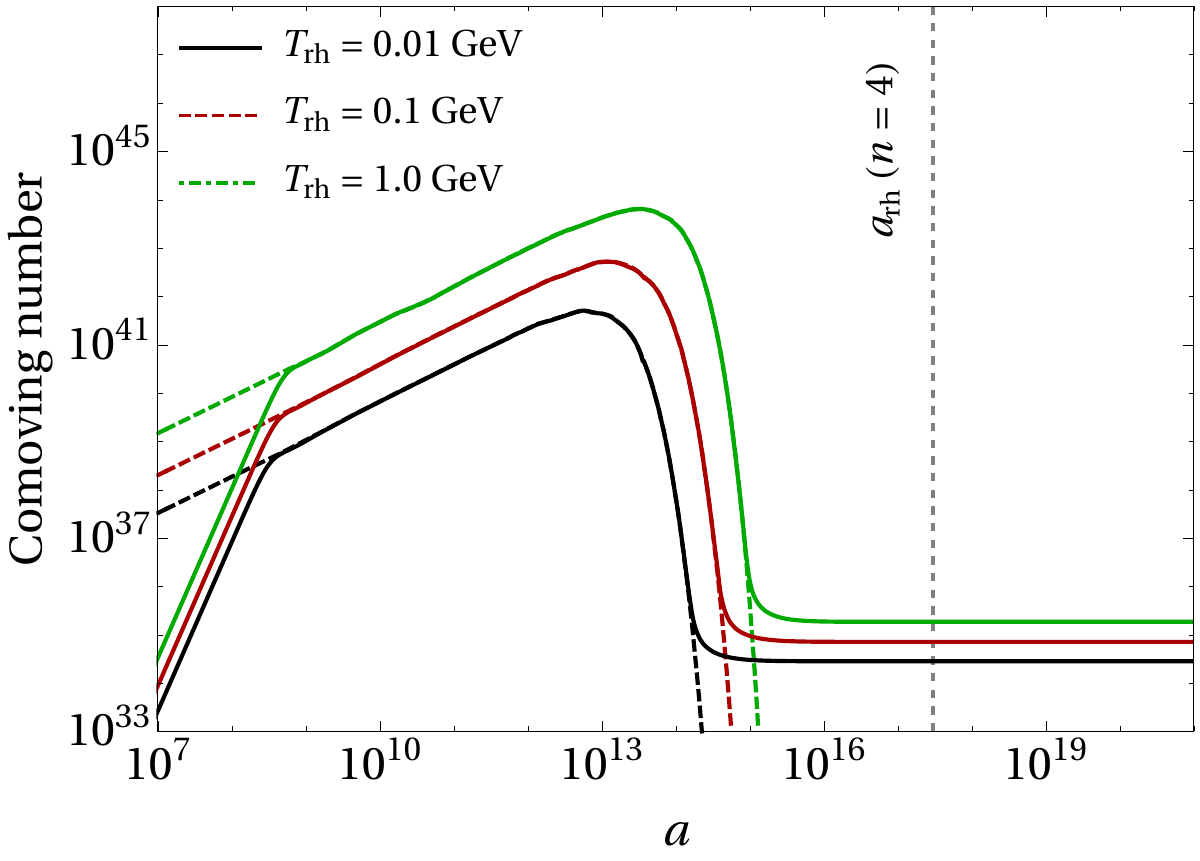}\\[10pt]
\includegraphics[scale=0.375]{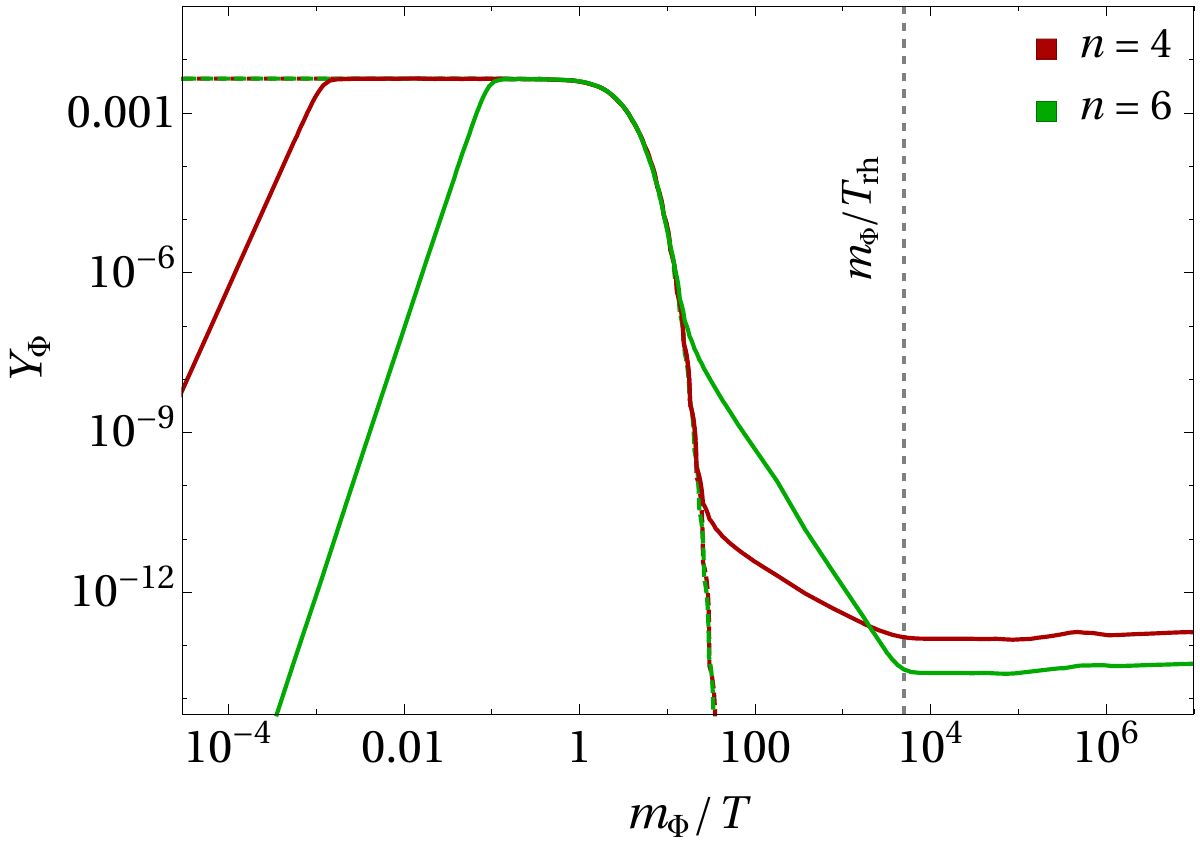}~\includegraphics[scale=0.375]{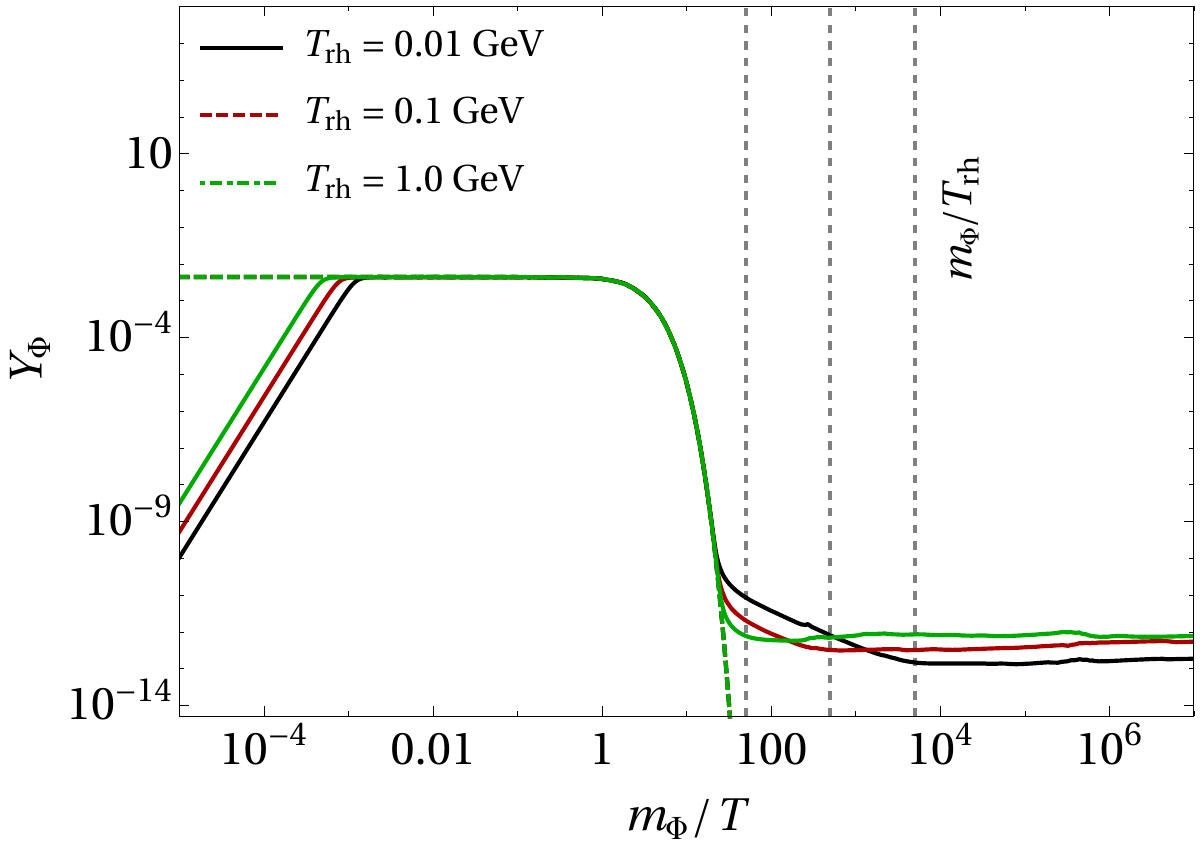}
\caption{Freeze-out yield of the DM as a function of the scale-factor (upper panel) and temperature (lower panel) during reheating, considering $\mathcal{O}_{D\Phi}$ operator. We have included both bosonic scattering and bosonic decay channels for the reheating dynamics. The red and green dashed curves in the top left panel show equilibrium yield corresponding to $n=4$ and $n=6$, respectively. The gray dashed vertical lines denote the end of reheating. We have fixed $\{m_\Phi,\,\Trh,\,\lNP\}=\{50\,\text{GeV},\,10\,\text{MeV},\,0.5\,\text{TeV}\}$ for left panel and $\{n,\,m_\Phi,\,\lNP\}=\{4,\,50\,\text{GeV},\,0.5\,\text{TeV}\}$ for right panel.}
\label{fig:ODPhi_FOyield}
\end{figure}
\begin{figure}[htb!]
\centering
\includegraphics[scale=0.375]{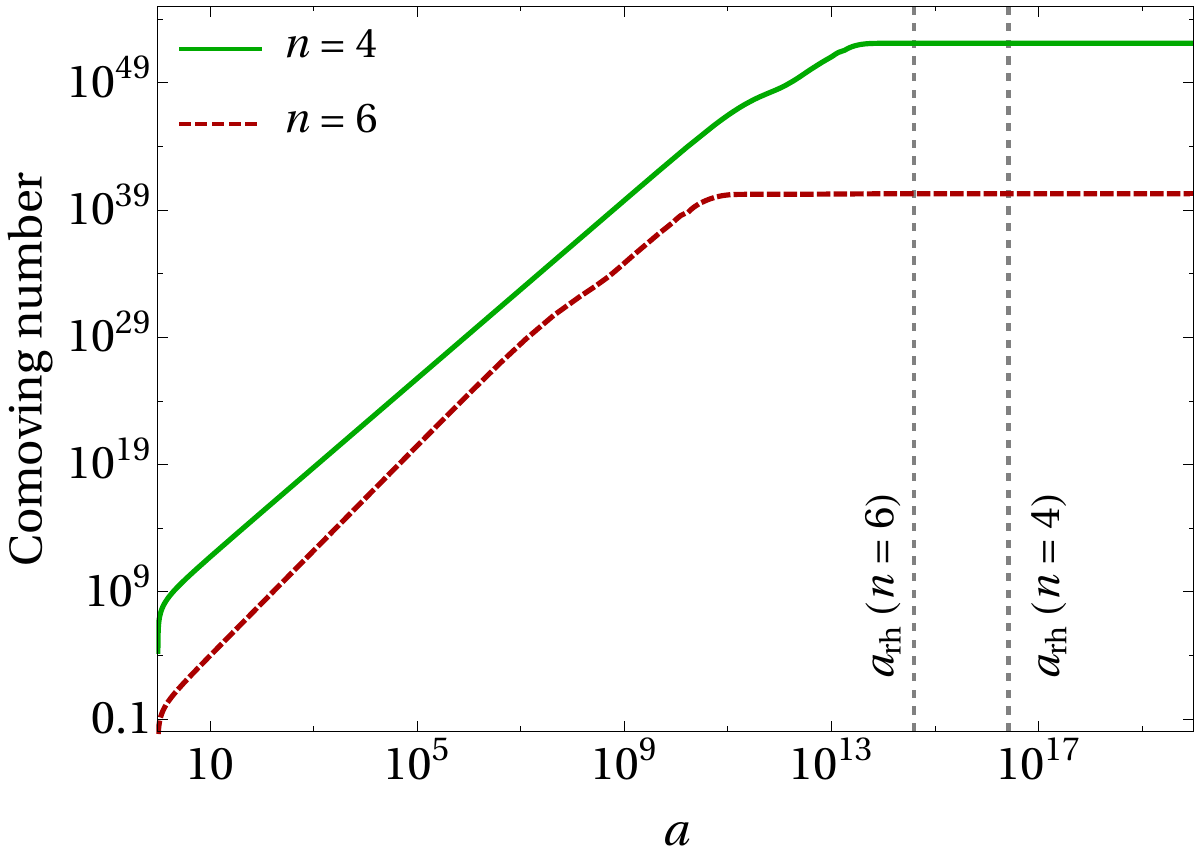}~\includegraphics[scale=0.375]{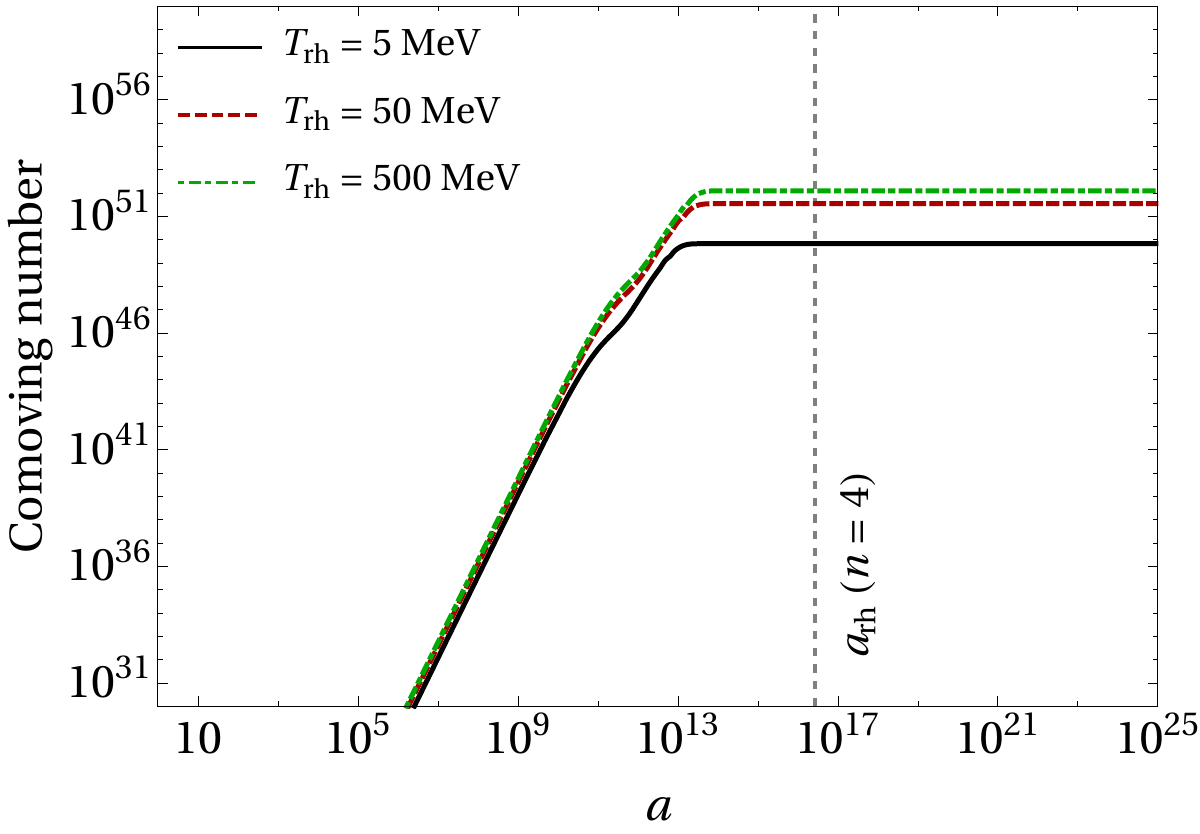}\\[10pt]
\includegraphics[scale=0.375]{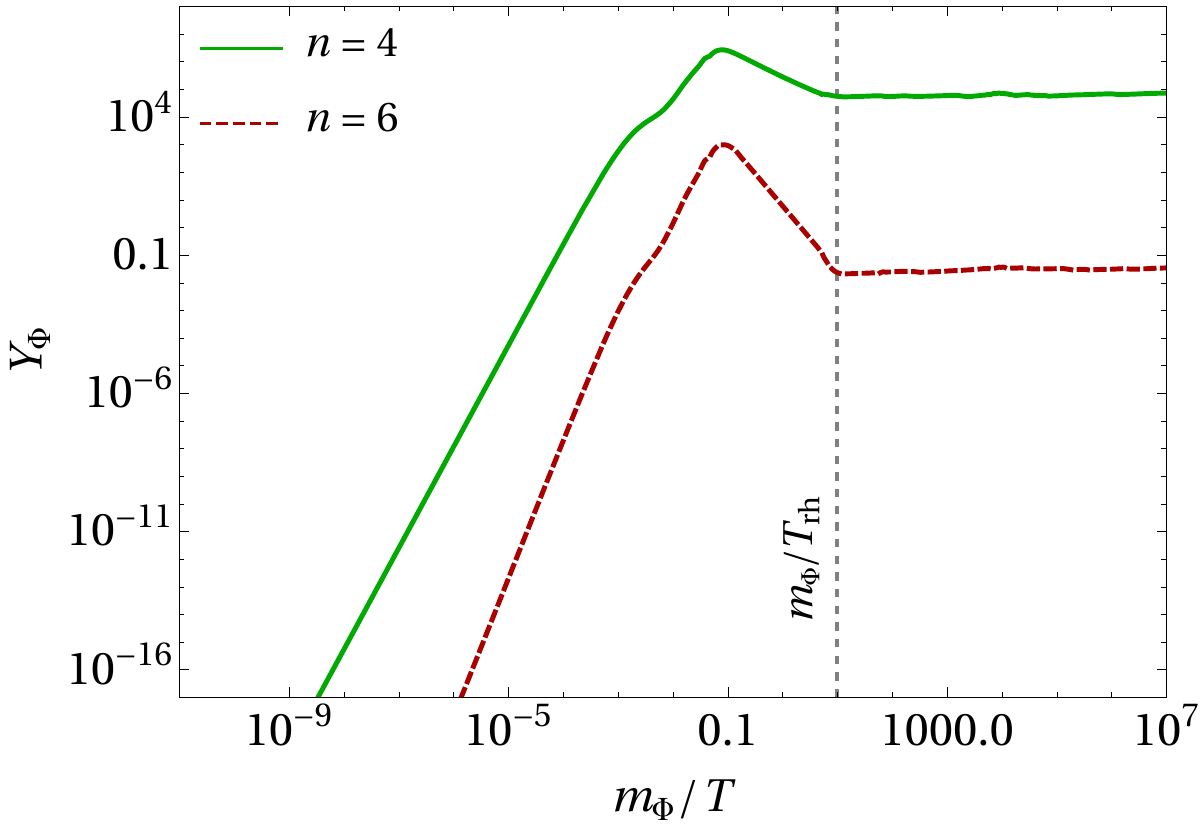}~\includegraphics[scale=0.375]{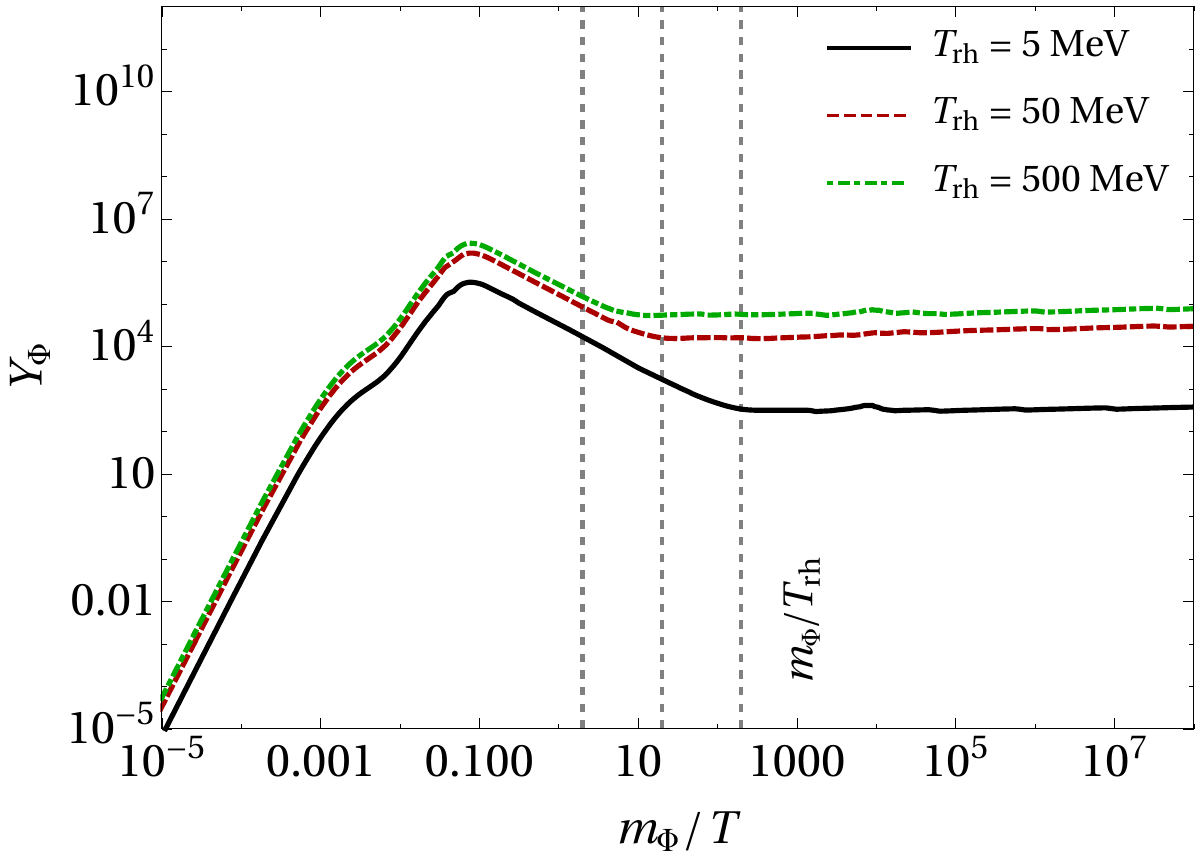}
\caption{Same as Fig.~\ref{fig:ODPhi_FOyield}, but considering freeze-in production during reheating. We have fixed $\{m_\Phi,\,\Trh,\,\lNP\}=\{1\,\text{GeV},\,0.1\,\text{GeV},\,2\,\text{TeV}\}$ for left panel and $\{n,\,m_\Phi,\,\lNP\}=\{4,\,1\,\text{GeV},\,2\,\text{TeV}\}$ for right panel.}
\label{fig:ODPhi_FIyield}
\end{figure}
\subsubsection{Freeze-in during reheating}
If the DM abundance is
initially negligible and the production cross section is sufficient small that DM remains out of chemical equilibrium with the SM bath $(\Ndm\ll N_{\rm eq})$, then DM is produced via freeze-in from the SM bath via 2-to-2 annihilation. In this case, depending on the DM mass, two cases may arise: (i) For $\mdm\ll\Trh$, the DM is produced at the end of reheating, with a DM comoving number
\begin{align}\label{eq:Ndm-FI1}
& \frac{\Ndm(\arh)}{\arh^3}\simeq\frac{g_{\rm dm}^2\,\mathcal{C}^2\,\zeta(3)^2}{\pi^4}\,\frac{\Trh^6}{\lNP^2\,\Hrh}\,\left(\frac{\Trh}{\lNP}\right)^k\,(n+2)
\nonumber\\&
\times\begin{dcases}
\frac{2}{3\,(k+2-4n)}\,\left[1-\left(\frac{a_I}{\arh}\right)^\frac{6n-15k-3}{n+2}\right]\,, & \text{for bosonic decay}\,,\\[10pt]
\frac{2}{3\,(3k-4n+14)}\,\left[1-\left(\frac{a_I}{\arh}\right)^\frac{12n-9k-42}{2n+4}\right]\,, & \text{for bosonic scattering}\,,
\end{dcases}
\end{align}
where we have considered a relativistic equilibrium number density $n_{\rm eq}=\mathcal{C}\,g_{\rm dm}\,\zeta(3)\,T^3/\pi^2$, with $\mathcal{C}=1\,(3/4)$, for bosonic (fermionic) DM. Also, the DM number at the beginning of reheating is set to zero or negligibly small, i.e., $\Ndm(a_I)\simeq0$; (ii) on the other hand, if $\Trh\ll\mdm\ll\Tmax$, then DM is produced {\it during} reheating. In this case one finds,
\begin{align}\label{eq:Ndm-FI2}
& \frac{\Ndm(\adm)}{\arh^3}\simeq\frac{g_{\rm dm}^2\,\mathcal{C}^2\,\zeta(3)^2}{\pi^4}\,\frac{\Trh^6}{\lNP^2\,\Hrh}\,\left(\frac{\Trh}{\lNP}\right)^k\,(n+2)
\nonumber\\&
\times\begin{dcases}
\left(\mdm/\Trh\right)^{k+2-4n}\,\frac{2}{3\,(k+2-4n)}\,\left[1-\left(\frac{a_I}{\adm}\right)^\frac{6n-15k-3}{n+2}\right]\,, & \text{for bosonic decay}\,,\\[10pt]
\left(\mdm/\Trh\right)^\frac{4n-3k-14}{3}\,\frac{2}{3\,(3k-4n+14)}\,\left[1-\left(\frac{a_I}{\adm}\right)^\frac{12n-9k-42}{2n+4}\right]\,, & \text{for bosonic scattering}\,,
\end{dcases}
\end{align}
where, $\adm=\arh\,\left(\Trh/\mdm\right)^{1/\xi}$. The final yeld at the end of reheating is then given by $Y(\arh)= Y(\adm)\,\left[\left(\gss(\mdm)/\gss(\Trh)\right)^{1/3}\,\left(\mdm/\Trh\right)\right]^3\,\left(\adm/\arh\right)^3$, considering the entropy injection between $\adm$ and $\arh$. Note that, in both cases, we see a strong dependence on $\Trh$, reflecting the typical UV freeze-in nature of the DM yield. For DM production during reheating, one can derive an upper bound on the effective scale $\Lambda_{\rm NP}$, below which the DM may thermalize before the completion of reheating. This bound is obtained by equating the interaction rate, $n_{\rm eq}\,\langle\sigma v\rangle$, with the Hubble expansion rate during reheating, 
\begin{align}\label{eq:lambda}
& \Lambda_{\rm NP}<\left[\frac{g_{\rm dm}\,M_P}{2\sqrt{2}\,\pi^{3/2}}\,\mdm^{3/2}\,T^{\frac{3}{2}+k-\frac{3n}{(2+n)\xi}}\,T_{\rm rh}^{\frac{3n}{(2+n)\xi}-2}\,
e^{-\mdm/T}\right]^{\frac{1}{k+2}}\,.
\end{align}
Evaluating this condition at $T = \mdm=1$ GeV yields $\lNP\lesssim 3.6\times 10^5$ GeV, for $n=4$ and $\Trh=10$ GeV, considering reheating via bosonic decay. For reheating via bosonic scattering this bound is more stringent, $\lNP\lesssim 1.6\times 10^4$ GeV. This condition guarantees that the interaction rate exceeds the Hubble expansion rate at some stage during reheating, thereby allowing the DM to reach thermal equilibrium.

The evolution of the DM yield during reheating, shown as a function of the scale factor (or equivalently the temperature), is illustrated in Fig.~\ref{fig:ODPhi_FOyield} for freeze-out production, and in Fig.~\ref{fig:ODPhi_FIyield} for freeze-in production. In all cases we consider $\mathcal{O}_{D\Phi}$ as the only operator contributing to DM abundance. We have considered both perturbative decay and scattering of inflaton condensate present at the same time, in order to solve the full BEQ numerically.  In the {\it freeze-out} case, the characteristic behavior is clearly visible: the DM yield initially tracks its equilibrium value and subsequently departs from it once the interaction rate drops below the Hubble expansion rate during reheating, leading to freeze-out. For fixed values of the DM mass, $\lNP$, $\Trh$, and $k$, the freeze-out occurs earlier for larger values of $n$. This behavior can be understood from the fact that increasing $n$ leads to a faster Hubble expansion at a fixed temperature (scale factor) during reheating, since $\mathcal{H}\propto T^{3n/\xi\,(n+2)}$. As a result, for a given interaction rate, the condition $\Gamma\simeq\mathcal{H}$ is satisfied at an earlier epoch, causing freeze-out to occur sooner. Moreover, an earlier freeze-out implies that the subsequent dilution due to entropy injection acts over a longer period, thereby reducing the final DM abundance. A similar effect is observed when $\Trh$ is reduced: a longer reheating phase enhances the dilution, leading to a smaller relic abundance. For {\it freeze-in} production, increasing $n$ again suppresses the final DM yield. In this case, a larger $n$ causes the reheating phase to complete more rapidly (as $\mathcal{H}^2\propto\rp\propto a^{-6n/(n+2)}$), reducing the time available for DM production and consequently lowering the final abundance. Finally, contrary to the behavior in the RD era, after chemical freeze-out, the comoving number density for WIMPs is diluted due to injection of entropy, until $T\sim\Trh$. The same dilution occurs for FIMPs after their freeze-in. It is important to mention here that for other operators, qualitatively, we checked that the yields follow the same pattern since it is controlled dominantly by the background cosmology.

A summary of the parameter space for the $\mathcal{O}_{f\Phi}$ operator is shown in Fig.~\ref{fig:Of3_relic-scan}, for two benchmark choices $n=4$ and $n=6$, displayed in the upper and lower panels, respectively. Here we restrict our analysis to reheating through the perturbative {\it decay} of the inflaton into bosonic final states. This choice is motivated by the fact that reheating via scattering leads to a higher SM bath temperature than the decay-dominated scenario [cf.~Fig.~\ref{fig:rho-T}]. As a result, freeze-in DM production is significantly enhanced in the scattering case. Reproducing the observed relic abundance would then require a substantially larger $\lNP$, thereby suppressing collider accessibility. Therefore, in order to retain the collider prospects, we focus exclusively on DM production during reheating via perturbative decay of the inflaton into bosonic final states. We first focus on the contours that satisfy the observed DM abundance, shown by the green, orange and blue points corresponding to $\Trh=\{0.05,\,0.5,\,10\}$ GeV, respectively. For a fixed DM mass, increasing $\lNP$ causes the DM to depart from thermal equilibrium and its production mechanism transitions from freeze-out to freeze-in, as discussed earlier around Eq.~\eqref{eq:lambda}. Furthermore, increasing $\Trh$ requires a larger value of $\lNP$ to reproduce the observed DM abundance. This behavior reflects the UV-dominated nature of freeze-in production: a higher $\Trh$ enhances DM production, which must be compensated by a larger $\lNP$ to avoid overabundance. Consequently, smaller $\Trh$ values are more favorable for collider production prospects. As noted earlier, the effective description remains valid only when $\lNP$ is the highest scale of the theory. Therefore, for a given $\Trh$, not all values of $\lNP$ are allowed. For $\Trh=\{0.05,\,0.5,\,10\}$ GeV, corresponding $\lNP$ must lie above $\Tmax\simeq\{768,\, 3930,\, 36561\}$ GeV for $n=4$, and $\lNP>\Tmax\simeq\{30,\, 197,\, 2371\}$ GeV for $n=6$. For the same reason, the gray shaded region is excluded since it corresponds to $\lNP<m_\Phi$, violating the validity of the effective field theory description. Additional shaded regions arise from various observational constraints, which will discuss in the following sections. It is worth noting that, for a fixed $\Trh$, a larger value of $n$ leads to a smaller $\Tmax$, as evident from the right panels of Fig.~\ref{fig:rho-T}. As a result, for $n=6$ one can accommodate smaller values of $\lNP$ while still satisfying the condition $\lNP > \Tmax$, thereby improving the observational prospects. For the operator $\mathcal{O}_{D\Phi}$, the viable parameter space is shown in Fig.~\ref{fig:ODPhi_relic-scan}, for the same set of $\Trh$ as before. Note that, in this case, the contours show a resonant peak at $\mdm=m_Z/2$, which appears due to $Z$-mediated DM annihilation channel. As $n$ increases, the Hubble expansion rate at a fixed temperature becomes larger. Consequently, for $n=6$, the DM--SM interaction departs from equilibrium earlier than in the $n=4$ case. When $\Tfo > \Trh$ ($\afo<\arh$), the DM abundance is further subject to entropy dilution due to the decay of the inflaton. This dilution can be estimated as: $\Delta S = S(a_{\rm fo})/S(a_{\rm rh})\propto \left(\Tfo/\Trh\right)^{\frac{9n+16}{3n+6}}$.
For a fixed $\Trh$, a larger $n$ therefore leads to a stronger dilution effect. As a result, in order to reproduce the observed relic abundance via freeze-out, a larger value of $\Lambda_{\rm NP}$ is required for $n=6$ compared to $n=4$ such that the the interaction rate becomes smaller to counter the effect of entropy dilution. In contrast, for freeze-in, the earlier decoupling combined with the subsequent entropy dilution suppresses the final DM abundance, since DM is produced from the thermal bath. Achieving the correct relic density in this case therefore necessitates a smaller $\Lambda_{\rm NP}$ as $n$ increases. 

We shall now move on to the discussion on the origin of various non-collider constraints appearing in our summary plots given by Figs.~\ref{fig:Of3_relic-scan} and \ref{fig:ODPhi_relic-scan}. 
\section{Non-collider searches \& constraints}
\label{sec:noncollider}
\subsection{Direct search}
Direct detection~\cite{PhysRevD.31.3059,Lewin:1996,Jungman:1995df} is one of the key dedicated DM search experiments, offering a laboratory-based probe of DM–matter interactions within the Milky Way. The goal is to measure rare DM scattering events off target nuclei in ultra-low-background environments. In conventional experiments targeting WIMPs with masses above a few GeV, the expected signal manifests as nuclear recoils. Despite decades of increasingly sensitive searches, no conclusive DM signal has been observed; instead, these experiments have placed stringent upper limits on DM–SM interaction cross-sections.

The nucleon matrix elements at zero momentum transfer are given by~\cite{Shifman:1978zn,Drees:1993bu},
\begin{align}
& \langle\mathcal{N}\big|m_q\,\bar{q}q\big|\mathcal{N}\rangle = m_\mathcal{N}\,f_{T_{q}}^{(\mathcal{N})}\,,\qquad\qquad
\langle\mathcal{N}\big|m_Q\,\bar{Q}Q\big|\mathcal{N}\rangle = \frac{2}{27}\,m_\mathcal{N}\,f_{T_{G}}^{(\mathcal{N})}\,,
\end{align}
where $q=\{u,d,s\}$ and $Q=\{c,b,t\}$ denote the light and heavy quark flavours, respectively, and $\mathcal{N}\in\{n,p\}$ represents the neutron or proton. The quantities $f_{T_{q}}^{(\mathcal{N})}$ parametrize the scalar quark content of the nucleon, while
\begin{align}
f_{T_{G}}^{(\mathcal{N})} = 1-\sum_q f_{T_{q}}^{(\mathcal{N})}
\end{align}
accounts for the gluonic contribution induced through heavy-quark loops. An analogous relation holds for operators involving the gluon field-strength tensor. In particular, the nucleon matrix element of the gluon operator reads \cite{Bishara:2017pfq,Hoferichter:2017olk},
\begin{align}
& \langle \mathcal{\mathcal{N}}\big|\alpha_s\,G_{\mu\nu}\,G^{\mu\nu}\big|\mathcal{N}\rangle
= -\frac{8\pi}{9}\,m_\mathcal{N}\,f_{T_{G}}^{(\mathcal{N})}\,,
\end{align}
where the result follows from the QCD trace anomaly. In the present analysis we have utilized the form factors present in {\texttt micrOMEGAs}~\cite{Alguero:2023zol}. Using the matrix
elements, the effective DM-nucleon spin independent (SI) cross section can be expressed as,
\begin{align}\label{eq:sig-DM-nuc}
& \sigma_{\rm SI}\simeq\frac{\mu_{\mathcal{N}\Phi}^2}{\pi}\,\left[Z\,f_p+(A-Z)\,f_n\right]^2\,,
\end{align}
where,
\begin{align}
\frac{f_\mathcal{N}}{m_\mathcal{N}}=\sum_q\,f_{T_{q}}^{(\mathcal{N})}\,\frac{\lambda_q}{m_q}+\frac{2}{27}\,\sum_Q\,f_{T_{G}}^{(\mathcal{N})}\,\frac{\lambda_Q}{m_Q}-\frac{8\pi}{9\alpha_s}\,\lambda_G\,f_{T_{G}}^{(\mathcal{N})}\,,    
\end{align}
with $\lambda_i$'s are operator dependent coefficients and $\mu_{\mathcal{N}\Phi}$ is the DM-nucleus reduced mass. Note that, none of the operators give rise to spin-dependent direct detection cross-section at the tree-level.

In Fig.~\ref{fig:Of3_relic-scan}, we display the most stringent constraints on DM–nucleon scattering from current experiments, namely PandaX-4T~\cite{PandaX:2024qfu} and LZ-2025~\cite{LZ:2024zvo}, along with the projected sensitivity of the future XLZD experiment~\cite{XLZD:2024nsu}. It is important to emphasize that these experiments probe different DM mass ranges. In particular, PandaX-4T is primarily sensitive to DM masses in the range $1 \lesssim \mdm \lesssim 10$ GeV, whereas XLZD (projected) and LZ-2025 cover a broader region of $10 \lesssim \mdm \lesssim 10^4$ GeV. As a result, the corresponding exclusion limits in Fig.~\ref{fig:Of3_relic-scan} appear in distinct DM mass intervals. Furthermore, as evident from Eq.~\eqref{eq:sig-DM-nuc}, for a fixed DM mass, increasing $\lNP$ suppresses the scattering cross-section, thereby weakening the experimental bounds. This leads to stringent constraints on the freeze-out (WIMP) parameter space, typically disfavoring $\lNP\lesssim\mathcal{O}(\text{1 TeV})$, while the freeze-in (FIMP) region remains largely unconstrained, depending on the DM mass and reheating temperature $\Trh$. For $\Trh=0.05$ GeV, PandaX-4T rules out freeze-in parameter space for DM in the mass range $0.1\lesssim\mdm\lesssim 10$ GeV,  while LZ-2025 excludes a significant portion of the FIMP parameter space for $\Trh = 10$ GeV, for DM mass $\gtrsim 10$ GeV.

Since in the EFT description the DM couples universally to all SM states, the DM–electron scattering channel cannot be neglected. However, it is well known that such scattering typically provides stringent constraints for sub-GeV DM. This arises because the small mass of the electron, together with the inelastic nature of the scattering with bound electrons (in detector materials), allows DM particles to transfer a significant fraction of their kinetic energy to the electron. As a result, even DM with masses as small as $\sim 1$ MeV can produce observable ionization signals. Following~\cite{Essig:2011nj,Essig:2013vha,Essig:2015cda}, the DM–electron scattering cross section can be expressed in a model-independent manner in terms of a reference cross section, $\bar{\sigma}_e$, and a DM form factor:
\begin{align}
& \bar{\sigma}_e \equiv \frac{\mu_{e\Phi}^2}{16\pi\,\mdm^2\,m_e^2}\,\left|\overline{\mathcal{M}(q)}\right|_{e\Phi}^2\Bigg|_{q^2=\alpha_e^2\,m_e^2}\,,
\end{align}
with
\begin{align}
& \left|\overline{\mathcal{M}(q)}\right|_{e\Phi}^2
= \left|\overline{\mathcal{M}(q)}\right|_{e\Phi}^2\Bigg|_{q^2=\alpha_e^2\,m_e^2}\times \left|F_{\rm DM}(q)\right|^2\,.
\end{align}
Here, $\bar{\sigma}_e$ corresponds to the non-relativistic elastic DM–electron scattering cross section evaluated at a fixed momentum transfer $q=\alpha_e\,m_e$, where $\alpha_e$ denotes the fine-structure constant. The quantity $\left|\overline{\mathcal{M}(q)}\right|_{e\Phi}^2 \propto 1/\Lambda_{\rm NP}^4$ represents the spin-averaged squared matrix element for DM–electron scattering. Once again, different DM-electron experiments probe complementary regions of the DM mass. Current experiments such as DAMIC-M~\cite{DAMIC-M:2025luv} are sensitive to masses in the range $1 \lesssim \mdm \lesssim 10^3$ MeV, while PandaX-4T~\cite{PandaX:2022xqx} covers approximately $0.05 \lesssim \mdm \lesssim 5$ GeV. Future experiments like Oscura~\cite{Oscura:2023qik} are expected to extend sensitivity to sub-MeV DM masses. As in previous cases, smaller values of $\Lambda_{\rm NP}$ lead to stronger experimental constraints, thereby potentially excluding regions of parameter space consistent with the observed relic abundance from freeze-out. Thus, present and future constraints from DM-nucleon and DM-electron experiments are capable of constraining not only WIMP scenario, but the FIMP case as well.    
\subsection{Indirect detection}
\subsubsection{From CMB}
The observation of secondary products from DM annihilation in astrophysical environments provides a powerful means to infer, or constrain in the absence of a signal, key DM properties such as its mass and annihilation cross section. These constraints, however, depend on the uncertain distribution of DM within the chosen astrophysical targets. A detailed analysis shows that, when the CMB is used as the target, the dominant contribution from DM annihilation arises at redshift $z \sim 600$~\cite{2012PhRvD..85d3522F}. At this epoch, DM has not yet significantly clustered into structures, making the CMB a particularly clean and competitive probe for indirect detection. Consequently, the bounds are largely insensitive to typical astrophysical uncertainties such as halo density profiles, sub-halo distributions, or the minimum halo mass. Additional systematic effects impacting CMB constraints have also been studied~\cite{Galli:2013dna,Weniger:2013hja}. To derive CMB constraints on the parameter of interest $\lNP$, one defines the `annihilation parameter' as,
\begin{eqnarray}
p_{\rm ann} (z=600) = \sum_j f_j(z, \mdm)\,\frac{\langle \sigma v \rangle_j (z)}{\mdm}\,,
\end{eqnarray}
where $\langle \sigma v \rangle_j (z)$ denotes the thermally averaged annihilation cross section into the $j^{\rm th}$ channel at redshift $z$, and $f_j(z, \mdm)$ represents the fraction of injected energy that is deposited into the plasma at redshift $z$~\cite{Liu:2016cnk}. The parameter $p_{\rm ann}$ is constrained by Planck TT, TE, EE, and lowP data as~\cite{Planck:2015fie},
\begin{eqnarray}\label{eq:pann}
p_{\rm ann} < 4.1 \times 10^{-28} \, \frac{\text{cm}^3}{\text{s GeV}} \quad \text{at 95\% C.L.}
\end{eqnarray}
This already rules out thermal s-wave annihilation cross sections for $\mdm\lesssim 10$ GeV~\cite{Galli:2009zc,Slatyer:2009yq,2012PhRvD..85d3522F,2011PhRvD..84b7302G,Liu:2016cnk}. 

Now, in the case of p-wave annihilation, relevant for the present scenario, the velocity dependence of the cross section induces a strong suppression at recombination, when DM is highly non-relativistic. Since CMB constraints are primarily sensitive to annihilation around this epoch, the resulting bounds are considerably weaker than those for s-wave processes. To quantify this effect, we consider the operator $\mathcal{O}_{f\Phi}$, for which
\begin{align}
\langle\sigma v\rangle_j=\frac{m_\Phi^2\,v^2}{4\pi\,\lNP^4}\,\mathcal{C}_j\,,
\end{align}
where $\mathcal{C}_j \simeq 2/3$ for annihilation into $e^\pm$, and $v$ denotes the dark matter velocity at the relevant epoch. The annihilation parameter then becomes
\begin{align}
p_{\rm ann}
=\sum_j f_j\,\frac{1}{m_\Phi}\left(\frac{m_\Phi^2\,v^2}{4\pi\,\lNP^4}\,\mathcal{C}_j\right)
=\frac{m_\Phi\,v^2}{4\pi\,\lNP^4}\sum_j f_j\,\mathcal{C}_j
\equiv \frac{m_\Phi\,v^2}{4\pi\,\lNP^4}\,f_{\rm eff}\,.
\end{align}
Using Eq.~\eqref{eq:pann}, we obtain the bound,
\begin{align}
\lNP \gtrsim 220\,\text{GeV}\,\left(f_{\rm eff}\,m_\Phi\,v^2\right)^{1/4}\,.
\end{align}
At redshift $z \sim 600$, the DM is already non-relativistic, leading to a highly suppressed velocity. Substituting a representative value we find,
\begin{align}
\lNP \gtrsim 7\times 10^{-2}\,\text{GeV}\,
\left[\frac{m_\Phi}{100\,\text{GeV}}\,
\left(\frac{v}{10^{-8}\,c}\right)^2\right]^{1/4},
\end{align}
where we have taken $f_{\rm eff}=1$ to obtain a conservative estimate. Evidently, due to the $v^{1/2}$-dependence, the constraint on $\lNP$ is extremely weak.
\subsubsection{Other indirect searches}
Constraints on DM annihilation cover a broad range of masses and are derived from a variety of astrophysical observations. For light DM with masses around $\mathcal{O}(100)$ MeV, X-ray measurements from NuSTAR, Suzaku, INTEGRAL, and XMM-Newton have been used to constrain p-wave suppressed annihilation into $e^+e^-$ final states~\cite{Cirelli:2023tnx}. In this context, INTEGRAL imposes an upper limit of $\langle\sigma v\rangle \lesssim 10^{-27}\,\text{cm}^3/\text{s}$ at $\mdm \simeq 100$ MeV, while NuSTAR provides stronger sensitivity for $\mdm \simeq 20$ MeV. For heavier DM, a joint analysis of gamma-ray data from the MAGIC Cherenkov telescopes and the Fermi Large Area Telescope (LAT), focusing on dwarf satellite galaxies, constrains the annihilation cross section over the mass range $10~\text{GeV}$ to $100~\text{TeV}$, depending on the annihilation channels~\cite{MAGIC:2016xys}. Furthermore, measurements of cosmic-ray antimatter by AMS-02~\cite{PhysRevLett.110.141102,AGUILAR20211} provide some of the most stringent bounds on the DM annihilation cross section for masses in the range $10\,\text{GeV} \lesssim \mdm \lesssim 1\,\text{TeV}$~\cite{Giesen:2015ufa,Jin:2015sqa,Evoli:2015vaa,Cuoco:2016eej,Cui:2016ppb,DiMauro:2021qcf,Calore:2022stf}. In principle, DM annihilation or decays (or other processes) could produce any SM particles. Most of those SM
particles will subsequently decay on short timescales. The signatures we can hope to detect are the stable particles
at the end of those decay chains, {\it viz.,} electrons, positrons, protons, antiprotons, photons and neutrinos.

In our setup, the thermally averaged DM pair-annihilation cross-sections into the relevant SM final states are given by  
\begin{align}
& \langle\sigma v\rangle\simeq\dfrac{m_\Phi^2\,v^2}{4\pi\,\lNP^4}
\begin{dcases}
\dfrac{2\,N_c}{3}\,\left(1+\frac{m_f^2}{2\,m_\Phi^2}\right)\,\left(1-\dfrac{m_f^2}{m_{\Phi}^2}\right)^{1/2}\,, & \text{to quarks \& leptons}\,,    
\\[10pt]
1\,,\ & \text{to neutrinos}\,,
\end{dcases}
\end{align}
for the $\mathcal{O}_{f\Phi}$ operator, where $N_c=1\,(3)$ for leptons (quarks). The DM velocity dispersion is denoted by $v\sim\{10^{-3},\,10^{-5},\,10^{-8},\,10^{-2}\}\,c$ for the Milky Way halo, dwarf spheroidal galaxies, the recombination epoch, and galaxy clusters, respectively~\cite{Slatyer:2015jla,Pinzke:2011ek,Voit:2004ah}. For $\mathcal{O}_{D\Phi}$, the annihilation processes are mediated by the $Z$ boson, leading to  
\begin{align}
& \langle\sigma v\rangle\simeq\frac{v^2}{\lNP^4}
\begin{dcases}
\dfrac{m_Z^6}{12\,\pi}\,\dfrac{N_c\,(m_\Phi/m_Z)^2}{\left(4\,m_\Phi^2-m_Z^2\right)^2+\Gamma_Z^2\,m_Z^2}\left(2-2\,\cos\theta_w+\cos4\theta_w\right)\,,\quad \text{to quarks \& leptons}\,, \\[10pt]
\dfrac{m_Z^6}{6\pi}\,\dfrac{r^2\,\left(1-\cos^2\theta_w/r^2\right)^{3/2}}{\left(4\,m_\Phi^2-m_Z^2\right)^2+\Gamma_Z^2\,m_Z^2}\,\left(4\,r^4+20\,r^2\,\cos^2\theta_w+3\cos^4\theta_w\right)\,, \quad \text{to}\,W^\pm\,,
\end{dcases}
\end{align}
where $\theta_w$ denotes the Weinberg angle and $r\equiv m_\Phi/m_Z$. In the first line, the final states are assumed to be massless for simplicity. As anticipated, the thermally averaged cross-sections for both operators exhibit an explicit $v^2$ dependence, indicating p-wave suppression. Consequently, while thermal relics with velocity-independent (s-wave) annihilation cross-sections are strongly constrained by current experiments and observations up to masses of order $\sim$ TeV, the present scenario remains largely unconstrained due to this p-wave suppression. To illustrate this quantitatively, for a DM mass of 100 GeV, the AMS-02 experiment sets an upper bound $\langle\sigma v\rangle_{\text{DM}\,\text{DM}\to e^+\,e^-}\lesssim 6\times 10^{-27}\,\text{cm}^3\,\text{s}^{-1}$. In comparison, we obtain $\langle\sigma v\rangle_{\Phi\Phi^\star\to e^+\,e^-}\simeq\{6.2\times 10^{-33},\,1.3\times 10^{-51}\}\,\text{cm}^3\,\text{s}^{-1}$ for $\{\mathcal{O}_{f\Phi},\,\mathcal{O}_{D\Phi}\}$, respectively, taking $\lNP=1$ TeV and $v\sim 10^{-3}\,c$. Similarly, to compare with the INTEGRAL bound on MeV-scale DM, discussed in the beginning of this subsection, we find $\langle\sigma v\rangle_{\Phi\Phi^\star\to e^+\,e^-}\simeq\{10^{-37},\,2\times 10^{-56}\}\,\text{cm}^3\,\text{s}^{-1}$ for $\{\mathcal{O}_{f\Phi},\,\mathcal{O}_{D\Phi}\}$, respectively, for a DM mass of 100 MeV and $\lNP=1$ TeV. Notably, even for smaller values of $\lNP$, these bounds remain weak because of $v^2$-dependence. Therefore, the predictions of our model lie safely below current experimental limits. Similar conclusions hold for other indirect detection constraints as well.
\subsection{Supernova bounds}
The supernova (SN) bound is based on the fact that if particles beyond the Standard Model are sufficiently light, they can be produced in astrophysical environments and potentially escape, leading to additional cooling. Such an energy loss could spoil the agreement between theoretical cooling predictions and observational data. Since the SN core contains a large abundance of electron-positron ($e^\pm$) pairs and nucleons ($\mathcal{N}$), these particles provide important production channels and scattering processes for DM inside the SN core. Furthermore, during the core-collapse process of SN, a proto-neutron star (PNS) is formed, where the extremely hot and dense environment can also lead to a significant enhancement of the muon population~\cite{Bollig:2017lki, Fischer:2020vie}. Consequently, the relevant production processes of DM inside the SN core are,
\begin{align*}
& e^+\,e^- \to \Phi\,\Phi^\star\,,\qquad\qquad \nu\,\overline{\nu} \to \Phi\,\Phi^\star\,, \qquad\qquad \mu^+\,\mu^- \to \Phi\,\Phi^\star\,,\qquad\qquad \mathcal{N}\,\mathcal{N} \to \mathcal{N}\,\mathcal{N}\,\Phi\,\Phi^\star\,,
\end{align*}
where the last one corresponds to a bremsstrahlung process. 

Once produced, the DM particles can freely stream out of the SN provided their mean free path $\bar{\lambda}_f$ satisfies the optical depth condition~\cite{Dreiner:2003wh, Dreiner:2013mua}
\begin{align}
& \int_{r_0}^{R_c} \frac{dr}{\bar{\lambda}_f} \leq \frac{2}{3}\,,
\end{align}
which determines whether the DM produced at a radius $r_0$ can escape from a supernova core of radius $R_c = \mathcal{O}(10)\,\text{km}$. The thermal average mean free path is given by~\cite{Manzari:2023gkt, DeRocco:2019jti}
\begin{align}\label{eq:freetream}
& \bar{\lambda}_f =\langle\lambda_f\rangle\approx \frac{\langle v_{\Phi}^{}\rangle}{\sum_{j}\,n_{j}\langle_{}\sigma^{}_{\Phi j}\,v_{\rm{m}\text{\o}\rm{l}}\rangle}\,,
\end{align}
where $n_j$ denotes the number density of target particles in the SN medium with $j\in\{\mathcal{N},\mu,e,\nu\}$, and $\sigma_{\Phi j}^{}$ are the scattering cross sections for the processes
\begin{align*}
& \Phi\,e \to \Phi\,e\,, \qquad\qquad\qquad\Phi\nu\to\Phi\nu\,,\qquad\qquad\qquad\Phi\mu\to\Phi\mu\,,\qquad\qquad\qquad\, \Phi\,\mathcal{N} \to \Phi\,\mathcal{N}\,.
\end{align*}
In the expression for $\bar{\lambda}_f$, $v_{\rm{m}\text{\o}\rm{l}}$ is the relative M\o ller velocity~\cite{Gondolo:1990dk} of the DM written as
\begin{align}\label{eq:vmoller}
& v_{\rm m\text{\o} l}=\sqrt{|\vec{v}_\Phi^{}-\vec{v}_j^{}|^2-|\vec{v}_\Phi^{}\times\vec{v}_j^{}|^2}=\frac{\sqrt{(p_\phi \cdot p_j)^2 - m_\Phi^2 m_j^2 }}{E_\Phi E_j}\,,    
\end{align}
and $\langle v_\Phi\rangle$ denotes the average velocity of the DM particles\footnote{It is important to note that, in the non-relativistic regime where DM scatters off non-relativistic particles such as nucleons and muons (when $m_\mu\gg m_{\Phi}$) inside the SN core, one has $v_{\rm{m}\text{\o}\rm{l}}\simeq v_{\rm rel}\approx v_{\Phi}^{}$ ($v_{\rm rel}\equiv|\vec{v}_\Phi^{}-\vec{v}_{k}|$ is the relative velocity of the DM) and consequently the free-streaming length becomes effectively independent of $v_\Phi$. However, this approximation is no longer valid for relativistic scattering processes of DM with $e^\pm$ and/or $\nu(\bar{\nu})$, for which $v_{\rm{m}\text{\o}\rm{l}}\neq v_{\rm rel}\neq v_{\Phi}^{}$. As a result, Eq.~\eqref{eq:freetream} retains an explicit dependence on $v_\Phi$.},
\begin{align}
    \langle v_\Phi^{}\rangle=\frac{\mathfrak{g}_{\Phi}^{}}{n_{\Phi}^{}}\int\frac{dp^3_{\Phi}}{(2\pi)^3}f_{\Phi}^{}\frac{p_{\Phi}^{}}{E_{\Phi}^{}}\,,
\end{align}
where $\mathfrak{g}_\Phi$, $n_\Phi$ and $f_\Phi$ represent the internal DoFs, the number density and the DM distribution function.  Furthermore, $\langle_{}\sigma^{}_{\Phi j}\,v_{\rm{m}\text{\o}\rm{l}}\rangle$ is the thermal average cross-section evaluated over the complete initial-state phase space, and is defined as
\begin{align}
    \langle_{}\sigma^{}_{\Phi j}\,v_{\rm{m}\text{\o}\rm{l}}\rangle=\frac{\mathfrak{g}_{\Phi}^{}\,\mathfrak{g}_{k}^{}}{n_{\Phi}^{}n_{k}^{}}\int\frac{dp^3_{\Phi}}{(2\pi)^3}f_{\Phi}^{}\int\frac{dp^3_{k}}{(2\pi)^3}f_{k}^{}\,\sigma(p_{\Phi}^{},p_{k}^{})v_{\rm{m}\text{\o}\rm{l}}\,.
\end{align}
To constrain the  SN energy loss due to the DM emission, we employ the Raffelt criterion~\cite{Raffelt:1996wa},
\begin{align}
& \dot{\mathcal{E}} \equiv \dot{\mathcal{E}}^{\rm max} < 10^{19}\,\text{erg g}^{-1}\,\text{s}^{-1}\simeq 7.3\times 10^{-27}\,\text{GeV}\,,
\end{align}
 $\dot{\mathcal{E}}$ denotes the emissivity, i.e., the energy emitted by the SN per unit time and per unit mass. Although a bound based on the total emitted energy is more robust, the Raffelt criterion provides a simpler and widely used estimate. Following the prescription of~\cite{Dreiner:2003wh}, the emissivity can be written as
\begin{align}\label{eq:emissivity}
& \dot{\mathcal{E}}(m_\Phi,\,T_c,\,\eta) = \frac{1}{\rho_{\rm SN}}\int \frac{d^3p_1}{(2\pi)^3}\, f_1 \int \frac{d^3p_2}{(2\pi)^3}\,f_2\, (E_1 + E_2)\, \sigma \widetilde{v}_{\rm{m}\text{\o}\rm{l}}\,,
\end{align}
where $E_1 + E_2$ is the total energy of the incoming SM particles, $f_i=1/\left[e^{(E_i\pm\upmu_i)/T_c}+1\right]$ are the Fermi-Dirac distribution functions. Here, $\rho_{\rm SN}$ and $T_c$ denote the density and temperature of the SN core, respectively. The quantity $\sigma \widetilde{v}_{\rm{m\o l}}$ denotes the DM production cross section multiplied by the relative M{$\o$}ller velocity $\widetilde{v}_{\rm{m\o l}}$ of the colliding particles. The expression for $\widetilde{v}_{\rm{m\o l}}$ can be obtained from Eq.~\eqref{eq:vmoller} by replacing $\Phi$ with $\bar{j}$, where $\bar{j}\in\{\bar{\mu},\bar{e},\bar{\nu}\}$. The parameter $\eta \equiv \upmu/T_c$ denotes the degeneracy parameter, with $\upmu$ being the chemical potential of the incoming particles. In the above expression, we have neglected the Bose enhancement factors for the final-state DM particles. In case of DM production from the non-relativistic particle $\mu$ inside the SN core, the expression for emissivity Eq.~\eqref{eq:emissivity} can be written as
\begin{align}
\dot{\mathcal{E}}_{\mu}=&\frac{\mathfrak{g}_{\mu}^2}{16\pi^4\rho_{\rm SN}}\int_{m_{\mu}}^{\infty} dE_{1}\sqrt{E_{1}^2-m_{\mu}^2}\int_{m_{\mu}}^{\infty} dE_{2}\sqrt{E_{2}^2-m_{\mu}^2}(E_{1}+E_{2})\,f_{1}\,f_{2}
\nonumber\\&
\int_{-1}^{1} d\cos{\theta}\,s\,\sqrt{1-\frac{4m_{\mu}^2}{s}}\,\sigma(s)\,.
\end{align}
Here, we employ the following expression for the relative M{\o}ller velocity of the muon pair,
$\widetilde{v}_{\rm m\text{\o}l}=\left(s\sqrt{1-4m_{\mu}^2/s}\right)/2E_{1}E_{2}$. It should also be noted that the angular integration entering the annihilation cross section depends explicitly on the scattering angle $\theta$ between the incoming muon and anti-muon through the Mandelstam variable $s$, evaluated in the SN core frame as $s = 2\left(m_\mu^2 + E_1 E_2 - |\mathbf{p}_1||\mathbf{p}_2| \cos\theta \right).$ Furthermore, for dark matter production processes involving relativistic particles such as electrons and neutrinos in the SN core, the corresponding emissivity expression can be expressed in the form
\begin{align}
\dot{\mathcal{E}}_{\nu/e}=\frac{\mathfrak{g}_{\nu/e}^2}{8\pi^4\rho_{\rm SN}}\int_{m_{\nu/e}}^{\infty} dE_{1}E_{1}^2\int_{m_{\nu/e}}^{\infty} dE_{2}E_{2}^2(E_{1}+E_{2})f_{1}f_{2}\int_{-1}^{1} d\cos{\theta}\,\sigma(s)(1-\cos{\theta})\,,
\end{align}
where we use the expression for relative M\o ller velocity for neutrinos as $\widetilde{v}_{\rm{m}\text{\o}\rm{l}}=(1-\cos{\theta})$ and the expression for center of mass energy as $s = 2E_1 E_2\left( 1 -  \cos\theta \right).$ Below, we summarize the SN core parameters employed in the evaluation of the SN constraints on the effective operators $\mathcal{O}_{f\Phi}$ and $\mathcal{O}_{D\Phi}$. The corresponding input data are adopted from~\cite{Manzari:2023gkt}, where they obtained the data from the simulation of SN model SFHo-18.8~\cite{Garching}
\begin{equation}
\boxed{
\arraycolsep=1.4pt
\def\arraystretch{1.5}
\begin{array}{c}
\text{\bf SN core conditions at typical core radius $R_c\simeq10$ km} \\[2mm]
T_c \simeq 30~\mathrm{MeV}, \qquad
\rho_{\rm SN} \simeq 2\times10^{14}~\mathrm{g\,cm^{-3}}, \\[1mm]
\upmu_\mu \simeq 100~\mathrm{MeV}, \qquad
\upmu_e \simeq 130~\mathrm{MeV}, \\[1mm]
\upmu_{\nu_\mu} \simeq -10~\mathrm{MeV}, \qquad
\upmu_{\nu_e} \simeq 20~\mathrm{MeV}.
\end{array}
}
\label{eq:typicalPNS}
\end{equation}
Now, as it has been explained in~\cite{Friman:1979ecl} (and subsequently in~\cite{Dreiner:2003wh,Timmermans:2002hc}), In the non-relativistic soft-radiation limit, the polar-vector contribution to the $\mathcal{N}\,\mathcal{N}\to\mathcal{N}\,\mathcal{N}+ X$ matrix element vanishes at {\it lowest order} for purely central (Landau-type) nucleon--nucleon interactions, independent of the detailed form of the Landau parameters. This suppression arises from the spin-operator structure of the interaction together with conservation of the nucleon vector current, rather than from the specific nature of the emitted final-state particles. We thus drop this process for obtaining the SN bounds. 

As explained above, we restrict our analysis to the dominant 2-to-2 processes contributing to the emissivity and to obtain an approximate analytical estimate, we further assume Maxwell--Boltzmann distributions for the incoming particles. Under these assumptions, Eq.~\eqref{eq:emissivity} simplifies to  
\begin{align}\label{eq:emissivity2}
\dot{\mathcal{E}}\simeq 2\pi^2\,T \int ds\, s^2 \sigma(s)\, K_1\!\left(\sqrt{s}/T\right)\,,
\end{align}
where $K_i$ denotes the modified Bessel function of the second kind of order $i$. In the limit $s\gg m_\Phi^2$, the cross section scales parametrically as $\sigma(s)\sim s/\left(\pi\,\lNP^4\right)$. Substituting this into Eq.~\eqref{eq:emissivity2}, we obtain,  
\begin{align}
\dot{\mathcal{E}}\approx 3150\,\pi^2\,\frac{T^9}{\Lambda_{\rm NP}^4}\,.
\end{align}
Applying the Raffelt energy-loss criterion then yields the approximate lower bound  
\begin{align}
\Lambda_{\rm NP}\gtrsim 1.7\times10^4\,\text{GeV}\,
\left(\frac{T}{30\,\text{MeV}}\right)^{9/4},
\end{align}
where we take the reference temperature to be the supernova core temperature, $T=T_c\sim 30\,\text{MeV}$. Because of the non-renormalizable nature of the DM-SM interaction, the DM production at the SN core increases with the increase in the core temperature. Thus, the emissivity rises with $T_c$, and hence a larger $\lNP$ is required to satisfy the Raffelt criterion.
\subsection{Inflationary gravitational waves}
Another tantalizing observational aspect that the present scenario offers is through the blue-tilted spectrum of inflationary gravitational waves (GWs). During inflation, quantum fluctuations generically generate a nearly scale-invariant spectrum of tensor perturbations on super-Hubble scales. In the standard post-inflationary history, these modes re-enter the horizon during the radiation-dominated (RD) era, preserving this scale invariance. However, the presence of an early \textit{stiff} phase prior to radiation domination modifies this picture. In such a scenario, modes re-entering the horizon during the stiff epoch acquire a significant blue tilt, leading to an enhancement of the primordial gravitational wave (PGW) spectrum at high frequencies. This effect has been extensively studied in various contexts, see, e.g., Refs.~\cite{Giovannini:1998bp,Giovannini:1999bh,Riazuelo:2000fc,Seto:2003kc,Boyle:2007zx,Stewart:2007fu,Li:2021htg,Artymowski:2017pua,Caprini:2018mtu,Bettoni:2018pbl,Figueroa:2019paj,Opferkuch:2019zbd,Bernal:2020ywq,Caldwell:2022qsj,Gouttenoire:2021jhk,Haque:2021dha,Maity:2024cpq}. This feature can be exploited as a probe of our scenario. In the following, we briefly review how a background equation of state stiffer than radiation leads to a blue-tilted PGW spectrum.
\begin{figure}[htb!]
\centering
\includegraphics[scale=0.57]{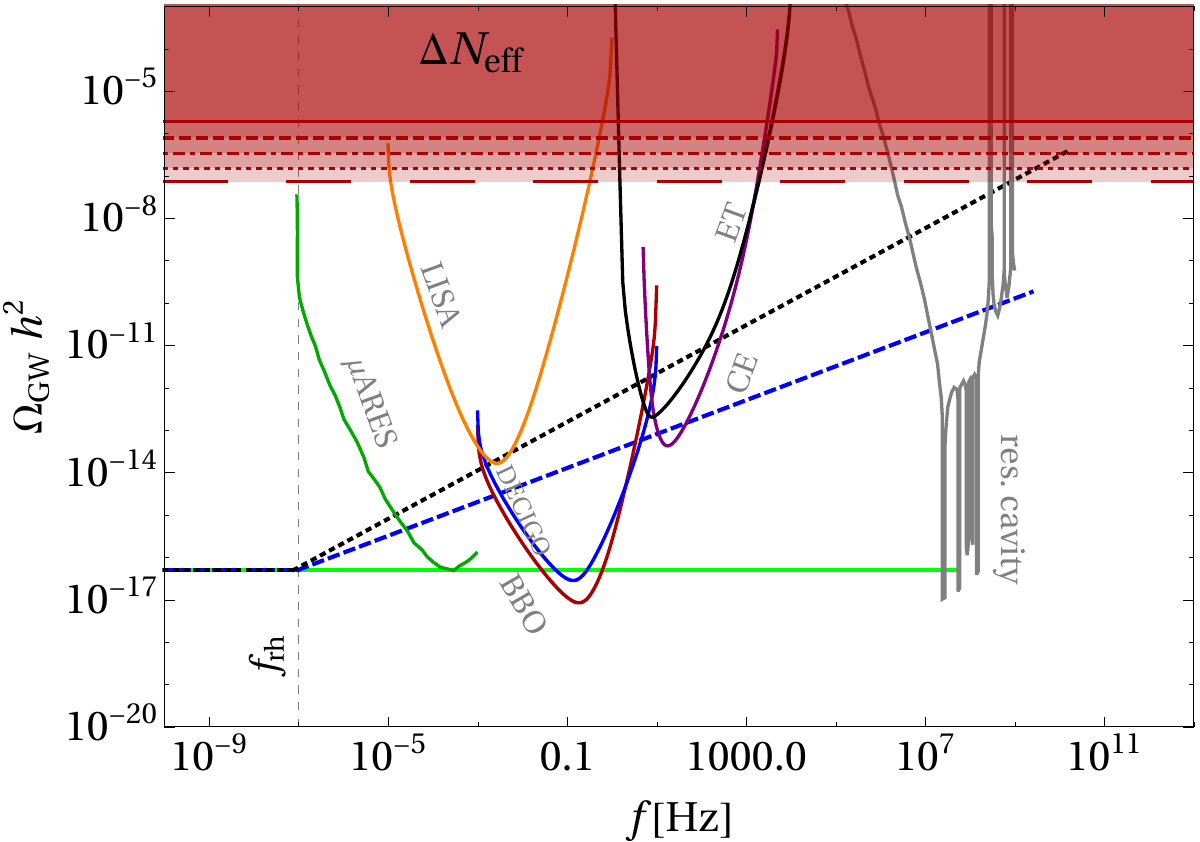}\\
\includegraphics[scale=0.57]{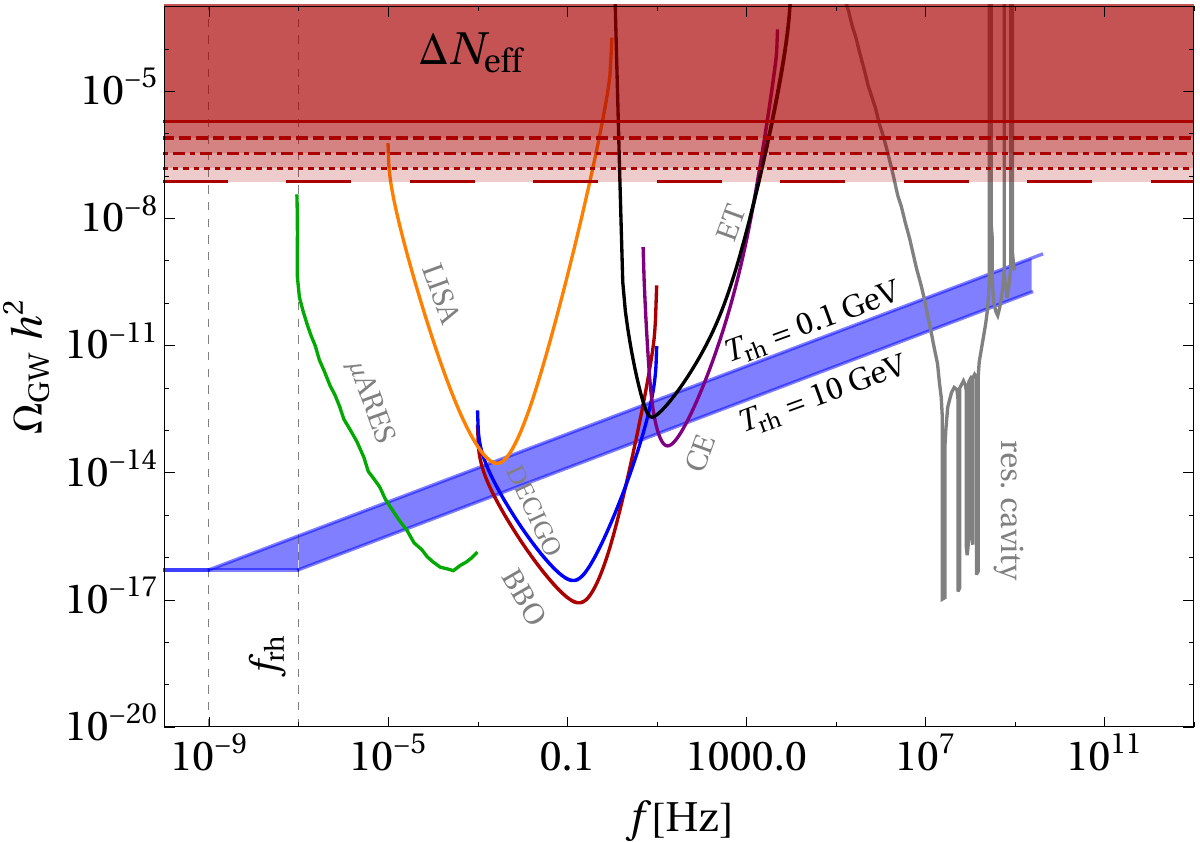}
\caption{Top: The PGW spectrum for $n=\{4,\,6,\,8\}$, shown via green solid, blue dashed and black dotted curves, respectively. We project sensitivity curves from several future experiments, for example, the Big Bang Observer (BBO)~\cite{Crowder:2005nr, Corbin:2005ny}, ultimate DECIGO (uDECIGO)~\cite{Seto:2001qf, Kudoh:2005as}, LISA~\cite{LISA:2017pwj}, $\mu$Ares~\cite{Sesana:2019vho}, the cosmic explorer (CE)~\cite{Reitze:2019iox}, the Einstein Telescope (ET)~\cite{Hild:2010id, Punturo:2010zz, Sathyaprakash:2012jk, Maggiore:2019uih}, and from resonant cavities~\cite{Herman:2022fau}. Here we have fixed $\Trh=10$ GeV, corresponding $f_{\rm rh}$ [cf. Eq.~\eqref{eq:frh}] is indicated via the vertical gray dashed line. The dark red shaded regions are discarded from bounds on $\DNeff$ (see text for details). Bottom: Same as top, but for $n=6$. The blue shaded region corresponds to different choices of $\Trh$ that can provide right DM abundance, with the boundaries fixed at $\Trh=10$ GeV (lower) and $\Trh=0.1$ GeV (upper), following Figs.~\ref{fig:Of3_relic-scan} and \ref{fig:ODPhi_relic-scan}.}
\label{fig:pgw}
\end{figure}
The PGW spectrum is defined as the energy density per logarithmic wavenumber interval,
\begin{equation}
\Omega_{\rm GW}(a,k)=\frac{1}{\rho_c}\frac{d\rho_{\rm GW}}{d\ln k}\,, 
\qquad \rho_c=3H^2 M_P^2\,.
\end{equation}
It can be written as~\cite{Maggiore:1999vm, Watanabe:2006qe, Saikawa:2018rcs, Caprini:2018mtu}
\begin{equation}
\Omega_{\rm GW}(a,k)=\frac{1}{12}\left(\frac{k}{aH}\right)^2 \mathcal{P}_{T,\rm prim}\,\mathcal{T}(a,k)\,.
\end{equation}
The primordial tensor spectrum is parametrized as~\cite{Saikawa:2018rcs, Caprini:2018mtu}
\begin{equation}
\mathcal{P}_{T,\rm prim}=r\,\mathcal{P}_\zeta(k_\star)\left(\frac{k}{k_\star}\right)^{n_T}\,,
\end{equation}
with $k_\star=0.05\,{\rm Mpc}^{-1}$, $\mathcal{P}_\zeta(k_\star)\simeq2.1\times10^{-9}$, and $n_T\simeq -r/8$. Given $r<0.036$, we set $n_T=0$. The transfer function is defined as~\cite{Boyle:2005se,Saikawa:2018rcs,Caprini:2018mtu},
\begin{equation}
\mathcal{T}(a,k)=\frac12\left(\frac{a_{\rm hc}}{a}\right)^2\,,
\end{equation}
where $a_{\rm hc}$ satisfies $a_{\rm hc}H(a_{\rm hc})=k$. At present,
\begin{equation}
\Omega_{\rm GW}(k)=\frac{1}{24}\left(\frac{k}{a_0 H_0}\right)^2 \mathcal{P}_{T,\rm prim}\left(\frac{a_{\rm hc}}{a_0}\right)^2\,.
\end{equation}
Accounting for horizon reentry,
\begin{align}\label{eq:ogw}
\Omega_{\rm GW}(a_{\rm hc}) \simeq &\ \Omega_\gamma^{(0)}\,\frac{\mathcal{P}_{T,\rm prim}}{24}\,\frac{g_*(T_{\rm hc})}{2}
\left(\frac{g_{*s}(T_0)}{g_{*s}(T_{\rm hc})}\right)^{4/3}
\nonumber\\
&\times
\begin{dcases}
\frac{g_*(T_{\rm rh})}{g_*(T_{\rm hc})}
\left(\frac{g_{*s}(T_{\rm hc})}{g_{*s}(T_{\rm rh})}\right)^{4/3}
\left(\frac{a_{\rm rh}}{a_{\rm hc}}\right)^{\frac{2(n-4)}{n+2}}\,, 
& a_I<a_{\rm hc}\le a_{\rm rh}\,,\\[10pt]
1\,, & a_{\rm rh}\le a_{\rm hc}\le a_{\rm eq}\,,
\end{dcases}
\end{align}
where $\Omega_\gamma^{(0)}=2.47\times10^{-5}h^{-2}$ is the present photon abundance and $a_{\rm eq}$ is the scale factor corresponding to matter-radiation equality $T_{\rm eq}\simeq 0.7$ eV. The GW frequency at present epoch is given by,
\begin{align}
f(a_{\rm hc})=\frac{k}{2\pi a_0}
=\mathcal{F}(g_*,g_{*s})\,\frac{T_0\sqrt{H_I M_P}}{M_P}\sqrt{H_r}
\times
\begin{dcases}
\left(\frac{a_{\rm rh}}{a_{\rm hc}}\right)^{\frac{2(n-1)}{n+2}}\,, 
& a_I<a_{\rm hc}\le a_{\rm rh}\,,\\[10pt]
\frac{a_{\rm rh}}{a_{\rm hc}}\,, 
& a_{\rm rh}\le a_{\rm hc}\le a_{\rm eq}\,,
\end{dcases}
\end{align}
with $H_r\equiv H(a_{\rm rh})/H_I$, where $H_I\simeq 4.4\times 10^{13}$ GeV, corresponding to the maximum value for the tensor-to-scalar ratio $r=0.035$~\cite{BICEP:2021xfz} and $\mathcal{F}(\gs,\,\gss)=\frac16 \sqrt{\frac{\gs(\Trh)}{10}} \left(\frac{\gss(T_0)}{\gss(\Trh)}\right)^\frac13\,\left(\frac{90}{\pi^2\,\gs(\Trh)}\right)^{1/4}$. Expressed in frequency,
\begin{align}
\Omega_{\rm GW}(f)\simeq &\ \Omega_\gamma^{(0)}\,\frac{\mathcal{P}_{T,\rm prim}}{24}\,\frac{g_*(T_{\rm hc})}{2}
\left(\frac{g_{*s}(T_0)}{g_{*s}(T_{\rm hc})}\right)^{4/3}\times
\nonumber\\&
\begin{dcases}
\frac{g_*(T_{\rm rh})}{g_*(T_{\rm hc})}
\left(\frac{g_{*s}(T_{\rm hc})}{g_{*s}(T_{\rm rh})}\right)^{4/3}
\left(\frac{f}{f_{\rm rh}}\right)^{\frac{n-4}{n-1}}\,, 
& f_{\rm rh}\le f<f_{\rm max}\,,\\[10pt]
1\,, & f_{\rm eq}\le f\le f_{\rm rh}\,,
\end{dcases}
\end{align}
where
\begin{equation}\label{eq:frh}
f_{\rm rh}=\mathcal{F}(g_*,g_{*s})\,T_0\,\sqrt{\frac{H_I\,H_r}{M_P}}\,,
\end{equation}
and
\begin{equation}\label{eq:fmax}
f_{\rm max}=f(a_I)=\frac{H_I}{2\pi}\frac{a_I}{a_0}\,.
\end{equation}
Modes with $f<f_{\rm rh}$ retain the primordial scaling (scale-invariant), while modes with $f_{\rm rh}<f<f_{\rm max}$ are enhanced for $n>4$, as $\Omega_{\rm GW}\propto f^{\frac{n-4}{n-1}}$, and modes with $f>\fmax$ are never created. In Fig.~\ref{fig:pgw} we show the PGW spectral energy density at present epoch, as a function of the frequency $f$, for $\Trh=10$ GeV. For $n=4$ (corresponding background EoS $w=1/3$) we see a scale-invariant spectrum following Eq.~\eqref{eq:ogw}, while the spectrum is blue-tilted for $n>4$. For a given $\Trh$ (and $H_I$) since $H(\arh)$ is also fixed, hence we see for all curves $f_{\rm rh}\simeq 9.8\times 10^{-8}$ Hz is also the same. However, $\fmax$ differs for different $n$-values since $\fmax\propto H_r^{(n+2)/(3n)}$, following Eq.~\eqref{eq:fmax}.

Since the energy density of PGWs redshifts like radiation, measurements of $\DNeff$, which quantify any excess relativistic degrees of freedom during BBN or CMB epochs, can be used to place an upper bound on the present-day GW abundance. Within the $\Lambda$CDM framework, the Planck legacy data yields $N_{\text{eff}} = 2.99 \pm 0.34$ at 95\% CL~\cite{Planck:2018vyg}, shown as the solid orange horizontal line in our plots. Including baryon acoustic oscillation (BAO) data tightens this to $N_{\text{eff}} = 2.99 \pm 0.17$. A combined BBN+CMB analysis gives $N_{\text{eff}} = 2.880 \pm 0.144$~\cite{Yeh:2022heq}, represented by the dashed line. Future CMB experiments such as CMB-S4~\cite{Abazajian:2019eic} and CMB-HD~\cite{CMB-HD:2022bsz} are expected to reach sensitivities of $\Delta N_{\text{eff}} \simeq 0.06$ and $0.027$, respectively, while next-generation missions like COrE~\cite{COrE:2011bfs} and Euclid~\cite{EUCLID:2011zbd} may further improve this to $\Delta N_{\text{eff}} \lesssim 0.013$. For frequencies in the range $f_{\text{BBN}} < f < f_{\max}$, where $f_{\text{BBN}}$ corresponds to modes entering the horizon at BBN, the constraint can be expressed as, $\Omega_{\text{GW}}\,h^2(f) \lesssim 5.62 \times 10^{-6}\,\Delta N_{\text{eff}}$~\cite{Boyle:2007zx,Kuroyanagi:2014nba,Caprini:2018mtu,Figueroa:2019paj}. Notably, this bound is frequency-independent. In Fig.~\ref{fig:pgw}, the corresponding exclusion is shown as a horizontal dark red shaded region, relevant for $n>8$, where the GW spectrum is highly blue-tilted and tends to exceed the $\DNeff$ limits inferred from BBN.

For $\Trh = 10~\text{GeV}$, we have already delineated the viable parameter space in Figs.~\ref{fig:Of3_relic-scan} and \ref{fig:ODPhi_relic-scan} for $\mathcal{O}_{f\Phi}$ and $\mathcal{O}_{D\Phi}$, respectively, considering $n=4$ and $n=6$ in each cases. As discussed earlier, non-collider constraints largely exclude the freeze-out region of the parameter space, while a substantial and viable region persists for the freeze-in scenario. In this context, a future detection of a stochastic gravitational wave (GW) signal in the frequency range $f \simeq [10^{-3},\,10^{-1}]~\text{Hz}$ by experiments such as $\mu$ARES, DECIGO, or BBO could be indicative of a bosonic reheating phase with $\Trh = 10~\text{GeV}$ and $n=4$. Such an observation would, in turn, provide a suggestive link to DM production via freeze-in through an effective dimension-6 operator. On the other hand, a broader GW detection spanning frequencies $f \simeq [10^{-3},\,10^{7}]~\text{Hz}$, potentially accessible to $\mu$ARES, LISA, DECIGO, BBO, CE, and ET, would instead point toward the existence of a pre-BBN stiff epoch. Beyond $n=6$, as one can see from Fig.~\ref{fig:pgw}, $\DNeff$ could further provide strong constraint. As shown in the bottom panel, for $n=6$, the $\Trh$ corresponding to right DM abundance (for a given $\lNP$ and $\mdm$), fall within the reach of the sensitivity of several future GW detectors, over a frequency range of mHz to GHz. Collectively, these would offer a complementary avenue to probe scenarios where DM is produced dominantly via freeze-in during bosonic reheating. It is important to emphasize, however, that such GW signatures do not constitute a unique probe of the scenario under consideration. Rather, they should be viewed as complementary to other searches for new physics, collectively enhancing our ability to test the underlying framework.   
\section{Collider searches \& constraints}
\label{sec:collider}
Collider searches for DM mainly rely on missing energy signals, since the DM particles escape the detector without interacting. The key strategy is to look for events with large missing transverse momentum ($\slashed{E}_T$) recoiling against visible particles such as jets, photons, or electroweak gauge bosons (mono-$X$ searches). A major strength of collider searches is their link to the cosmological origin of DM. If the collider-produced particle accounts for today’s relic abundance, collider measurements can be directly connected to the parameters that governed its production in the early Universe. Combining collider data with relic density constraints therefore provides a powerful test of the DM production mechanism and helps narrow down the viable parameter space~\cite{Barman:2024nhr,Barman:2024tjt,Bhattacharya:2025wef,Ghosh:2024boo,Ghosh:2024nkj,Ghosh:2025agw,Bernal:2025qkj,C:2026bqd}. This connection is particularly important when DM production depends on the reheating era, since the same interactions probed at colliders can also determine the relic abundance. In such scenarios, collider results can offer indirect insight into the reheating temperature and the thermal history of the Universe. Thus, collider searches probe not only DM particle properties but also aspects of early-Universe cosmology. In the following, we examine the relevant collider bounds together with their cosmological implications.
\subsection{Invisible meson decays}
Searches for invisible decays of light mesons constitute some of the most sensitive low-energy tests of new physics in the sub-GeV mass regime. Experimentally, invisible decays of pseudoscalar mesons have been investigated for multiple species: the NA62 experiment established the bound $\mathcal{B}(\pi^0\to\text{invisible})<4.4\times10^{-9}$ at 90\% C.L.\ using $4\times10^9$ tagged $\pi^0$ mesons~\cite{NA62:2020}, while BESIII reported $\mathcal{B}(\eta\to\text{invisible})<2.4\times10^{-5}$ at 90\% C.L.~\cite{ParticleDataGroup:2024cfk}. In the vector meson sector, BESIII searched for invisible decays of the $\omega$ and $\phi$ through $J/\psi\to V\eta$ and observed no excess, thereby setting $\mathcal{B}(\omega\to\text{invisible})<7.3\times10^{-5}$ and $\mathcal{B}(\phi\to\text{invisible})<1.7\times10^{-4}$ at 90\% C.L.~\cite{BESIII:2018}. Similarly, BaBar probed the bottomonium sector using $\Upsilon(3S)\to\pi^+\pi^-\Upsilon(1S)$ with an invisibly decaying $\Upsilon(1S)$, deriving the constraint $\mathcal{B}(\Upsilon(1S)\to\text{invisible})<3.0\times10^{-4}$ at 90\% C.L.~\cite{BaBar:2009}. Meanwhile, flavour-changing neutral current (FCNC) processes provide especially powerful probes for such searches because they are highly suppressed within the SM by the Glashow–Iliopoulos–Maiani (GIM) mechanism. The decay mode $K_L\to\pi^0\nu\bar\nu$ is among the theoretically cleanest channels for exploring new physics in the kaon sector, and the KOTO experiment has recently set the stringent direct limit $\mathcal{B}(K_L\to\pi^0\nu\bar\nu)<2.2\times10^{-9}$~\cite{KOTO2025}. In the $B$-meson sector, Belle-II places the bound $\mathcal{B}(B^0\to\pi^0\nu\bar\nu)<9\times10^{-6}$~\cite{ParticleDataGroup:2024cfk}.
\begin{table}[t]
\centering
\begin{tabular}{c c c c c }
\hline\hline
Operators & Decay modes & Flavour & $\lNP^{\rm max}$ (GeV) & $m_\Phi$ (GeV) \\ 
 & & & & \\
\hline
$\mathcal{O}_{f\Phi}$, $\frac{2m_Z^2}{g_Z g_V v^2}\mathcal{O}_{D\Phi}$ & $\Upsilon(1S) \to \text{inv}$ &  $(\bar{f}_3\gamma^\mu f_3)\,(\Phi^\star i \overleftrightarrow{\partial_\mu}\Phi)$ & 115 & $\lesssim$ 4.5\\
$\mathcal{O}_{f\Phi}$ & $K_L \to \pi^0 +  \text{inv}$ & $(\bar{f}_2\gamma^\mu f_1)\,(\Phi^\star i \overleftrightarrow{\partial_\mu}\Phi)$ & $ 5.2 \times 10^4$ & $\lesssim$ 0.18\\
$\mathcal{O}_{f\Phi}$ & $B^0 \to \pi^0 +  \text{inv}$ & $(\bar{f}_3\gamma^\mu f_1)\,(\Phi^\star i \overleftrightarrow{\partial_\mu}\Phi)$ & $6.8 \times 10^3$ & $\lesssim$ 2.5\\
\hline\hline
\end{tabular}
\caption{Lower limit on $\lNP$ and DM mass obtained from experimental limits on invisible decay branching fractions of vector and psuedoscalar mesons.}
\label{tab:inv_decay_bounds}
\end{table}

\begin{figure}
    \centering    \includegraphics[width=0.65\linewidth]{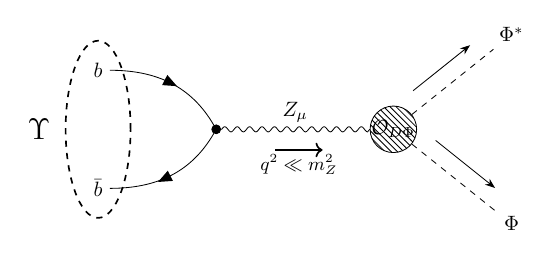}
    \caption{Upsilon invisible decay process generated by $\mathcal{O}_{D\Phi}$ operator}
    \label{fig:Upsilon_decayDphi}
\end{figure}

\begin{figure}
    \centering    \includegraphics[width=0.52\linewidth]{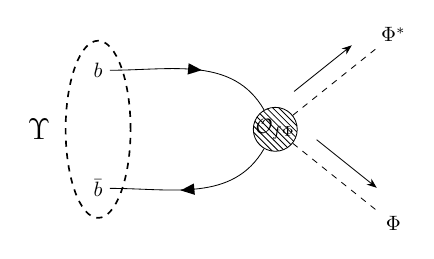}
    \caption{Upsilon invisible decay process generated by $\mathcal{O}_{f\Phi}$ operator}
    \label{fig:Upsilon_decayfphi}
\end{figure}

\begin{figure}
    \centering    \includegraphics[width=0.65\linewidth]{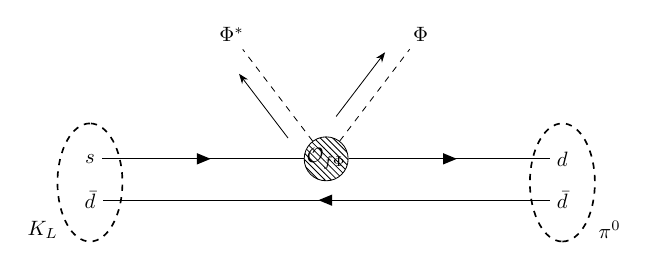}
    \caption{$K_L \to \pi^0 + \text{inv.}$  process generated by $\mathcal{O}_{f\Phi}$ operator}
    \label{fig:K_decay}
\end{figure}

\begin{figure}
    \centering    \includegraphics[width=0.65\linewidth]{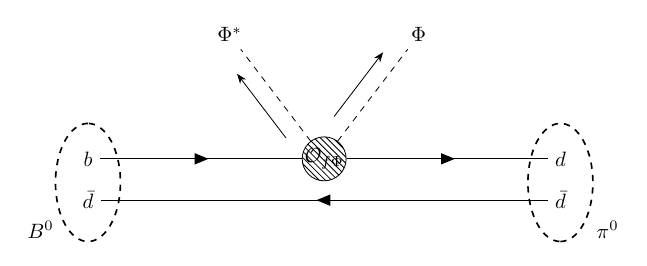}
    \caption{$B^0 \to \pi^0 + \text{inv.}$ process generated by $\mathcal{O}_{f\Phi}$ operator}
    \label{fig:B_decay}
\end{figure}

Despite this broad experimental landscape, vector and pseudoscalar mesons probe qualitatively distinct operators within the low-energy effective field theory. Our focus here is on the phenomenology of the dimension-6 effective operators,
\begin{eqnarray}
\mathcal{O}_{f\Phi} &=& \frac{1}{\Lambda^2_{\rm NP}}\sum_{i,j=1,2,3}\bigl(\bar{f}_i\,\gamma^{\mu} f_j\bigr) \bigl(\Phi^{\star} i\overleftrightarrow{\partial_\mu} \Phi\bigr) \ ,\nonumber \\ 
\mathcal{O}_{D\Phi} & = & \frac{1}{\Lambda^2_{\rm NP}} \bigl(H^{\dagger} i\overleftrightarrow{D_\mu} H\bigr)\bigl(\Phi^{\star} i\overleftrightarrow{\partial_\mu} \Phi\bigr) \nonumber \\ 
&\supset& \frac{g_Z}{2 m_Z^2}\frac{v^2}{\Lambda^2_{\rm NP}} \sum_{i=1,2,3}\bigl(\bar{f}_i\,\gamma^{\mu} ( g_V - g_A \gamma^5)f_i\bigr) \bigl(\Phi^{\star} i\overleftrightarrow{\partial_\mu} \Phi\bigr) \Bigg|_{q^2 \ll m_Z^2, \, \text{tree-level}}\,,
\end{eqnarray}
where $g_Z$, $g_V$ and $g_A$ are the respective $Z$-boson couplings with the SM Higgs and fermions. In general, the quark flavour structure of $\mathcal{O}_{f\Phi}$ can include flavour-violating quark currents, thereby inducing tree-level flavour changing neutral currents (FCNC). In contrast, processes generated by $\mathcal{O}_{D\Phi}$ are mediated through $Z$-boson exchange and therefore cannot produce tree-level FCNCs. For vector mesons with $J^{PC}=1^{--}$, the vector current matrix element $\langle 0|\bar{f}\gamma^\mu f|V(k,\varepsilon)\rangle = f_V m_V\varepsilon^\mu$ is non-zero, with the decay constant $f_V$ determined from the measured leptonic partial width. For pseudoscalar mesons $P,M$ ($J^{PC}=0^{-+}$), the relevant non-vanishing quark vector current matrix element is
\begin{eqnarray}
\langle P(p_P)|\,\bar{f}_i\gamma^\mu f_j\,|M(p_M)\rangle = f_+(s_{12})(p_M+p_P)^\mu \ ,
\end{eqnarray}
where $s_{12}=(p_M-p_P)^2$ and $f_+(s_{12})$ denotes the vector form factor. Although $\mathcal{O}_{D\Phi}$ can also generate the pseudoscalar axial-vector matrix element $\langle 0|\,\bar{f}_i\gamma^\mu \gamma^5 f_j\,|M\rangle$, the decay $M\to \Phi \Phi^{\star}$ vanishes for equal-mass scalar fields. Consequently, at $q^2\ll m_Z$, both $\mathcal{O}_{f\Phi}$ and $\mathcal{O}_{D\Phi}$ effectively contain the structure $(\bar{f}_i\gamma^\mu f_j)(\Phi^\star i\overleftrightarrow{\partial_\mu}\Phi)$, with the crucial distinction that $\mathcal{O}_{D\Phi}$ cannot generate flavour-violating processes at tree level because the Z boson does not mediate tree-level FCNCs. Therefore, decays of the form $M \to P + \text{invisible}$ constrain only $\mathcal{O}_{f\Phi}$, whereas invisible vector meson decays $V \to \text{invisible}$ constrain both $\mathcal{O}_{f\Phi}$ and $\mathcal{O}_{D\Phi}$. Nevertheless, both operators couple to vector and pseudoscalar meson sectors, and are constrained from experimental searches for $\omega$, $\phi$, $J/\psi$, and $\Upsilon(1S)$ invisible decays. Vector meson decay like $\Upsilon(1S)\to \text{invisible}$ constraints both $\mathcal{O}_{D\Phi}$ and flavour diagonal parts of $\mathcal{O}_{f\Phi}$. The processes are shown in Fig.~\ref{fig:Upsilon_decayDphi} and Fig.~\ref{fig:Upsilon_decayfphi}, respectively. While, $\mathcal{O}_{f\Phi}$ with flavour violating quark currents is constrained by decays like $K_L \to \pi^0 +\text{invisible}$ and $B^0 \to \pi^0 +\text{invisible}$. These processes are shown in Fig.~\ref{fig:K_decay} and Fig.~\ref{fig:B_decay}. Below, we compute the decay widths and constraint $\lNP$ for the above operator. 
\subsubsection{Vector meson decays}
For vector mesons $J^{PC}=1^{--}$, the relevant hadronic matrix element, obtained through matching to the chiral Lagrangian at $\mathcal{O}(p^2)$, is given by
\begin{equation}
  \langle 0\,|\,\bar{q}\gamma^{\mu}q\,|\,V(k,\epsilon)\rangle
  = f_V m_V \epsilon^{\mu}\,,
  \label{eq:me}
\end{equation}
where $\epsilon^{\mu}$ denotes the transverse polarization vector of the meson and $f_V$ is the corresponding vector decay constant, determined from the experimentally measured leptonic partial width. The resulting decay amplitude then takes the form,
\begin{equation}
  i\mathcal{M}
  = \frac{i}{\Lambda^2_{\rm NP}}\,f_V m_V\,\epsilon^{\mu}(p_1-p_2)_{\mu}.
  \label{eq:amp}
\end{equation}
Provided that $m_\Phi < m_V/2$, the vector meson can undergo the invisible decay process $V\to\Phi\Phi^{\star}$. After summing over the physical polarization states, the squared amplitude becomes
\begin{equation}
\sum_\lambda|\mathcal{M}|^2
= \frac{f_V^2 m_V^2}{\Lambda^4_{\rm NP}}\left[-(p_1-p_2)^2+\frac{\left((p_1-p_2)\cdot k\right)^2}{m_V^2}\right]\,.
\end{equation}
The second term identically vanishes because, for on-shell final-state particles of equal mass, $(p_1-p_2)\cdot k = p_1^2-p_2^2 = 0$. Using the kinematic identity $-(p_1-p_2)^2 = m_V^2-4m_\Phi^2$, and incorporating the standard two-body phase-space factor, one obtains the decay width
\begin{equation}  \Gamma(V\to\Phi\,\Phi^{\star})
= \frac{f_V^2\,m_V^3}{48\pi\,\Lambda^4_{\rm NP}}    \left(1-\frac{4\,m_\Phi^2}{m_V^2}\right)^{\!\!3/2}\,.
\label{eq:width_V}
\end{equation}
Among invisible decays of vector mesons, the most stringent constraints emerge from $\Upsilon(1S)$, primarily due to the strong mass dependence of the decay width. For our numerical analysis, we adopt $f_V = 715$ MeV for $\Upsilon(1S)$ and present the strongest lower bounds on $\lNP$ derived from the BaBar limit $\mathcal{B}(\Upsilon(1S)\to\text{invisible})<3.0\times10^{-4}$ at 90\% C.L.~\cite{BaBar:2009} in Tab.~\ref{tab:inv_decay_bounds}. Although the limits are comparatively weaker than $\omega$, and $\phi$ meson decays, its larger mass can enhance sensitivity. The resulting constraints obtained from $\Upsilon$ decay is illustrated in Fig.~\ref{fig:Of3_relic-scan}.
\subsubsection{Pseudoscalar meson decays}
We investigate here the three-body flavour-changing neutral current decays $K_L\to\pi^0\Phi\Phi^{\star}$ and $B^0\to\pi^0\Phi\Phi^{\star}$, both mediated by the operator $\mathcal{O}_{f\Phi}$. In contrast, processes induced by $\mathcal{O}_{D\Phi}$ are absent at tree level, since the $Z$ boson does not mediate tree-level FCNC transitions. Experimentally, the final-state signature of a single $\pi^0$ recoiling against missing energy is indistinguishable from the SM channels $K_L\to\pi^0\nu\bar{\nu}$ and $B^0\to\pi^0\nu\bar{\nu}$. Consequently, the most sensitive searches for these rare SM processes can be directly reinterpreted as constraints on $\mathcal{O}_{f\Phi}$. The KOTO experiment at J-PARC currently provides the world’s most stringent direct limit, $\mathcal{B}(K_L\to\pi^0\nu\bar{\nu})<2.2\times10^{-9}$~\cite{KOTO2025}, while Belle~II sets the bound $\mathcal{B}(B^0\to\pi^0\nu\bar{\nu})<9\times10^{-6}$~\cite{ParticleDataGroup:2024cfk}, both at 90\% C.L. Translating these limits into lower bounds on $\Lambda_{\rm NP}$ as a function of $m_\Phi$, we find that kaon decays probes $\lNP\gtrsim 50$ TeV for $m_\Phi\lesssim 100$ MeV, whereas the $B^0$ channel retains sensitivity up to $m_\Phi\approx 2.6$ GeV. The relevant scalar and hadronic matrix elements are
\begin{align}  & \langle\Phi(k_1)\,\Phi^{\star}(k_2)|  \,\Phi^\star i\overleftrightarrow{\partial_\mu}\Phi\,|0\rangle=(k_1-k_2)_\mu, \nonumber \\&\langle P(p_P)\,|\,\bar{f}_i\gamma^\mu f_j\,|\,M(p_M)\rangle
=f^{MP}_+(s_{12})\,(p_M+p_P)^\mu\,, 
\end{align}
where $s_{12}=(k_1+k_2)^2$ denotes the invariant mass-squared of the $\Phi\Phi^{\star}$ system, and $f^{MP}_+(q^2)$ is the corresponding vector form factor. The scalar form factor is neglected here due to its suppressed contribution. Contracting these two currents gives the squared matrix element
\begin{equation}
  |\mathcal{M}|^2
  = \left(\frac{1}{\Lambda^2_{\rm NP}}\right)^{\!2}
    f^{MP}_+(s_{12})^2\,(s_{13}-s_{23})^2,
  \label{eq:M2_FCNC}
\end{equation}
where $s_{13}=m_M^2+m_P^2+2m_\Phi^2-s_{12}-s_{23}$ and $s_{23}=(p_P+k_2)^2$ are the standard Dalitz variables. The corresponding three-body decay width is given by,
\begin{equation}
  \Gamma = \frac{1}{256\,\pi^3 m_M^3}
  \left(\frac{1}{\Lambda^2_{\rm NP}}\right)^{\!2}
  \int_{4m_\Phi^2}^{(m_M-m_P)^2}\!\!d s_{12}
  \int_{s_{23}^-}^{s_{23}^+}\!\!ds_{23}\;
  f^{MP}_+(s_{12})^2\,(s_{13}-s_{23})^2,
  \label{eq:width_FCNC}
\end{equation}
with phase-space boundaries
$s_{23}^{\pm}=(E_2^*+E_3^*)^2 -\bigl(\sqrt{E_2^{*2}-m_\Phi^2}\mp\sqrt{E_3^{*2}-m_P^2}\bigr)^2$. The multidimensional phase-space integration is performed numerically using the {\tt VegasFlow} Monte Carlo framework~\cite{Carrazza:2020rdn, vegasflow_package}.

For the $K_L\to\pi^0\Phi\Phi^{\star}$ channel, we adopt the vector form factor
\begin{equation}
f^{K\pi}_+(s) = \frac{1}{1-\lambda^\prime_+ s/m_{\pi^+}^2},
\end{equation}
where $m_{\pi^+}$ is the charged pion mass and $\lambda^\prime_+ = 0.0245$~\cite{Bijnens:1994me, Bijnens:2003uy,Boito:2010me}, normalized such that $f^{K\pi}_+(0)=1$. For $B^0\to\pi^0\Phi\Phi^{\star}$, we employ the BCL parametrization of the vector form factor~\cite{Bourrely:2008za},
\begin{equation}
  f^{B\pi}_+(s) = \frac{1}{1-s/M_{B^*}^2}
  \sum_{k=0}^{K-1} b_k\,\bigl[z(s)^k - (-1)^{k-K}\tfrac{k}{K}\,z(s)^K\bigr],
  \label{eq:ffB_BCL}
\end{equation}
where $M_{B^*}=5.325\GeV$ is the lowest-lying $B^*$ pole and
\begin{equation}
  z(s) = \frac{\sqrt{t_+-s}-\sqrt{t_+-t_0}}{\sqrt{t_+-s}+\sqrt{t_+-t_0}},
  \quad
  t_+ = (M_B+M_\pi)^2,
  \quad
  t_0 = (M_B+M_\pi)\!\left(\sqrt{M_B}-\sqrt{M_\pi}\right)^{\!2} \ .
  \label{eq:z}
\end{equation}
We truncate the expansion at $K=3$ and use the coefficients $b_0 = 0.421,\, b_1 = -0.476,\, b_2 = -0.399$, corresponding to the normalization $f_+^B(0)=0.252$~\cite{Bourrely:2008za}. Despite its comparatively restricted phase space, the $K_L$ channel delivers the strongest bound at low $m_\Phi$, primarily because the KOTO upper limit on $\Gamma_{K_L}^\mathrm{tot}\times\mathcal{B}$ is substantially more restrictive than the corresponding Belle~II bound. Kinematically, the kaon mode closes at $m_\Phi=(m_{K_L}-m_{\pi^0})/2\approx 181$ MeV, whereas the $B^0$ channel remains open up to $m_\Phi\approx 2.56$ GeV. Together, these two decay modes offer highly complementary sensitivity: kaon decays dominate in the light-scalar regime and probe exceptionally high $\lNP$, while $B$ meson decays extend this reach deep into the GeV-scale scalar mass window. The resulting maximal lower bounds on the $\lNP$ derived from current experimental constraints are summarized in Tab.~\ref{tab:inv_decay_bounds}, while the full excluded parameter space is presented in Fig.~\ref{fig:Of3_relic-scan}.
\subsection{Constraints from \texorpdfstring{$K^0-\bar{K^0}$}{K0-K0bar} Oscillations}
\label{sec:kaon_oscillations}
The neutral kaon system has been one of the most sensitive probes of flavour-changing neutral currents and CP violation. The experimental measurable, $K_L$--$K_S$ mass difference $\Delta m^{exp}_K = m_{K_L} - m_{K_S} = (3.484\pm0.006)\times10^{-15}\,\mathrm{GeV}$ and indirect CP-violation parameter $|\varepsilon_K| = (2.228\pm0.011)\times10^{-3}$~\cite{PDG}, constraint the new-physics efficiently. In extensions of the SM, with new degrees of freedom at a scale $\Lambda_{\rm NP}\gg m_W$, integrating out the heavy states produces a $\Delta S=2$ effective Hamiltonian~\cite{Gabbiani:1996hi,Ciuchini:1998ix},
\begin{equation}
  \label{eq:Heff}  \mathbb{H}_\mathrm{eff}^{\Delta S=2}=\sum_{i=1}^{5} C_i(\Lambda_{\rm NP})\,O_i+\sum_{i=1}^{3}\tilde{C}_i(\Lambda_{\rm NP})\,\tilde{O}_i
  + \mathrm{h.c.}\,,
\end{equation}
where the $O_i$ and $\tilde{O}_i$ are the eight independent four-quark operators that can appear beyond the SM, defined in SUSY basis~\cite{Buras:2000if},
\begin{eqnarray}
  \label{eq:basis}
  O_1 &=& [\bar{s}^\alpha\gamma^\mu P_L d^\alpha]
         [\bar{s}^\beta\gamma_\mu P_L d^\beta], \quad
  O_2 = [\bar{s}^\alpha P_L d^\alpha]
         [\bar{s}^\beta P_L d^\beta], \quad
  O_3 = [\bar{s}^\alpha P_L d^\beta]
         [\bar{s}^\beta P_L d^\alpha], \nonumber\\
  O_4 &=& [\bar{s}^\alpha P_L d^\alpha]
         [\bar{s}^\beta P_R d^\beta], \quad
  O_5 = [\bar{s}^\alpha P_L d^\beta]
         [\bar{s}^\beta P_R d^\alpha]\,,
\end{eqnarray}
together with the parity-conjugate operators $\tilde{O}_{1,2,3}$. Here $\alpha, \beta$ are the summed over colour indices. The total new-physics contribution to the Kaon mass splitting and CP violation is determined by the real and imaginary parts of the off-diagonal element ($M_{12}$) of the effective Hamiltonian in Eq.~\eqref{eq:Heff}, given by,
\begin{equation}
  \label{eq:M12}
  M_{12} = \frac{1}{2m_K}  \langle\bar{K^0}|\mathbb{H}_\mathrm{eff}^{\Delta S=2}|K^0\rangle\,,
\end{equation}
and the two physical observables are related to $M_{12}$ by,
\begin{equation}
  \label{eq:obs}
  \Delta m_K \propto \mathrm{Re}\,M_{12},
  \qquad
  \varepsilon_K \propto \mathrm{Im}\,M_{12}\,.
\end{equation}
Since we are working with a real Wilson coefficients the constraint comes from $\Delta m_K$, while a complex phase
in the flavour-violating coupling would get constrained by
$\varepsilon_K$ bound, which is typically an order of magnitude more stringent.

\begin{figure}
    \centering    \includegraphics[width=0.65\linewidth]{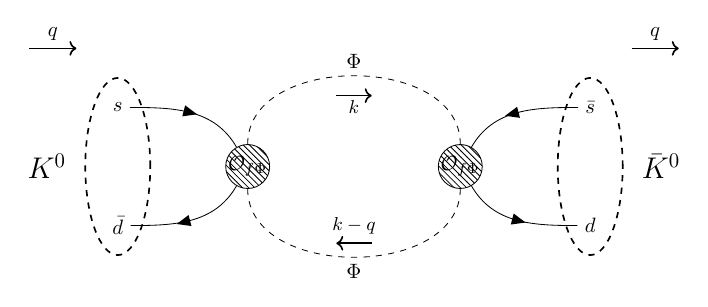}
    \caption{Kaon oscillation generated by $\mathcal{O}_{f\Phi}$.}
    \label{fig:kaon_oscillation}
\end{figure}

Here we derive the constraint imposed by $K^0$-$\bar{K^0}$ mixing on the cut-off scale of a theory containing the dimension-six operator,
\begin{equation}
  \label{eq:op}
  \mathcal{O} = \frac{1}{\Lambda_{\rm NP}^2}
  \sum_{q=u,d,s} (\bar{q}\,\gamma^\mu q)\,
  \bigl(\Phi^\star\, i\overleftrightarrow{\partial_\mu} \Phi\bigr),
\end{equation}
which in the presence of flavour-violating interactions the quark current acquires off-diagonal elements. 
Kaon oscillation requires a $\Delta S=2$ transition, which is generated at one loop by two insertions of the above operator connected by a $\Phi$ loop as shown in Fig.\ref{fig:kaon_oscillation}. 
The amplitude for the process, now becomes, 
\begin{equation}
  \label{eq:amp2}
  i\mathcal{M} = \frac{1}{\Lambda_{\rm NP}^4}
  \,(\bar{d}^\alpha\,\gamma^\mu s^\alpha)(\bar{s}^\beta\,\gamma_\nu d^\beta) \, \mathcal{I}_{\mu \nu}(q),
\end{equation}
where $\alpha,\beta$ are colour indices and $\mathcal{I}_{\mu \nu}(q)$ is the loop integral given by,
\begin{eqnarray}
    \mathcal{I}_{\mu \nu}(q)=\int\frac{d^4k}{(2\pi)^4}
  \frac{(2k+q)_\mu\,(2k+q)_\nu}
  {(k^2-m_\Phi^2)\bigl[(k+q)^2-m_\Phi^2\bigr]} \approx \frac{i\,m_\Phi^2}{8\pi^2}\,g_{\mu\nu}.
\end{eqnarray}
upto the leading finite term.
where the $m_\Phi^2$ in the numerator is a direct consequence of the derivative structure of the vertices.
The $\Delta S=2$ amplitude therefore becomes,
\begin{equation}
  \label{eq:amp3}
  i\mathcal{M} = \frac{i\,m_\Phi^2}{8\pi^2\Lambda_{\rm NP}^4}
  \,(\bar{d}^\alpha\,\gamma^\mu s^\alpha)(\bar{s}^\beta\,\gamma_\mu d^\beta) \ .
\end{equation}
To match onto the SUSY operator basis~\cite{Bertone:2012cu} we decompose the vector currents into chirality components,
\begin{eqnarray}
  (\bar{d}\,\gamma^\mu s)(\bar{s}\,\gamma_\mu d)
  &=& (\bar{d}_L\gamma^\mu s_L)(\bar{s}_L\gamma_\mu d_L)
  + (\bar{d}_R\gamma^\mu s_R)(\bar{s}_R\gamma_\mu d_R) \nonumber \\
  && + (\bar{d}_L\gamma^\mu s_L)(\bar{s}_R\gamma_\mu d_R)
  + (\bar{d}_R\gamma^\mu s_R)(\bar{s}_L\gamma_\mu d_L) \ .
\end{eqnarray}
Upon Fierzing the $LL$ sector $(\bar{d}_L^\alpha\gamma^\mu s_L^\alpha)(\bar{s}_L^\beta\gamma_\mu d_L^\beta)$ it gets mapped onto,
\begin{equation}
  (\bar{d}_L^\alpha\gamma^\mu s_L^\alpha)(\bar{s}_L^\beta\gamma_\mu d_L^\beta)
  = -\frac{1}{4}\,O_1 - \frac{1}{12}\,O_3 \ .
\end{equation}
By the same argument with $L\leftrightarrow R$, the $RR$ sector gives,
\begin{equation}
  (\bar{d}_R^\alpha\gamma^\mu s_R^\alpha)(\bar{s}_R^\beta\gamma_\mu d_R^\beta)
  = -\frac{1}{4}\,\tilde{O}_1 - \frac{1}{12}\,\tilde{O}_3 \ .
\end{equation}
And for the $LR$ and $RL$ sectors, the Fierz transformation generates,
\begin{equation}
  (\bar{d}_L^\alpha\gamma^\mu s_L^\alpha)(\bar{s}_R^\beta\gamma_\mu d_R^\beta)
  +(\bar{d}_R^\alpha\gamma^\mu s_R^\alpha)(\bar{s}_L^\beta\gamma_\mu d_L^\beta)
  = -\frac{1}{8}\,O_4 - \frac{1}{24}\,O_5 \ .
\end{equation}
Combining all four chirality sectors, the complete Fierz decomposition of the loop amplitude~Eq.~\eqref{eq:amp} is,
\begin{equation}
  \label{eq:fierz}
  (\bar{d}^\alpha\gamma^\mu s^\alpha)(\bar{s}^\beta\gamma_\mu d^\beta)
  = -\frac{1}{4}\,O_1 - \frac{1}{12}\,O_3
    -\frac{1}{4}\,\tilde{O}_1 - \frac{1}{12}\,\tilde{O}_3
    -\frac{1}{8}\,O_4 - \frac{1}{24}\,O_5.
\end{equation}
Due to chiral enhancement for the left-right operators, the contribution of $O_4$ operator, to the total hadronic matrix, dominates over the rest of the operators. Thus, it is sufficient to study the hadronic matrix element of $K^0-\bar{K^0}$ oscillation generated by the $\mathcal{O}_4$ operator, given as, 
\begin{equation}
  \langle\bar{K^0}|O_4(\mu)|K^0\rangle
  = \xi_4\,B_4(\mu)\left(\frac{m_K^2\,f_K}{m_s(\mu)+m_d(\mu)}\right)^2,
\end{equation}
with $\xi_4=2$, and $B_4=0.78$. Thus, the new-physics contribution to the kaon mass splitting becomes, 
\begin{equation}
  \Delta m_K^\mathrm{NP}
  = \frac{1}{m_K}\,\mathrm{Re}\,    \langle\bar{K^0}|\mathbb{H}^{\Delta S=2}_\mathrm{eff}|K^0\rangle
  = \frac{m_\Phi^2}{8\pi^2\Lambda_{\rm NP}^4\,m_K}\,\frac{1}{8}\,\xi_4 B_4\left(\frac{m_K^2 f_K}{m_s+m_d}\right)^2\,,
\end{equation}
which sets to the lower bound on the scale of new physics as,
\begin{equation}
  \label{eq:bound}
  \Lambda_{\rm NP} \gtrsim
  \left(
    \frac{m_\Phi^2}{8\pi^2\,m_K\,\Delta m_K^\mathrm{NP}}\,\frac{1}{8}\,\xi_4 B_4\left(\frac{m_K^2 f_K}{m_s+m_d}\right)^2
  \right)^{1/4}\,.
\end{equation}
To compute the limits on the new-physics scale, taking, $\Delta m_K^\mathrm{NP} = \Delta m_K^\mathrm{exp} =3.484\times10^{-15}\,\mathrm{GeV}$, and using $f_K=113\,\mathrm{MeV}$, $m_K=497.6\,\mathrm{MeV}$,
$m_s+m_d=100\,\mathrm{MeV}$, $B_4=0.78$ the bound becomes,
\begin{equation}
  \Lambda_{\rm NP} \gtrsim 1.93 \times 10^4\,\mathrm{GeV}
  \qquad(m_\Phi = 10^3\,\mathrm{GeV})\, .
\end{equation}
The result is shown in Fig.~\ref{fig:Of3_relic-scan}.
\subsection{Constraints from LEP}
\label{sec:collider1}
The Large Electron-Positron (LEP) collider operated at CM energies up to $\sqrt{s} \sim 209~\text{GeV}$ with high-precision control over the initial state. Measurements at the $Z$-pole, particularly of the invisible decay width, tightly constrained additional light degrees of freedom beyond the three SM neutrinos. Any deviation from the precisely measured $Z$ width would have indicated contributions from light DM states coupling to the $Z$ boson, relevant for our case in context of the $\mathcal{O}_{D\Phi}$ operator. In addition, mono-photon searches, $e^{+}e^{-} \to \gamma + \slashed{E}$, were extensively performed to probe pair production of invisible particles. These analyses provided stringent limits on DM couplings by accurately accounting for SM neutrino backgrounds, leaving only a small window for potential DM contributions.
\subsubsection{$Z$ decay constraints}
\label{sec:collider11}
Precision measurements of the invisible decay width of the $Z$ boson provide a stringent probe of new physics. The current experimental determinations at $95\%$ confidence level (CL) are given by
\begin{equation}
{\Gamma}(Z\to \text{invisible})<
\begin{dcases}
499.0\pm1.5~{\rm MeV}\,, & \text{LEP~\cite{ALEPH:2005ab}}\,,\\[5pt]
523\pm3\pm16~{\rm MeV}\,, & \text{CMS~\cite{CMS:2022ett}}\,,\\[5pt]
506\pm2\pm12~{\rm MeV}\,, & \text{ATLAS~\cite{ATLAS:2023ynf}}\,.
\end{dcases}
\label{eq:Z-invisible}
\end{equation}
In the presence of the effective operator $\mathcal{O}_{D\Phi}$, the $Z$ boson acquires an additional invisible decay channel into DM, $Z \to \Phi \Phi^\star$, in addition to the Standard Model (SM) contribution from neutrinos. The corresponding partial decay widths are \cite{Carpenter:2013xra},
\begin{gather}
\Gamma_{Z\to \nu\bar\nu}=\frac{3\,m_Z^3}{8\pi v^2}\,, \qquad
\Gamma_{Z\to \Phi\Phi^\star}=\frac{m_W^2\,m_Z^3}{48\pi^2 \alpha_e\,\lNP^4}\,\left(1-\frac{m_W^2}{m_Z^2}\right)\,\left(1-\frac{4\,m_\Phi^2}{m_Z^2}\right)^{3/2}\,,
\end{gather}
where the factor of $3$ in $\Gamma_{Z\to \nu\bar\nu}$ accounts for the three active neutrino flavours. The electromagnetic fine-structure constant evaluated at the $Z$-pole is taken as $\alpha_e(m_Z)\simeq 1/127.95$~\cite{Martin:2000by}. The additional decay mode into DM increases the total invisible width of the $Z$ boson. Requiring consistency with the experimental measurements in Eq.~\eqref{eq:Z-invisible} therefore constrains the size of $\Gamma_{Z\to \Phi\Phi^\star}$. Using the LEP result, we translate this into a lower bound on the NP scale $\Lambda_{\rm NP}$ for $m_\Phi < m_Z/2$, as shown in Fig.~\ref{fig:ODPhi_relic-scan}. In this low mass regime, $\Lambda_{\rm NP} \lesssim 1$ TeV is restricted if the $Z$-portal operator is considered. We note that the operator $\mathcal{O}_{f\Phi}$ can also contribute to the invisible width via the four-body decay $Z \to \nu \bar{\nu} \Phi \Phi^\star$, but this contribution is kinematically suppressed and subdominant. Along with the \(Z\)-invisible constraint, Higgs-invisible decay bounds are also applicable through the process $h\to \Phi\Phi^\star Z$ for $\ODPhi$ operator with $4\mPhi > m_h > 2\mPhi + m_Z$. The current upper limit on the Higgs-invisible decay branching fraction from ATLAS \cite{ATLAS:2023tkt} and CMS \cite{CMS:2023sdw} is approximately \(10\%\). Owing to this comparatively weaker branching-fraction limit and the additional three-body phase-space suppression, the resulting bound on \(\lNP\) is expected to be weaker than that obtained from the \(Z\)-invisible decay constraint.
\subsubsection{Mono-$\gamma$ search at LEP}
\label{sec:collider12}
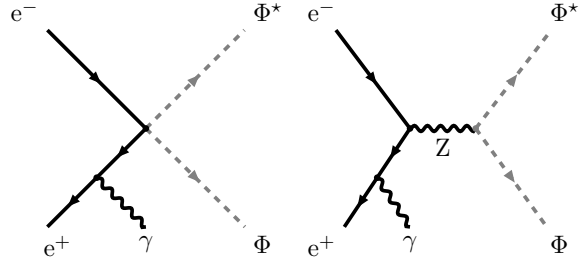
\begin{figure}[htb!]
\centering
\begin{adjustbox}{width=0.5\textwidth}
\begin{tikzpicture}[baseline={(current bounding box.center)},style={scale=1.0, transform shape}]
\begin{feynman}
\vertex (a);
\vertex [right =1.5 cm of a] (b);
\vertex [below =1.5 cm of a] (c);
\vertex [below left=0.75 cm and 0.75cm of a] (a0);
\vertex [above right=1.5 cm and 1.5cm of a] (b1);
\vertex [below right=1.5 cm and 1.5cm of a] (b2){\(\rm \color{black}{\Phi}\)};
\vertex [above left=1.5 cm and 1.5cm of a] (a1);
\vertex [below left=1.5 cm and 1.5cm of a] (a2);
\diagram*{
(a1) -- [line width=0.25mm, fermion, arrow size=1.2pt, style=black,ultra thick] (a) -- [line width=0.25mm, fermion, arrow size=1.2pt, style=black,ultra thick] (a0) -- [line width=0.25mm, fermion, arrow size=1.2pt, style=black,ultra thick] (a2),
(a0) -- [line width=0.25mm, boson, arrow size=1.2pt, style=black,ultra thick] (c),
(a) -- [line width=0.25mm, charged scalar, arrow size=1.2pt, style=gray,ultra thick] (b1),
(a) -- [line width=0.25mm, charged scalar, arrow size=1.2pt, style=gray, ultra thick] (b2)};
\vertex [below left=1.5 cm and 1.0cm of a] (a2){\(\rm \color{black}{e^+}\)};
\vertex [below =1.5 cm of a] (c){\(\rm \color{black}{\gamma}\)};
\vertex [above right=1.5 cm and 1.5cm of a] (b1){\(\rm \color{black}{\Phi^\star}\)};
\vertex [above left=1.5 cm and 1.5cm of a] (a1){\(\rm \color{black}{e^-}\)};
\node at (a)[circle,fill,style=black,inner sep=1pt]{};
\node at (a0)[circle,fill,style=black,inner sep=1pt]{};
\end{feynman}
\end{tikzpicture}
\begin{tikzpicture}[baseline={(current bounding box.center)},style={scale=1.0, transform shape}]
\begin{feynman}
\vertex (a);
\vertex [right =1.0 cm of a] (b);
\vertex [below =1.5 cm of a] (c);
\vertex [below left=0.75 cm and 0.5cm of a] (a0);
\vertex [above right=1.5 cm and 1.0cm of b] (b1){\(\rm \color{black}{\Phi^\star}\)};
\vertex [below right=1.5 cm and 1.0cm of b] (b2){\(\rm \color{black}{\Phi}\)};
\vertex [above left=1.5 cm and 1.0cm of a] (a1){\(\rm \color{black}{e^-}\)};
\vertex [below left=1.5 cm and 1.0cm of a] (a2);
\diagram*{
(a1) -- [line width=0.25mm, fermion, arrow size=1.2pt, style=black,ultra thick] (a) -- [line width=0.25mm, fermion, arrow size=1.2pt, style=black,ultra thick] (a0) -- [line width=0.25mm, fermion, arrow size=1.2pt, style=black,ultra thick] (a2),
(a0) -- [line width=0.25mm, boson, arrow size=1.2pt, style=black,ultra thick] (c),
(a) -- [line width=0.25mm, boson, arrow size=1.2pt, style=black,ultra thick,edge label'={\(\rm\color{black}{Z}\)}] (b),
(b) -- [line width=0.25mm, charged scalar, arrow size=1.2pt, style=gray,ultra thick] (b1),
(b) -- [line width=0.25mm, charged scalar, arrow size=1.2pt, style=gray, ultra thick] (b2)};
\vertex [below left=1.5 cm and 1.0cm of a] (a2){\(\rm \color{black}{e^+}\)};
\vertex [below =1.5 cm of a] (c){\(\rm \color{black}{\gamma}\)};
\node at (a)[circle,fill,style=black,inner sep=1pt]{};
\node at (b)[circle,fill,style=gray,inner sep=1pt]{};
\node at (a0)[circle,fill,style=black,inner sep=1pt]{};
\end{feynman}
\end{tikzpicture}
\end{adjustbox}
\caption{The Feynman diagrams relevant to DM production via mono-$\gamma$ searches for the $\OfPhi$ (\textit{left}) and $\ODPhi$ (\textit{right}) operators at LEP and FCC-ee.}
\label{feyn:mono-photon_LEP}
\end{figure}
Mono-$\gamma$ searches at LEP probe invisible particle production through the process $e^{+}e^{-} \to \gamma + \slashed{E}$ (shown in Fig.~\ref{feyn:mono-photon_LEP}), where the photon arises predominantly from initial-state radiation. The signal is characterized by a single energetic photon with large missing energy and momentum, while the dominant SM background originates from $e^{+}e^{-} \to \nu \bar{\nu} \gamma$. These analyses exploited both the photon energy spectrum and angular distributions to discriminate potential NP contributions from the well-understood neutrino background. In Ref.~\cite{Fox:2011fx}, LEP mono-$\gamma$ data (primarily from the DELPHI experiment) were recast in a model-independent framework using effective operators describing DM-electron interactions. The recast closely follows the experimental selection by imposing cuts on the photon energy fraction and polar angle, and by reproducing the observed photon spectrum. Signal events are simulated within the EFT framework and compared to data to derive limits on the suppression scale $\Lambda$ as a function of the DM mass. In our setup, the operator $\mathcal{O}_{f\Phi}$ maps directly onto the leptophilic operators considered in Ref.~\cite{Fox:2011fx}, allowing us to straightforwardly translate their bounds on $\Lambda_{\rm NP}$, as shown in Fig.~\ref{fig:Of3_relic-scan}. Fow the low mass regime i.e. $m_{\Phi} \lesssim 100$ GeV, $\Lambda_{\rm NP} \lesssim 500$ GeV is restricted if the $\mathcal{O}_{f\Phi}$ contact interaction is considered. Although a similar recast can be performed for $\mathcal{O}_{D\Phi}$, the resulting constraints are weaker than those obtained from the $Z$-pole invisible width measurements, and hence remain subdominant in our analysis.
\subsection{DM search at the LHC}
\label{sec:collider2}
The LHC provides a powerful environment to probe DM through high-energy $pp$ collisions, where DM particles can be produced in association with visible SM objects. Since the longitudinal momentum of the initial state is not known, these searches rely on $\slashed{E}_T$ as the key observable. The most sensitive probes typically arise from mono-$j$ signatures, where a high-$p_T$ jet recoils against large $\slashed{E}_T$. The dominant SM backgrounds, primarily from $Z(\nu\bar{\nu})+$jets and $W(\ell\nu)+$jets, are well understood and controlled using data-driven techniques.
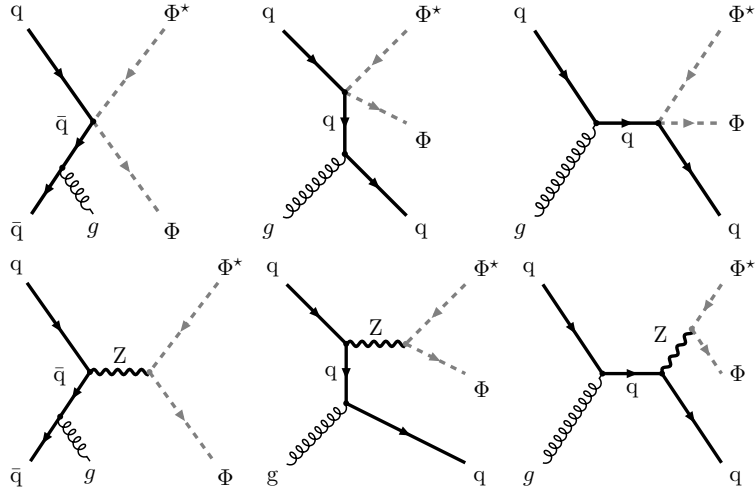
\begin{figure}[htb!]
\centering
\begin{adjustbox}{width=0.65\textwidth}
\begin{tikzpicture}[baseline={(current bounding box.center)},style={scale=1.0, transform shape}]
\begin{feynman}
\vertex (a);
\vertex [right =1.0 cm of a] (b);
\vertex [below =1.5 cm of a] (c);
\vertex [below left=0.75 cm and 0.5cm of a] (a0);
\vertex [above right=1.5 cm and 1.0cm of a] (b1){\(\rm \color{black}{\Phi^\star}\)};
\vertex [below right=1.5 cm and 1.0cm of a] (b2){\(\rm \color{black}{\Phi}\)};
\vertex [above left=1.5 cm and 1.0cm of a] (a1){\(\rm \color{black}{q}\)};
\vertex [below left=1.5 cm and 1.0cm of a] (a2);
\diagram*{
(a1) -- [line width=0.25mm, fermion, arrow size=1.2pt, style=black,ultra thick] (a) -- [line width=0.25mm, fermion, arrow size=1.2pt, style=black,ultra thick,edge label'={\(\rm\color{black}{\bar{q}}\)}] (a0) -- [line width=0.25mm, fermion, arrow size=1.2pt, style=black,ultra thick] (a2),
(a0) -- [line width=0.25mm, gluon, arrow size=1.2pt, style=black] (c),
(b1) -- [line width=0.25mm, charged scalar, arrow size=1.2pt, style=gray,ultra thick] (a) -- [line width=0.25mm, charged scalar, arrow size=1.2pt, style=gray, ultra thick] (b2)};
\vertex [below left=1.5 cm and 1.0cm of a] (a2){\(\rm \color{black}{\bar{q}}\)};
\vertex [below =1.5 cm of a] (c){\(\color{black}{g}\)};
\node at (a)[circle,fill,style=black,inner sep=1pt]{};
\node at (a0)[circle,fill,style=black,inner sep=1pt]{};
\end{feynman}
\end{tikzpicture} \qquad
\begin{tikzpicture}[baseline={(current bounding box.center)},style={scale=1.0, transform shape}]
\begin{feynman}
\vertex (a);
\vertex [below =1.0 cm of a] (b);
\vertex [above left=1.0 cm and 1.0cm of a] (a1);
\vertex [above right=1.0 cm and 1.0cm of a] (a2);
\vertex [below right=0.5 cm and 1.0cm of a] (a3);
\vertex [below left=1.0 cm and 1.0cm of b] (b1){\(\color{black}{g}\)};
\vertex [below right=1.0 cm and 1.0cm of b] (b2);
\diagram*{
(a1) -- [line width=0.25mm, fermion, arrow size=1.2pt, style=black,ultra thick] (a)  -- [line width=0.25mm, fermion, arrow size=1.2pt, style=black,ultra thick, edge label'={\(\rm\color{black}{q}\)}] 
(b)  -- [line width=0.25mm, fermion, arrow size=1.2pt, style=black,ultra thick] (b2),
(b1) -- [line width=0.25mm, gluon, arrow size=1.2pt, style=black] (b),
(a2) -- [line width=0.25mm, charged scalar, arrow size=1.2pt, style=gray,ultra thick] (a) -- [line width=0.25mm, charged scalar, arrow size=1.2pt, style=gray, ultra thick] (a3)};
\vertex [above left=1.0 cm and 1.0cm of a] (a1){\(\rm \color{black}{q}\)};
\vertex [above right=1.0 cm and 1.0cm of a] (a2){\(\rm \color{black}{\Phi^\star}\)};
\vertex [below right=0.5 cm and 1.0cm of a] (a3){\(\rm \color{black}{\Phi}\)};
\vertex [below right=1.0 cm and 1.0cm of b] (b2){\(\rm \color{black}{q}\)};
\node at (a)[circle,fill,style=black,inner sep=1pt]{};
\node at (b)[circle,fill,style=black,inner sep=1pt]{};
\end{feynman}
\end{tikzpicture} \qquad
\begin{tikzpicture}[baseline={(current bounding box.center)},style={scale=1.0, transform shape}]
\begin{feynman}
\vertex (a);
\vertex [right =1.0 cm of a] (b);
\vertex [below left=1.5 cm and 1.0cm of a] (a1);
\vertex [above left=1.5 cm and 1.0cm of a] (a2);
\vertex [above right=1.5 cm and 1.0cm of b] (b1);
\vertex [right=1.0 cm of b] (b2);
\vertex [below right=1.5 cm and 1.0cm of b] (b3);
\diagram*{
(a2) -- [line width=0.25mm, fermion, arrow size=1.2pt, style=black,ultra thick] (a)  -- [line width=0.25mm, fermion, arrow size=1.2pt, style=black,ultra thick, edge label'={\(\rm\color{black}{q}\)}] 
(b)  -- [line width=0.25mm, charged scalar, arrow size=1.2pt, style=gray,ultra thick] (b2),
(a1) -- [line width=0.25mm, gluon, arrow size=1.2pt, style=black] (a),
(b1) -- [line width=0.25mm, charged scalar, arrow size=1.2pt, style=gray,ultra thick] (b) -- [line width=0.25mm, fermion, arrow size=1.2pt, style=black, ultra thick] (b3)};
\vertex [below left=1.5 cm and 1.0cm of a] (a1){\(\color{black}{g}\)};
\vertex [above left=1.5 cm and 1.0cm of a] (a2){\(\rm \color{black}{q}\)};
\vertex [above right=1.5 cm and 1.0cm of b] (b1){\(\rm \color{black}{\Phi^\star}\)};
\vertex [right= 1.0cm of b] (b2){\(\rm \color{black}{\Phi}\)};
\vertex [below right=1.5 cm and 1.0cm of b] (b3){\(\rm \color{black}{q}\)};
\node at (a)[circle,fill,style=black,inner sep=1pt]{};
\node at (b)[circle,fill,style=black,inner sep=1pt]{};
\end{feynman}
\end{tikzpicture}
\end{adjustbox}
\\
\begin{adjustbox}{width=0.65\textwidth}
\begin{tikzpicture}[baseline={(current bounding box.center)},style={scale=1.0, transform shape}]
\begin{feynman}
\vertex (a);
\vertex [right =1.0 cm of a] (b);
\vertex [below =1.5 cm of a] (c);
\vertex [below left=0.75 cm and 0.5cm of a] (a0);
\vertex [above right=1.5 cm and 1.0cm of b] (b1){\(\rm \color{black}{\Phi^\star}\)};
\vertex [below right=1.5 cm and 1.0cm of b] (b2){\(\rm \color{black}{\Phi}\)};
\vertex [above left=1.5 cm and 1.0cm of a] (a1){\(\rm \color{black}{q}\)};
\vertex [below left=1.5 cm and 1.0cm of a] (a2);
\diagram*{
(a1) -- [line width=0.25mm, fermion, arrow size=1.2pt, style=black,ultra thick] (a) -- [line width=0.25mm, fermion, arrow size=1.2pt, style=black,ultra thick,edge label'={\(\rm\color{black}{\bar{q}}\)}] (a0) -- [line width=0.25mm, fermion, arrow size=1.2pt, style=black,ultra thick] (a2),
(a0) -- [line width=0.25mm, gluon, arrow size=1.2pt, style=black] (c),
(a) -- [line width=0.25mm, boson, arrow size=1.2pt, style=black,ultra thick,edge label={\(\rm\color{black}{Z}\)}] (b),
(b1) -- [line width=0.25mm, charged scalar, arrow size=1.2pt, style=gray,ultra thick] (b) -- [line width=0.25mm, charged scalar, arrow size=1.2pt, style=gray, ultra thick] (b2)};
\vertex [below left=1.5 cm and 1.0cm of a] (a2){\(\rm \color{black}{\bar{q}}\)};
\vertex [below =1.5 cm of a] (c){\(\color{black}{g}\)};
\node at (a)[circle,fill,style=black,inner sep=1pt]{};
\node at (a0)[circle,fill,style=black,inner sep=1pt]{};
\node at (b)[circle,fill,style=gray,inner sep=1pt]{};
\end{feynman}
\end{tikzpicture}
\begin{tikzpicture}[baseline={(current bounding box.center)},style={scale=1.0, transform shape}]
\begin{feynman}
\vertex (a);
\vertex [right =1.0 cm of a] (c);
\vertex [below =1.0 cm of a] (b);
\vertex [above left=1.0 cm and 1.0cm of a] (a1);
\vertex [above right=1.0 cm and 1.0cm of c] (a2);
\vertex [below right=0.5 cm and 1.0cm of c] (a3);
\vertex [below left=1.0 cm and 1.0cm of b] (b1){\(\rm \color{black}{g}\)};
\vertex [below right=1.0 cm and 2.0cm of b] (b2);
\diagram*{
(a1) -- [line width=0.25mm, fermion, arrow size=1.2pt, style=black,ultra thick] (a)  -- [line width=0.25mm, fermion, arrow size=1.2pt, style=black,ultra thick, edge label'={\(\rm\color{black}{q}\)}] 
(b)  -- [line width=0.25mm, fermion, arrow size=1.2pt, style=black,ultra thick] (b2),
(b1) -- [line width=0.25mm, gluon, arrow size=1.2pt, style=black] (b),
(a) -- [line width=0.25mm, boson, arrow size=1.2pt,ultra thick, style=black,edge label={\(\rm\color{black}{Z}\)}] (c),
(a2) -- [line width=0.25mm, charged scalar, arrow size=1.2pt, style=gray,ultra thick] (c),
(c) -- [line width=0.25mm, charged scalar, arrow size=1.2pt, style=gray, ultra thick] (a3)};
\vertex [above left=1.0 cm and 1.0cm of a] (a1){\(\rm \color{black}{q}\)};
\vertex [above right=1.0 cm and 1.0cm of c] (a2){\(\rm \color{black}{\Phi^\star}\)};
\vertex [below right=0.5 cm and 1.0cm of c] (a3){\(\rm \color{black}{\Phi}\)};
\vertex [below right=1.0 cm and 2.0cm of b] (b2){\(\rm \color{black}{q}\)};
\node at (a)[circle,fill,style=black,inner sep=1pt]{};
\node at (b)[circle,fill,style=black,inner sep=1pt]{};
\node at (c)[circle,fill,style=gray,inner sep=1pt]{};
\end{feynman}
\end{tikzpicture}
\begin{tikzpicture}[baseline={(current bounding box.center)},style={scale=1.0, transform shape}]
\begin{feynman}
\vertex (a);
\vertex [right =1.0 cm of a] (b);
\vertex [above right=0.75 cm and 0.5cm of b] (c);
\vertex [below left=1.5 cm and 1.0cm of a] (a1);
\vertex [above left=1.5 cm and 1.0cm of a] (a2);
\vertex [above right=1.5 cm and 1.0cm of b] (b1);
\vertex [right=1.0 cm of b] (b2);
\vertex [below right=1.5 cm and 1.0cm of b] (b3);
\diagram*{
(a2) -- [line width=0.25mm, fermion, arrow size=1.2pt, style=black,ultra thick] (a)  -- [line width=0.25mm, fermion, arrow size=1.2pt, style=black,ultra thick, edge label'={\(\rm\color{black}{q}\)}] 
(b)  -- [line width=0.25mm, boson, arrow size=1.2pt, style=black, edge label={\(\rm\color{black}{Z}\)}, ultra thick] (c),
(b1)  -- [line width=0.25mm, charged scalar, arrow size=1.2pt, style=gray,ultra thick] (c) -- [line width=0.25mm, charged scalar, arrow size=1.2pt, style=gray,ultra thick] (b2),
(a1) -- [line width=0.25mm, gluon, arrow size=1.2pt, style=black] (a),
(b) -- [line width=0.25mm, fermion, arrow size=1.2pt, style=black, ultra thick] (b3)};
\vertex [below left=1.5 cm and 1.0cm of a] (a1){\(\color{black}{g}\)};
\vertex [above left=1.5 cm and 1.0cm of a] (a2){\(\rm \color{black}{q}\)};
\vertex [above right=1.5 cm and 1.0cm of b] (b1){\(\rm \color{black}{\Phi^\star}\)};
\vertex [right= 1.0cm of b] (b2){\(\rm \color{black}{\Phi}\)};
\vertex [below right=1.5 cm and 1.0cm of b] (b3){\(\rm \color{black}{q}\)};
\node at (a)[circle,fill,style=black,inner sep=1pt]{};
\node at (b)[circle,fill,style=black,inner sep=1pt]{};
\node at (c)[circle,fill,style=gray,inner sep=1pt]{};
\end{feynman}
\end{tikzpicture}
\end{adjustbox}
\caption{The Feynman diagrams relevant to DM production via mono-$j$ $(g,q,\overline{q})$ searches for the $\OfPhi$ (\textit{top} row) and $\ODPhi$ (\textit{bottom} row) operators at LHC and HL-LHC.}
\label{feyn:mono-jet_LHC}
\end{figure}
\subsubsection{Recasting LHC limits}
\label{sec:collider21}
Since no dedicated LHC analysis currently targets the effective operators considered in this work, we reinterpret existing searches to derive constraints on our parameter space. In particular, we recast the mono-jet $+ \slashed{E}_{T}$ analysis: \texttt{atlas\_2102\_10874} implemented in \textsc{CheckMATE2}~\cite{Dercks:2016npn}, corresponding to the ATLAS Run II dataset at $\sqrt{s}=13$~TeV with an integrated luminosity of $139~\text{fb}^{-1}$~\cite{ATLAS:2021kxv}. This search targets events with a high-$p_T$ jet recoiling against substantial missing transverse momentum, making it highly sensitive to DM pair production in association with initial-state radiation. The relevant Feynman diagrams are shown in Fig.~\ref{feyn:mono-jet_LHC}. The event selection requires at least one energetic jet, significant $\slashed{E}_T$, and vetoes on additional leptons, photons or excessive hadronic activity to suppress SM backgrounds. The analysis is organized into 24 $\slashed{E}_T$ signal regions, comprising 12 exclusive bins defined over closed intervals between $200~\text{GeV}$ and $1200~\text{GeV}$ (with an additional overflow bin beyond $1200~\text{GeV}$), and 12 inclusive regions defined by open intervals starting from $200~\text{GeV}$ up to the highest $\slashed{E}_T$ values. This binning strategy enhances sensitivity across a wide kinematic range, particularly in the high-$\slashed{E}_T$ regime where new physics contributions are expected to be more prominent.

To quantify the exclusion, we follow the standard prescription used in \textsc{CheckMATE2}, defining the $r$-value for each signal region as
\begin{equation}
    r = \left(\frac{S - 1.64\, \Delta S}{S_{95}}\right)\,,
\end{equation}
where $S$ is the predicted number of signal events, $\Delta S$ denotes its associated uncertainty, and $S_{95}$ is the experimentally determined upper limit on the signal yield at $95\%$ CL. A parameter point is considered excluded if $r > 1$ in at least one signal region. The model is implemented in \texttt{FeynRules}~\cite{Alloul:2013bka}, and the resulting UFO files are interfaced with \texttt{MG5\_aMC}~\cite{Alwall:2011uj} to generate parton-level events for the signal processes. The events are subsequently passed to \texttt{Pythia8}~\cite{Sjostrand:2014zea,Bierlich:2022pfr} for parton showering and hadronization. This procedure is repeated over the relevant parameter space, and the resulting event samples are analyzed using \textsc{CheckMATE2}, which applies the experimental selection criteria and evaluates the corresponding constraints. In our analysis, we scan over all signal regions and identify the most sensitive one to derive the final bounds. The resulting constraints on $\mathcal{O}_{f\Phi}$ and $\mathcal{O}_{D\Phi}$ are presented in Figs.~\ref{fig:Of3_relic-scan} and~\ref{fig:ODPhi_relic-scan}, respectively. From the recast analysis, we obtain lower bounds on the NP scale of $\Lambda_{\rm NP} \sim 1.8~\text{TeV}$ for $\mathcal{O}_{f\Phi}$ and $\Lambda_{\rm NP} \sim 400~\text{GeV}$ for $\mathcal{O}_{D\Phi}$. The stronger limit in the former case arises due to the direct coupling to quark currents, leading to enhanced production rates at hadron colliders, while the latter is comparatively suppressed.

It is important to emphasize that, for $\mathcal{O}_{D\Phi}$, the extracted bound lies close to (or even below) the typical momentum transfer accessible at the LHC, which can reach scales of $\mathcal{O}(1~\text{TeV})$. In such a regime, the validity of the EFT description becomes questionable, as higher-dimensional operators may no longer provide a reliable truncation of the underlying UV-complete theory. Consequently, the bound on $\Lambda$ for $\mathcal{O}_{D\Phi}$ should be interpreted with caution. Nevertheless, we report this limit as an indicative measure of the LHC sensitivity to the corresponding interaction strength. In other words, it reflects the collider reach in terms of the effective cross section associated with this operator, rather than a strict probe of the NP scale itself. A more robust interpretation would require embedding the operator within a UV-complete framework, which we leave for future investigation.
\subsubsection{Projections for HL-LHC}
\label{sec:collider22}
\begin{figure}[htb!]
\centering
\includegraphics[width=0.94\linewidth]{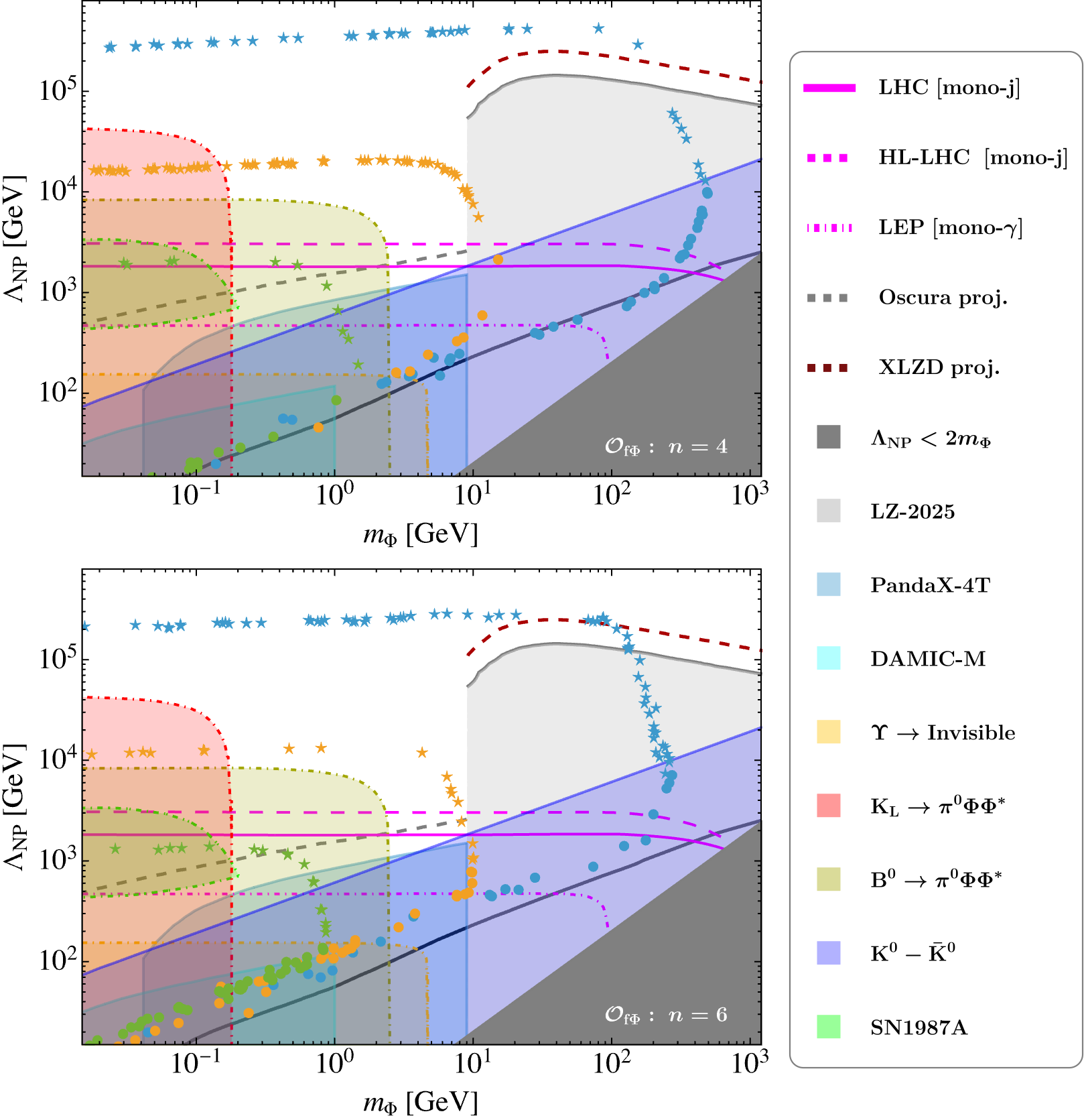}
\caption{Summary of the viable parameter space for $\mathcal{O}_{f\Phi}$, for $n=4$ and $n=6$, shown in the upper and in the lower panel, respectively, considering reheating via perturbative bosonic decay. Different coloured points satisfy the observed DM abundance via freeze-out and freeze-in for $\Trh=\{0.05,\,0.5,\,10\}$ GeV, shown via green, orange and blue, respectively. The filled {\it circular} points correspond to freeze-out and the filled {\it starred} points produce right DM abundance via freeze-in, both during reheating. The shaded regions are disallowed by DM-nucleon and DM-electron direct detection experiments, invisible meson decay measurements, as well as from the validity of EFT that demands $\lNP>m_\Phi$ (see subsection~\ref{sec:noncollider} for details). Along the solid black curve, right relic abundance is satisfied for freeze-out during RD. All projected bounds are shown via dashed curves.}
\label{fig:Of3_relic-scan}
\end{figure}

For the HL-LHC projection, we recast the mono-jet $+ \slashed{E}_{T}$ analysis \texttt{atlas\_phys\_2014\_007} implemented in \textsc{CheckMATE2}, corresponding to $\sqrt{s}=14$~TeV and an integrated luminosity of $3000~\text{fb}^{-1}$~\cite{ATLAS:2014jim}. The analysis is structured into 21 signal regions, categorized as exclusive (E) and inclusive (I), defined over $\slashed{E}_T$ intervals starting from $400~\text{GeV}$ and extending to the multi-TeV regime. The exclusive regions correspond to closed bins, while the inclusive regions include all events above a given threshold, thereby enhancing sensitivity to higher energy tails. We follow the same simulation and analysis pipeline as in the Run II recast. For each parameter point, the exclusion is determined using the $r$-value, and the most sensitive signal region is identified. The high-$\slashed{E}_T$ inclusive bins are found to dominate the sensitivity, particularly for larger values of $\Lambda_{\rm NP}$. The resulting projections indicate a substantial improvement in reach compared to current limits, with the HL-LHC probing significantly higher NP scales. The corresponding bounds are presented in Figs.~\ref{fig:Of3_relic-scan} and~\ref{fig:ODPhi_relic-scan}.

For a dedicated study of future collider sensitivity, we select representative benchmark points in the parameter space that are consistent with the observed relic abundance and lie close to the lowest values of $\Lambda_{\rm NP}$ allowed by cosmology. These choices are therefore particularly relevant for assessing collider reach, as they correspond to scenarios that are both phenomenologically viable and maximally accessible at high-energy experiments. The benchmark points are defined as
\begin{itemize}
    \item \textbf{BP($\boldsymbol{\mathcal{O}_{f\Phi}}$)}: $m_{\Phi} = 10$ GeV, $\Lambda_{\rm NP}=2.0$ TeV.
    \item \textbf{BP($\boldsymbol{\mathcal{O}_{D\Phi}}$)}: $m_{\Phi} = 10$ GeV, $\Lambda_{\rm NP}=1.0$ TeV.
\end{itemize}
These points are chosen to lie near the minimal $\Lambda_{\rm NP}$ values compatible with the relic density constraint in each scenario, and thus represent the most optimistic cases for collider observability.

From Fig.~\ref{fig:ODPhi_relic-scan}, it is evident that for the operator $\mathcal{O}_{D\Phi}$ the entire region within the projected sensitivity of the HL-LHC is already excluded, primarily due to a combination of constraints from the $Z$-boson invisible width and direct detection limits. As a result, no viable parameter space remains for a meaningful collider probe of this operator at the HL-LHC. The benchmark point considered for $\mathcal{O}_{D\Phi}$ therefore lies beyond the reach of hadron colliders and may instead be probed at future high-energy lepton colliders, as will be discussed in the next section. In contrast, the benchmark point BP$(\mathcal{O}_{f\Phi})$ remains within the HL-LHC sensitivity region. We therefore use the recasted analysis of the mono-jet $+\slashed{E}_T$ channel to estimate the expected event yields in each signal region. For this benchmark, signal and background event numbers are computed separately for both exclusive and inclusive regions, following the definitions of the HL-LHC analysis. These yields are then used to assess the potential observability of the signal at the HL-LHC by evaluating the statistical significance in the most sensitive signal regions. The corresponding event counts for signal ($S$) and background ($B$) in the inclusive and exclusive bins are presented below:
\begin{table}[htb!]
\centering
\begin{tabular}{|c|c|c|c|c|c|}
\hline
\multicolumn{2}{|c|}{\textbf{Signal Region}} & \textbf{$\slashed{E}_T$ range (GeV)} & \textbf{Background ($\boldsymbol{B}$)} & \textbf{Signal ($\boldsymbol{S}$)} & \textbf{$\boldsymbol{r}$-value} \\ \hline \hline
\multirow{11}*{\rotatebox{90}{\textbf{Exclusive Signal Regions}}}
&E040 & $400-450$   & 305000 & 5463 & 0.73 \\
&E045 & $450-500$   & 192500 & 3925 & 0.74 \\
&E050 & $500-600$   & 178400 & 5207 & 1.36 \\
&E060 & $600-700$   & 64500  & 2751 & 1.55 \\
&E070 & $700-800$   & 23600  & 1795 & 1.92 \\
&E080 & $800-900$   & 9770   & 858  & 1.81 \\
&E090 & $900-1000$  & 4570   & 562  & 1.86 \\
&E100 & $1000-1200$ & 3450   & 483  & 1.94 \\
&E120 & $1200-1400$ & 861    & 197  & 1.73 \\
&E140 & $1400-1600$ & 180    & 69   & 0.85 \\
&E160 & $>1600$     & 37     & 49   & 0.63 \\
\hline \hline
\multirow{11}*{\rotatebox{90}{\textbf{Inclusive Signal Regions}}}
&I040 & $>400$      & 783000 & 21359 & 0.89 \\
&I045 & $>450$      & 478000 & 15896 & 0.99 \\
&I050 & $>500$      & 285400 & 11971 & 1.48 \\
&I060 & $>600$      & 107000 & 6765  & 1.64 \\
&I070 & $>700$      & 42500  & 4013  & 1.88 \\
&I080 & $>800$      & 19000  & 2219  & 1.69 \\
&I090 & $>900$      & 9100   & 1361  & 1.93 \\
&I100 & $>1000$     & 4530   & 799   & 1.79 \\
&I120 & $>1200$     & 1080   & 316   & 1.56 \\
&I140 & $>1400$     & 216    & 118   & 1.35 \\
\hline
\end{tabular}
\caption{Signal and background event yields for exclusive (E) and inclusive (I) signal regions of the recasted mono-jet analysis, together with the corresponding missing transverse energy ($\slashed{E}_T$) ranges.}
\label{tab:sr_yields}
\end{table}

The results are summarized in Tab.~\ref{tab:sr_yields}, where we list the signal and background yields along with the corresponding $r$-values for each signal region. As expected, the sensitivity improves with increasing $\slashed{E}_T$ up to a point where the background is sufficiently suppressed while retaining a sizable signal rate. Among the exclusive signal regions, the most sensitive bin is found to be $\mathbf{E100}$, corresponding to the range $1000~\text{GeV} < \slashed{E}_T < 1200~\text{GeV}$, which yields the largest $r$-value. This indicates that intermediate-to-high $\slashed{E}_T$ regions provide the optimal balance between signal efficiency and background reduction. At even higher $\slashed{E}_T$, although the background continues to decrease, the signal rate also drops significantly, leading to a reduced sensitivity. For the inclusive signal regions, the strongest sensitivity arises from the $\mathbf{I090}$ region, defined by $\slashed{E}_T > 900~\text{GeV}$. This reflects the fact that inclusive regions efficiently accumulate signal events from higher $\slashed{E}_T$ tails while maintaining manageable background levels. Overall, the inclusive regions tend to outperform the exclusive ones, as they effectively integrate contributions from multiple kinematic regimes.These results demonstrate that the HL-LHC has significant discovery potential for $\mathcal{O}_{f\Phi}$, with multiple signal regions yielding $r > 1$, indicating that the benchmark point is within reach. In particular, the dominant sensitivity originates from high-$\slashed{E}_T$ inclusive regions, highlighting the importance of the boosted regime in probing such effective interactions.
\subsubsection{Mono-Higgs search projections}
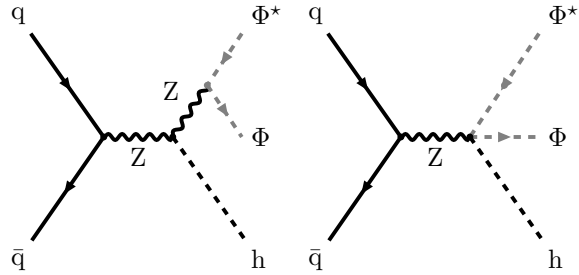
\begin{figure}[htb!]
\centering
\begin{adjustbox}{width=0.5\textwidth}
\begin{tikzpicture}[baseline={(current bounding box.center)},style={scale=1.0, transform shape}]
\begin{feynman}
\vertex (a);
\vertex [right =1.0 cm of a] (b);
\vertex [right =1.0 cm of b] (c);
\vertex [above right=0.75 cm and 0.5cm of b] (b0);
\vertex [above right=1.5 cm and 1.0cm of b] (b1);
\vertex [below right=1.5 cm and 1.0cm of b] (b2){\(\rm \color{black}{h}\)};
\vertex [above left=1.5 cm and 1.0cm of a] (a1){\(\rm \color{black}{q}\)};
\vertex [below left=1.5 cm and 1.0cm of a] (a2){\(\rm \color{black}{\bar{q}}\)};
\diagram*{
(a1) -- [line width=0.25mm, fermion, arrow size=1.2pt, style=black,ultra thick] (a) -- [line width=0.25mm, fermion, arrow size=1.2pt, style=black,ultra thick] (a2),
(a) -- [line width=0.25mm, boson, arrow size=1.2pt, style=black,ultra thick,edge label'={\(\rm\color{black}{Z}\)}] (b),
(b) -- [line width=0.25mm, boson, arrow size=1.2pt, style=black,ultra thick,edge label={\(\rm\color{black}{Z}\)}] (b0),
(b1) -- [line width=0.25mm, charged scalar, arrow size=1.2pt, style=gray,ultra thick] (b0),
(b0) -- [line width=0.25mm, charged scalar, arrow size=1.2pt, style=gray,ultra thick] (c),
(b) -- [line width=0.25mm, scalar, arrow size=1.2pt, style=black,ultra thick] (b2)};
\vertex [right =1.0 cm of b] (c){\(\rm \color{black}{\Phi}\)};
\vertex [above right=1.5 cm and 1.0cm of b] (b1){\(\rm \color{black}{\Phi^\star}\)};
\node at (a)[circle,fill,style=black,inner sep=1pt]{};
\node at (b)[circle,fill,style=black,inner sep=1pt]{};
\node at (b0)[circle,fill,style=gray,inner sep=1pt]{};
\end{feynman}
\end{tikzpicture}
\begin{tikzpicture}[baseline={(current bounding box.center)},style={scale=1.0, transform shape}]
\begin{feynman}
\vertex (a);
\vertex [right =1.0 cm of a] (b);
\vertex [right =1.0 cm of b] (c);
\vertex [above right=0.75 cm and 0.5cm of b] (b0);
\vertex [above right=1.5 cm and 1.0cm of b] (b1);
\vertex [below right=1.5 cm and 1.0cm of b] (b2){\(\rm \color{black}{h}\)};
\vertex [above left=1.5 cm and 1.0cm of a] (a1){\(\rm \color{black}{q}\)};
\vertex [below left=1.5 cm and 1.0cm of a] (a2){\(\rm \color{black}{\bar{q}}\)};
\diagram*{
(a1) -- [line width=0.25mm, fermion, arrow size=1.2pt, style=black,ultra thick] (a) -- [line width=0.25mm, fermion, arrow size=1.2pt, style=black,ultra thick] (a2),
(a) -- [line width=0.25mm, boson, arrow size=1.2pt, style=black,ultra thick,edge label'={\(\rm\color{black}{Z}\)}] (b),
(b) -- [line width=0.25mm, charged scalar, arrow size=1.2pt, style=gray,ultra thick] (c),
(b1) -- [line width=0.25mm, charged scalar, arrow size=1.2pt, style=gray,ultra thick] (b),
(b) -- [line width=0.25mm, scalar, arrow size=1.2pt, style=black,ultra thick] (b2)};
\vertex [right =1.0 cm of b] (c){\(\rm \color{black}{\Phi}\)};
\vertex [above right=1.5 cm and 1.0cm of b] (b1){\(\rm \color{black}{\Phi^\star}\)};
\node at (a)[circle,fill,style=black,inner sep=1pt]{};
\node at (b)[circle,fill,style=black,inner sep=1pt]{};
\node at (b0)[circle,fill,style=gray,inner sep=1pt]{};
\end{feynman}
\end{tikzpicture}
\end{adjustbox}
\caption{The Feynman diagrams relevant to DM production via mono-Higgs searches for the $\ODPhi$ operator at LHC and HL-LHC.}
\label{feyn:mono-Higgs}
\end{figure}
\begin{figure}[htb!]
\centering
\includegraphics[width=0.94\linewidth]{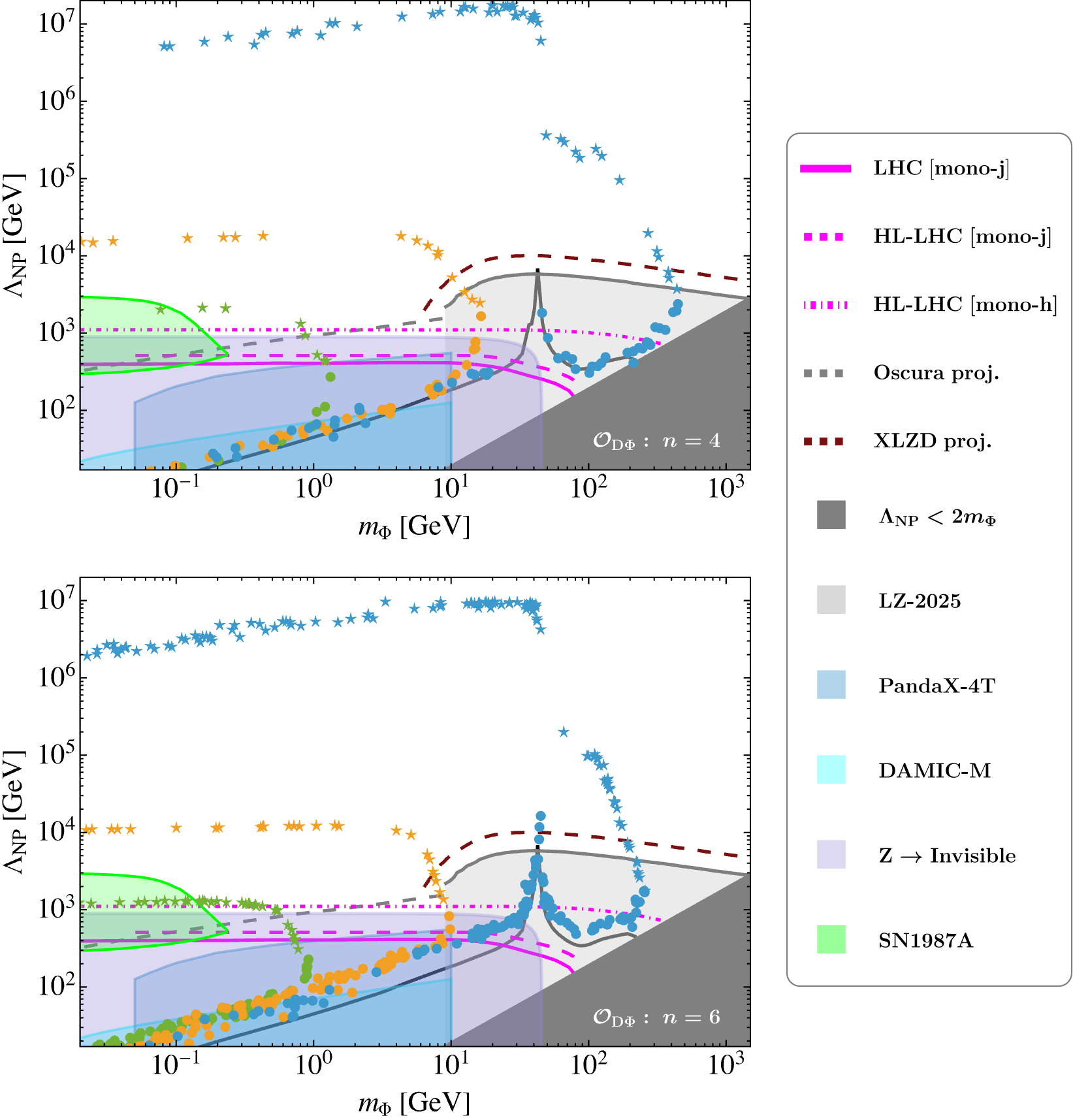}
\caption{Same as Fig.~\ref{fig:Of3_relic-scan}, but for $\mathcal{O}_{D\Phi}$. Though Upsilon invisible decay process places a lower bound on $\Lambda_{\rm NP}$, they are much weaker as given in Tab.\ref{tab:inv_decay_bounds} and hence not shown here. Moreover, flavour violating decay bounds do not apply here.}
\label{fig:ODPhi_relic-scan}
\end{figure}

Interestingly, $\mathcal{O}_{D\Phi}$ operator provides a tree-level $hZ\Phi^\star\Phi$ contact interaction, and $Z\Phi^\star\Phi$ vertex term.  Consequently, this operator can be probed via the mono-Higgs channel (shown in Fig.~\ref{feyn:mono-Higgs}) at the LHC through the following processes: $q\bar{q} \to Z^* h\to h \Phi^\star \Phi$, and $q\bar{q} \to Z^* \to h \Phi^\star \Phi$. The latter occurs through contact four-point interaction, yielding a definitive $h + \slashed{E}_T$ experimental signature. The projected sensitivities from Ref.~\cite{Carpenter:2013xra} at the $\sqrt{s} = 14$~TeV LHC with an integrated luminosity of $300\text{ fb}^{-1}$ in the diphoton plus missing transverse energy ($\gamma\gamma + \slashed{E}_T$) final state, are displayed in Fig.~\ref{fig:ODPhi_relic-scan}.  Although the $h \to \gamma\gamma$ branching ratio is significantly smaller than that of the dominant $h \to b\bar{b}$ mode, the exceptionally clean electromagnetic signature of the former provides a powerful handle for background reduction. Conversely, the $b\bar{b} + \slashed{E}_T$ channel suffers from overwhelming SM backgrounds, dominated primarily by $t\bar{t}$ production. This search reveals high sensitivity to the $\mathcal{O}_{D\Phi}$ operator, providing a lower limit on the new physics cutoff scale of $\Lambda_{\text{NP}} \sim 1$~TeV.

Additionally, we observe that the constraints obtained from the mono-Higgs channel are significantly more stringent than those derived from LHC mono-jet searches for the $\mathcal{O}_{D\Phi}$ operator. This is because the Higgs boson cannot be emitted as Initial State Radiation (ISR) from the incoming light quarks, due to their negligible Yukawa couplings. Instead, the mono-Higgs topology is entirely mediated by the hard, primary vertex that originates from the effective operator itself. As a result, the subsequent resonant $h \to \gamma\gamma$ decay provides a sharp kinematic peak and serves as a far superior probe.

\subsection{DM search at the FCC-ee}
\label{sec:collider3}
Future $e^{+}e^{-}$ colliders such as the FCC-ee offer a clean experimental environment with well-defined initial states, making them particularly suitable for probing light DM scenarios through missing energy signatures. In contrast to hadron colliders, the absence of large QCD backgrounds and the precise control over beam energy allow for an accurate reconstruction of kinematic observables. The dominant search channel in this context is the mono-$\gamma$ process\footnote{Additional search channels include mono-$Z$ and mono-$h$ signatures. However, these processes are typically too suppressed to yield a feasible signal significance at the FCC-ee.}, $e^{+}e^{-} \to \gamma + \slashed{E}$, where the missing energy is carried away by DM particles (Feynman diagrams same as LEP, shown in Fig.~\ref{feyn:mono-photon_LEP}). The primary SM background arises from neutrino pair production in association with a photon, $e^{+}e^{-} \to \nu\bar{\nu}\gamma$. The production cross sections for the signal benchmarks and background processes at different FCC-ee CM energies are presented in Tab.~\ref{tab:xsec-fcc}. A generation-level cut of  $p_T^\gamma > 10$ GeV is imposed to regulate infrared and collinear divergences, yielding a finite cross section. One observes that the background is comparatively larger at lower CM energies, except in the vicinity of the $Z$-pole. The SM production of $e^{+}e^{-} \to  \nu \bar{\nu} \gamma$ proceeds via two primary channels: $s$-channel $Z$-boson annihilation and $t$-channel $W$-exchange, the latter contributes exclusively to the $\nu_e \bar{\nu}_e$ final state. The relatively small cross section at $\sqrt{s}=91~\mathrm{GeV}$ arises because the emission of a hard photon ($p_T^\gamma > 10~\mathrm{GeV}$) shifts the invariant mass of the invisible system away from the $Z$-pole, thereby suppressing the resonant enhancement. The sharp increase in cross section at $\sqrt{s}=160~\mathrm{GeV}$ is a consequence of radiative return kinematics, where the system emits a hard ISR photon to bring the effective centre-of-mass energy down to the $Z$-pole, together with sizable $W$-exchange contributions. At higher energies, the cross section decreases as the process moves farther from the resonant region and follows the characteristic $1/s$ electroweak scaling, though this decline is tempered by the $t$-channel $W$-exchange contribution, which becomes increasingly prominent at higher energies. However, the $\sqrt{s}=91$ GeV run remains highly sensitive to the precise modeling of the $Z$ resonance, including effects from line-shape reconstruction and systematic uncertainties. A robust interpretation of potential new physics contributions in this regime therefore requires a dedicated treatment beyond the scope of this work. For this reason, we restrict our analysis to runs above the $Z$-pole.
\begin{table}[htb!]
    \centering
    \begin{tabular}{|c|c|c|c|}
    \hline
    \multirow{2}*{\textbf{$\boldsymbol{\sqrt{s}}$ (GeV)}} & \multicolumn{3}{c|}{\textbf{Production Cross Section (in fb)}} \\ \cline{2-4}
    & \textbf{Background} & \textbf{BP($\boldsymbol{{\mathcal{O}_{f\Phi}}}$)} & \textbf{BP($\boldsymbol{{\mathcal{O}_{D\Phi}}}$)} \\
    \hline
     91 & 248.2 & 0.0066 & 0.177 \\
     160 & 6502 & 0.0462 & 3.598 \\
     240 & 2896 & 0.1602 & 1.178 \\
     365 & 2140 & 0.5081 & 0.445 \\
    \hline
    \end{tabular}
    \caption{Production cross section of mono-$\gamma$ signal for DM benchmarks and background processes at different FCC-ee runs.}
    \label{tab:xsec-fcc}
\end{table}

We additionally observe that the benchmark BP($\mathcal{O}_{f\Phi}$) exhibits enhanced sensitivity at higher CM energies, which can be attributed to the contact-interaction nature of the operator, leading to a growth of the production rate with energy. In contrast, BP($\mathcal{O}_{D\Phi}$) is more effectively probed at lower energies, where the production is dominated by an $s$-channel mechanism that benefits from proximity to the resonance. Motivated by this complementarity, we perform our analysis for both the FCC-ee runs at $\sqrt{s}=160$ GeV and $\sqrt{s}=365$ GeV, which together provide optimal coverage of the parameter space for the operators under consideration.
\subsubsection{FCC-ee at 160 GeV}
\label{sec:collider31}
At $\sqrt{s}=160$~GeV, we define two mutually exclusive signal regions based on missing transverse energy ($\slashed{E}_T$) and total missing energy ($\slashed{E}$), as shown in Tab.~\ref{tab:cuts1}. These regions are designed to isolate kinematic regimes where the signal-to-background ratio is enhanced. Signal Region I targets events with relatively soft $\slashed{E}_T$ but large total missing energy, while Signal Region II focuses on harder $\slashed{E}_T$ configurations. The corresponding event yields and significances for an integrated luminosity of $3~\text{ab}^{-1}$ are presented in Tab.~\ref{tab:cuts1}. The signal significance $(\mathcal{S})$ is defined as
\begin{equation}
\mathcal{S} = \frac{S}{\sqrt{B}}\,,
\end{equation}
where $S$ and $B$ are signal and background event counts, respectively. For BP($\mathcal{O}_{f\Phi}$), the signal remains highly suppressed across both regions, resulting in negligible significance even at high luminosity. In contrast, BP($\mathcal{O}_{D\Phi}$) exhibits a comparatively stronger signal, particularly in Signal Region II, where the harder $\slashed{E}_T$ cut enhances the sensitivity. We further project the significance for higher luminosities of $10~\text{ab}^{-1}$ and $90~\text{ab}^{-1}$. While BP($\mathcal{O}_{f\Phi}$) remains beyond reach, BP($\mathcal{O}_{D\Phi}$) shows a significant improvement, reaching $\mathcal{S}\sim 6.4$ in Signal Region II at $90~\text{ab}^{-1}$. This demonstrates that the FCC-ee at $\sqrt{s}=160$~GeV can provide strong sensitivity to $\mathcal{O}_{D\Phi}$ in the high-luminosity regime.
\begin{table}[htb!]
    \centering
    \resizebox{\textwidth}{!}{
    \begin{tabular}{|c|c|c||c|c|c|}
    \hline
    \multicolumn{3}{|c||}{\textbf{Signal Region I}} & \multicolumn{3}{c|}{\textbf{Signal Region II}} \\
    \hline
     \multicolumn{3}{|c||}{$10\text{\;GeV}<\slashed{E}_{T}<45\text{\;GeV}$} & \multicolumn{3}{c|}{$45\text{\;GeV}<\slashed{E}_{T}<80\text{\;GeV}$} \\
     \multicolumn{3}{|c||}{$ 115\text{\;GeV}<\slashed{E}<150\text{\;GeV} $} & \multicolumn{3}{c|}{$ 80\text{\;GeV}<\slashed{E}<115\text{\;GeV} $} \\
    \hline
    \multicolumn{6}{c}{} \\
    \hline
    \textbf{Process} & \textbf{Events} & \textbf{Significance, $\boldsymbol{\mathcal{S}}$} & \textbf{Process} & \textbf{Events} & \textbf{Significance, $\boldsymbol{\mathcal{S}}$} \\
    \hline
    Background & 2349696 & $-$ & Background & 3684029 & $-$ \\
    BP(${\mathcal{O}_{f\Phi}}$) & 108 & 0.07 & BP(${\mathcal{O}_{f\Phi}}$) & 4 & $\sim 0$ \\
    BP(${\mathcal{O}_{D\Phi}}$) & 543 & 0.35 & BP(${\mathcal{O}_{D\Phi}}$) & 2242 & 1.17 \\
    \hline
    \multicolumn{6}{c}{} \\
    \hline
    \textbf{Benchmark} & \textbf{$\boldsymbol{\mathcal{S}\;(10\text{ ab}^{-1})}$} & \textbf{$\boldsymbol{\mathcal{S}\;(90\text{ ab}^{-1})}$} & \textbf{Benchmark} & \textbf{$\boldsymbol{\mathcal{S}\;(10\text{ ab}^{-1})}$} & \textbf{$\boldsymbol{\mathcal{S}\;(90\text{ ab}^{-1})}$} \\
    \hline
    BP(${\mathcal{O}_{f\Phi}}$) & 0.13 & 0.38 & BP(${\mathcal{O}_{f\Phi}}$) & $\sim 0$ & 0.01 \\
    BP(${\mathcal{O}_{D\Phi}}$) & 0.64 & 1.94 & BP(${\mathcal{O}_{D\Phi}}$) & 2.13 & 6.40 \\
    \hline
    \end{tabular}}
    \caption{Mutually exclusive signal regions (\textit{top table}), respective event counts and signal significance at the FCC-ee with CM energy of 160 GeV and integrated luminosity of 3 ab$^{-1}$ (\textit{middle table}), and, significance at higher luminosities (\textit{bottom table}).}
    \label{tab:cuts1}
\end{table}

\subsubsection{FCC-ee at 365 GeV}
\label{sec:collider32}
We again define two mutually exclusive signal regions at $\sqrt{s}=365$~GeV, optimized for the higher energy configuration, as shown in Tab.~\ref{tab:cuts2}. Compared to the $160$~GeV case, the kinematic reach is extended, allowing for a broader exploration of phase space. The event yields and significances at $3~\text{ab}^{-1}$ are listed in Tab.~\ref{tab:cuts2}. In this setup, BP($\mathcal{O}_{f\Phi}$) shows an enhanced signal in Signal Region I, benefiting from the increased production cross section at higher energies. The significance improves further with luminosity, reaching $\mathcal{S}\sim 3.8$ at $90~\text{ab}^{-1}$, indicating a promising discovery potential. On the other hand, BP($\mathcal{O}_{D\Phi}$) yields a suppressed signal in Signal Region I but shows moderate sensitivity in Signal Region II. The projected significance reaches $\mathcal{S}\sim 2.9$ at $90~\text{ab}^{-1}$, suggesting that while the sensitivity is weaker compared to the $160$~GeV run, it still provides a complementary probe.
\begin{table}[htb!]
    \centering
    \resizebox{\textwidth}{!}{
    \begin{tabular}{|c|c|c||c|c|c|}
    \hline
    \multicolumn{3}{|c||}{\textbf{Signal Region I}} & \multicolumn{3}{c|}{\textbf{Signal Region II}} \\
    \hline
     \multicolumn{3}{|c||}{$10\text{\;GeV}<\slashed{E}_{T}<95\text{\;GeV}$} & \multicolumn{3}{c|}{$95\text{\;GeV}<\slashed{E}_{T}<180\text{\;GeV}$} \\
     \multicolumn{3}{|c||}{$ 250\text{\;GeV}<\slashed{E}<350\text{\;GeV} $} & \multicolumn{3}{c|}{$ 150\text{\;GeV}<\slashed{E}<250\text{\;GeV} $} \\
    \hline
    \multicolumn{6}{c}{} \\
    \hline
    \textbf{Process} & \textbf{Events} & \textbf{Significance, $\boldsymbol{\mathcal{S}}$} & \textbf{Process} & \textbf{Events} & \textbf{Significance, $\boldsymbol{\mathcal{S}}$} \\
    \hline
    Background & 3369262 & $-$ & Background & 894621 & $-$ \\
    BP(${\mathcal{O}_{f\Phi}}$) & 1281 & 0.70 & BP(${\mathcal{O}_{f\Phi}}$) & 30 & 0.03 \\
    BP(${\mathcal{O}_{D\Phi}}$) & 25 & 0.01 & BP(${\mathcal{O}_{D\Phi}}$) & 497 & 0.52 \\
    \hline
    \multicolumn{6}{c}{} \\
    \hline
    \textbf{Benchmark} & \textbf{$\boldsymbol{\mathcal{S}\;(10\text{ ab}^{-1})}$} & \textbf{$\boldsymbol{\mathcal{S}\;(90\text{ ab}^{-1})}$} & \textbf{Benchmark} & \textbf{$\boldsymbol{\mathcal{S}\;(10\text{ ab}^{-1})}$} & \textbf{$\boldsymbol{\mathcal{S}\;(90\text{ ab}^{-1})}$} \\
    \hline
    BP(${\mathcal{O}_{f\Phi}}$) & 1.27 & 3.82 & BP(${\mathcal{O}_{f\Phi}}$) & 0.06 & 0.17 \\
    BP(${\mathcal{O}_{D\Phi}}$) & 0.02 & 0.08 & BP(${\mathcal{O}_{D\Phi}}$) & 0.96 & 2.88 \\
    \hline
    \end{tabular}}
    \caption{Mutually exclusive signal regions (top), respective event counts and signal significance at the FCC-ee with CM energy of 365 GeV and integrated luminosity of 3 ab$^{-1}$ (middle), and, significance at higher luminosities (bottom).}
    \label{tab:cuts2}
\end{table}
Overall, the FCC-ee demonstrates strong potential for probing light DM scenarios, with different energy stages offering complementary sensitivity to the effective operators considered in this work. 
\section*{Summary of the parameter space}
Finally, we summarize the allowed and excluded regions of the parameter space by identifying the ranges of DM mass and new physics scale constrained by current bounds. We also highlight the surviving regions that remain consistent with the observed DM relic abundance and can be probed in future dedicated DM and collider search experiments. Below, we discuss each operator scenario individually.
\begin{itemize}
\item[(i)] For the operator $\mathcal{O}_{f\Phi}$ (Fig.~\ref{fig:Of3_relic-scan}):
\begin{itemize}
\item[(a)] In the range $4~\mathrm{MeV} \lesssim \mPhi \lesssim 10~\mathrm{GeV}$, values of $\lNP \lesssim 1~\mathrm{TeV}$ are excluded by the combined constraints from $K_L \to \pi^0 \Phi \Phi^\star$, $B^0 \to \pi^0 \Phi \Phi^\star$, $K^0-\bar K^0$ mixing, and DAMIC-M experiment. In this region, the DM abundance is generated through the freeze-in mechanism, while the thermal freeze-out scenario is completely ruled out. Cosmological considerations further require $\Trh \gtrsim 0.1~\mathrm{GeV}$, to produce right DM abundance during reheating. The surviving parameter space can be explored in future direct detection experiments such as Oscura and through collider searches at the HL-LHC.
\item[(b)] For heavier DM masses, 
$10~\mathrm{GeV} \lesssim \mPhi < 4~\mathrm{TeV}$, direct detection constraints from LZ-2025 excludes $\lNP \lesssim 200~\mathrm{TeV}$. In this regime, the observed relic abundance is obtained via freeze-in mechanism for $\Trh \gtrsim 1~\mathrm{GeV}$. Future direct detection experiments, particularly XLZD, provide the best opportunity to probe the remaining parameter space. Compared to the $n=4$ case, $n=6$ scenario allows slightly more freedom in $\Trh$ and $\lNP$ along the relic-density contour.
\end{itemize}
\item[(ii)] For the operator $\mathcal{O}_{D\Phi}$ (Fig.~\ref{fig:ODPhi_relic-scan}):
\begin{itemize}
\item[(a)] For $4~\mathrm{MeV} \lesssim \mPhi \lesssim 10~\mathrm{GeV}$, $\lNP \lesssim 1~\mathrm{TeV}$ is excluded by the invisible decay width of the $Z$ boson. For $\mPhi \lesssim 200~\mathrm{MeV}$, SN1987A cooling bound further excludes $\lNP$ up to about $3~\mathrm{TeV}$. The most promising future probes in this mass range are HL-LHC mono-Higgs searches and DM--electron scattering experiments such as Oscura. The thermal DM scenario is once again disfavored, and the relic abundance requirement via freeze-in constrains the reheating temperature to $\Trh \lesssim 50~\mathrm{MeV}$ for $n=4$. In contrast, for $n=6$, values of $\lNP$ up to $1~\mathrm{TeV}$ remain allowed, whereas the $n=4$ case excludes $\lNP$ up to about $2~\mathrm{TeV}$.
\item[(b)] In the heavier mass range, 
$10~\mathrm{GeV} \lesssim \mPhi < 4~\mathrm{TeV}$, $\lNP\sim 5~\mathrm{TeV}$ is excluded by LZ-2025 direct search experiment. The thermal DM scenario survives only near the $Z$-resonance region; elsewhere, the observed relic density is achieved through non-thermal production during reheating. In this case, obtaining the correct FIMP relic density requires $\Trh \gtrsim 0.5~\mathrm{GeV}$ for $\lNP \gtrsim 5~\mathrm{TeV}$. The remaining viable parameter space can mainly be tested in future direct search experiments such as XLZD. 
\end{itemize}
\end{itemize}
\section{Conclusions}
\label{sec:concl}
At its core, this study addresses a fundamental question: to what extent can experimental data illuminate the otherwise inaccessible epoch preceding BBN? To investigate this possibility, we employ a minimal effective field theory framework that connects particle physics phenomenology with the cosmological dynamics of the pre-BBN Universe quite evidently. Within this setup, we examine dark matter production during the post-inflationary era, concentrating on scenarios governed at leading order by p-wave suppressed operators, which are generically less vulnerable to conventional direct and indirect detection constraints. For the operator classes considered here, current bounds from invisible decay searches, collider measurements, and dark matter direct detection together impose severe restrictions on freeze-out production during reheating, particularly excluding much of the parameter space associated with new physics scales below the TeV frontier. In addition, reheating temperatures below approximately $\mathcal{O}(10^{-2})$ GeV are effectively ruled out. By contrast, freeze-in production during reheating remains comparatively elusive under present constraints, largely because the extremely feeble couplings required for non-thermal generation naturally suppress observable signatures. Nevertheless, forthcoming direct detection facilities, combined with projected High-Luminosity LHC sensitivities, have the potential to substantially extend this reach, probing reheating temperatures up to roughly the GeV scale. Effective operators whose leading contributions proceed through s-wave channels would, in principle, be subject to even stronger scrutiny. The main results of our study are shown via Fig.~\ref{fig:Of3_relic-scan} and \ref{fig:ODPhi_relic-scan}, taking into account the existing and projected bounds from several different experiments that search for new physics beyond the Standard Model. On a broader perspective, the significance of this work lies not merely in constraining specific realizations of physics beyond the Standard Model, but in demonstrating that terrestrial and astrophysical searches for new physics can also function as indirect probes of the Universe’s earliest thermal history. In this sense, present and future experimental programs may offer a rare empirical pathway toward reconstructing aspects of the pre-BBN cosmological epoch.
\section*{Acknowledgments}
The authors would like to thank IIT, Hyderabad for their hospitality during WHEPP-2025, where this work was initiated during working group discussion sessions. We thank Sushant Yadav for being part of the project at the very early stage.

\begin{thebibliography}{100}
	
	\bibitem{Roszkowski:2017nbc}
	L.~Roszkowski, E.M.~Sessolo and S.~Trojanowski, \emph{{WIMP dark matter
			candidates and searches\textemdash{}current status and future prospects}},
	\href{https://doi.org/10.1088/1361-6633/aab913}{\emph{Rept. Prog. Phys.}
		{\bfseries 81} (2018) 066201}
	[\href{https://arxiv.org/abs/1707.06277}{{\ttfamily 1707.06277}}].
	
	\bibitem{Arcadi:2017kky}
	G.~Arcadi, M.~Dutra, P.~Ghosh, M.~Lindner, Y.~Mambrini, M.~Pierre et~al.,
	\emph{{The waning of the WIMP? A review of models, searches, and
			constraints}},
	\href{https://doi.org/10.1140/epjc/s10052-018-5662-y}{\emph{Eur. Phys. J. C}
		{\bfseries 78} (2018) 203}
	[\href{https://arxiv.org/abs/1703.07364}{{\ttfamily 1703.07364}}].
	
	\bibitem{Chang:2019xva}
	J.H.~Chang, R.~Essig and A.~Reinert, \emph{{Light(ly)-coupled Dark Matter in
			the keV Range: Freeze-In and Constraints}},
	\href{https://doi.org/10.1007/JHEP03(2021)141}{\emph{JHEP} {\bfseries 03}
		(2021) 141} [\href{https://arxiv.org/abs/1911.03389}{{\ttfamily
			1911.03389}}].
	
	\bibitem{Darme:2019wpd}
	L.~Darm\'e, A.~Hryczuk, D.~Karamitros and L.~Roszkowski, \emph{{Forbidden
			frozen-in dark matter}},
	\href{https://doi.org/10.1007/JHEP11(2019)159}{\emph{JHEP} {\bfseries 11}
		(2019) 159} [\href{https://arxiv.org/abs/1908.05685}{{\ttfamily
			1908.05685}}].
	
	\bibitem{Arcadi:2024ukq}
	G.~Arcadi, D.~Cabo-Almeida, M.~Dutra, P.~Ghosh, M.~Lindner, Y.~Mambrini et~al.,
	\emph{{The Waning of the WIMP: Endgame?}},
	\href{https://doi.org/10.1140/epjc/s10052-024-13672-y}{\emph{Eur. Phys. J. C}
		{\bfseries 85} (2025) 152}
	[\href{https://arxiv.org/abs/2403.15860}{{\ttfamily 2403.15860}}].
	
	\bibitem{McDonald:2001vt}
	J.~McDonald, \emph{{Thermally generated gauge singlet scalars as
			selfinteracting dark matter}},
	\href{https://doi.org/10.1103/PhysRevLett.88.091304}{\emph{Phys. Rev. Lett.}
		{\bfseries 88} (2002) 091304}
	[\href{https://arxiv.org/abs/hep-ph/0106249}{{\ttfamily hep-ph/0106249}}].
	
	\bibitem{Hall:2009bx}
	L.J.~Hall, K.~Jedamzik, J.~March-Russell and S.M.~West, \emph{{Freeze-In
			Production of FIMP Dark Matter}},
	\href{https://doi.org/10.1007/JHEP03(2010)080}{\emph{JHEP} {\bfseries 03}
		(2010) 080} [\href{https://arxiv.org/abs/0911.1120}{{\ttfamily 0911.1120}}].
	
	\bibitem{Bernal:2017kxu}
	N.~Bernal, M.~Heikinheimo, T.~Tenkanen, K.~Tuominen and V.~Vaskonen, \emph{{The
			Dawn of FIMP Dark Matter: A Review of Models and Constraints}},
	\href{https://doi.org/10.1142/S0217751X1730023X}{\emph{Int. J. Mod. Phys. A}
		{\bfseries 32} (2017) 1730023}
	[\href{https://arxiv.org/abs/1706.07442}{{\ttfamily 1706.07442}}].
	
	\bibitem{Elahi:2014fsa}
	F.~Elahi, C.~Kolda and J.~Unwin, \emph{{UltraViolet Freeze-in}},
	\href{https://doi.org/10.1007/JHEP03(2015)048}{\emph{JHEP} {\bfseries 03}
		(2015) 048} [\href{https://arxiv.org/abs/1410.6157}{{\ttfamily 1410.6157}}].
	
	\bibitem{Barman:2020plp}
	B.~Barman, D.~Borah and R.~Roshan, \emph{{Effective Theory of Freeze-in Dark
			Matter}}, \href{https://doi.org/10.1088/1475-7516/2020/11/021}{\emph{JCAP}
		{\bfseries 11} (2020) 021}
	[\href{https://arxiv.org/abs/2007.08768}{{\ttfamily 2007.08768}}].
	
	\bibitem{Giudice:2000ex}
	G.F.~Giudice, E.W.~Kolb and A.~Riotto, \emph{{Largest temperature of the
			radiation era and its cosmological implications}},
	\href{https://doi.org/10.1103/PhysRevD.64.023508}{\emph{Phys. Rev. D}
		{\bfseries 64} (2001) 023508}
	[\href{https://arxiv.org/abs/hep-ph/0005123}{{\ttfamily hep-ph/0005123}}].
	
	\bibitem{Bernal:2019mhf}
	N.~Bernal, F.~Elahi, C.~Maldonado and J.~Unwin, \emph{{Ultraviolet Freeze-in
			and Non-Standard Cosmologies}},
	\href{https://doi.org/10.1088/1475-7516/2019/11/026}{\emph{JCAP} {\bfseries
			11} (2019) 026} [\href{https://arxiv.org/abs/1909.07992}{{\ttfamily
			1909.07992}}].
	
	\bibitem{Bernal:2020qyu}
	N.~Bernal, J.~Rubio and H.~Veerm\"ae, \emph{{UV Freeze-in in Starobinsky
			Inflation}}, \href{https://doi.org/10.1088/1475-7516/2020/10/021}{\emph{JCAP}
		{\bfseries 10} (2020) 021}
	[\href{https://arxiv.org/abs/2006.02442}{{\ttfamily 2006.02442}}].
	
	\bibitem{Garcia:2020eof}
	M.A.G.~Garcia, K.~Kaneta, Y.~Mambrini and K.A.~Olive, \emph{{Reheating and
			Post-inflationary Production of Dark Matter}},
	\href{https://doi.org/10.1103/PhysRevD.101.123507}{\emph{Phys. Rev. D}
		{\bfseries 101} (2020) 123507}
	[\href{https://arxiv.org/abs/2004.08404}{{\ttfamily 2004.08404}}].
	
	\bibitem{Garcia:2020wiy}
	M.A.G.~Garcia, K.~Kaneta, Y.~Mambrini and K.A.~Olive, \emph{{Inflaton
			Oscillations and Post-Inflationary Reheating}},
	\href{https://doi.org/10.1088/1475-7516/2021/04/012}{\emph{JCAP} {\bfseries
			04} (2021) 012} [\href{https://arxiv.org/abs/2012.10756}{{\ttfamily
			2012.10756}}].
	
	\bibitem{Ahmed:2021try}
	A.~Ahmed and S.~Najjari, \emph{{Ultraviolet freeze-in dark matter through the
			dilaton portal}},
	\href{https://doi.org/10.1103/PhysRevD.107.055020}{\emph{Phys. Rev. D}
		{\bfseries 107} (2023) 055020}
	[\href{https://arxiv.org/abs/2112.14261}{{\ttfamily 2112.14261}}].
	
	\bibitem{Mambrini:2021zpp}
	Y.~Mambrini and K.A.~Olive, \emph{{Gravitational Production of Dark Matter
			during Reheating}},
	\href{https://doi.org/10.1103/PhysRevD.103.115009}{\emph{Phys. Rev. D}
		{\bfseries 103} (2021) 115009}
	[\href{https://arxiv.org/abs/2102.06214}{{\ttfamily 2102.06214}}].
	
	\bibitem{Kaneta:2021pyx}
	K.~Kaneta, P.~Ko and W.-I.~Park, \emph{{Conformal portal to dark matter}},
	\href{https://doi.org/10.1103/PhysRevD.104.075018}{\emph{Phys. Rev. D}
		{\bfseries 104} (2021) 075018}
	[\href{https://arxiv.org/abs/2106.01923}{{\ttfamily 2106.01923}}].
	
	\bibitem{Ahmed:2021fvt}
	A.~Ahmed, B.~Grzadkowski and A.~Socha, \emph{{Implications of time-dependent
			inflaton decay on reheating and dark matter production}},
	\href{https://doi.org/10.1016/j.physletb.2022.137201}{\emph{Phys. Lett. B}
		{\bfseries 831} (2022) 137201}
	[\href{https://arxiv.org/abs/2111.06065}{{\ttfamily 2111.06065}}].
	
	\bibitem{Barman:2022tzk}
	B.~Barman, N.~Bernal, Y.~Xu and {\'O}.~Zapata, \emph{{Ultraviolet freeze-in
			with a time-dependent inflaton decay}},
	\href{https://doi.org/10.1088/1475-7516/2022/07/019}{\emph{JCAP} {\bfseries
			07} (2022) 019} [\href{https://arxiv.org/abs/2202.12906}{{\ttfamily
			2202.12906}}].
	
	\bibitem{Haque:2023yra}
	M.R.~Haque, D.~Maity and R.~Mondal, \emph{{WIMPs, FIMPs, and Inflaton
			phenomenology via reheating, CMB and \ensuremath{\Delta}N$_{eff}$}},
	\href{https://doi.org/10.1007/JHEP09(2023)012}{\emph{JHEP} {\bfseries 09}
		(2023) 012} [\href{https://arxiv.org/abs/2301.01641}{{\ttfamily
			2301.01641}}].
	
	\bibitem{Becker:2023tvd}
	M.~Becker, E.~Copello, J.~Harz, J.~Lang and Y.~Xu, \emph{{Confronting dark
			matter freeze-in during reheating with constraints from inflation}},
	\href{https://doi.org/10.1088/1475-7516/2024/01/053}{\emph{JCAP} {\bfseries
			01} (2024) 053} [\href{https://arxiv.org/abs/2306.17238}{{\ttfamily
			2306.17238}}].
	
	\bibitem{Bernal:2024ykj}
	N.~Bernal, J.~Harz, M.A.~Mojahed and Y.~Xu, \emph{{Graviton- and
			Inflaton-mediated Dark Matter Production after Large Field Polynomial
			Inflation}},  \href{https://arxiv.org/abs/2406.19447}{{\ttfamily
			2406.19447}}.
	
	\bibitem{Barman:2024ujh}
	B.~Barman, A.~Basu, D.~Borah, A.~Chakraborty and R.~Roshan, \emph{{Testing
			cogenesis during reheating with primordial gravitational waves}},
	\href{https://arxiv.org/abs/2410.19048}{{\ttfamily 2410.19048}}.
	
	\bibitem{Beltran:2008xg}
	M.~Beltran, D.~Hooper, E.W.~Kolb and Z.C.~Krusberg, \emph{{Deducing the nature
			of dark matter from direct and indirect detection experiments in the absence
			of collider signatures of new physics}},
	\href{https://doi.org/10.1103/PhysRevD.80.043509}{\emph{Phys. Rev. D}
		{\bfseries 80} (2009) 043509}
	[\href{https://arxiv.org/abs/0808.3384}{{\ttfamily 0808.3384}}].
	
	\bibitem{Cao:2009uw}
	Q.-H.~Cao, C.-R.~Chen, C.S.~Li and H.~Zhang, \emph{{Effective Dark Matter
			Model: Relic density, CDMS II, Fermi LAT and LHC}},
	\href{https://doi.org/10.1007/JHEP08(2011)018}{\emph{JHEP} {\bfseries 08}
		(2011) 018} [\href{https://arxiv.org/abs/0912.4511}{{\ttfamily 0912.4511}}].
	
	\bibitem{Goodman:2010ku}
	J.~Goodman, M.~Ibe, A.~Rajaraman, W.~Shepherd, T.M.P.~Tait and H.-B.~Yu,
	\emph{{Constraints on Dark Matter from Colliders}},
	\href{https://doi.org/10.1103/PhysRevD.82.116010}{\emph{Phys. Rev. D}
		{\bfseries 82} (2010) 116010}
	[\href{https://arxiv.org/abs/1008.1783}{{\ttfamily 1008.1783}}].
	
	\bibitem{Buckley:2011kk}
	M.R.~Buckley, \emph{{Asymmetric Dark Matter and Effective Operators}},
	\href{https://doi.org/10.1103/PhysRevD.84.043510}{\emph{Phys. Rev. D}
		{\bfseries 84} (2011) 043510}
	[\href{https://arxiv.org/abs/1104.1429}{{\ttfamily 1104.1429}}].
	
	\bibitem{Abdallah:2015ter}
	J.~Abdallah et~al., \emph{{Simplified Models for Dark Matter Searches at the
			LHC}}, \href{https://doi.org/10.1016/j.dark.2015.08.001}{\emph{Phys. Dark
			Univ.} {\bfseries 9-10} (2015) 8}
	[\href{https://arxiv.org/abs/1506.03116}{{\ttfamily 1506.03116}}].
	
	\bibitem{Berlin:2014tja}
	A.~Berlin, D.~Hooper and S.D.~McDermott, \emph{{Simplified Dark Matter Models
			for the Galactic Center Gamma-Ray Excess}},
	\href{https://doi.org/10.1103/PhysRevD.89.115022}{\emph{Phys. Rev. D}
		{\bfseries 89} (2014) 115022}
	[\href{https://arxiv.org/abs/1404.0022}{{\ttfamily 1404.0022}}].
	
	\bibitem{Kahlhoefer:2017dnp}
	F.~Kahlhoefer, \emph{{Review of LHC Dark Matter Searches}},
	\href{https://doi.org/10.1142/S0217751X1730006X}{\emph{Int. J. Mod. Phys. A}
		{\bfseries 32} (2017) 1730006}
	[\href{https://arxiv.org/abs/1702.02430}{{\ttfamily 1702.02430}}].
	
	\bibitem{Bertuzzo:2017lwt}
	E.~Bertuzzo, C.J.~Caniu~Barros and G.~Grilli~di Cortona, \emph{{MeV Dark
			Matter: Model Independent Bounds}},
	\href{https://doi.org/10.1007/JHEP09(2017)116}{\emph{JHEP} {\bfseries 09}
		(2017) 116} [\href{https://arxiv.org/abs/1707.00725}{{\ttfamily
			1707.00725}}].
	
	\bibitem{Belyaev:2018pqr}
	A.~Belyaev, E.~Bertuzzo, C.~Caniu~Barros, O.~Eboli, G.~Grilli Di~Cortona,
	F.~Iocco et~al., \emph{{Interplay of the LHC and non-LHC Dark Matter searches
			in the Effective Field Theory approach}},
	\href{https://doi.org/10.1103/PhysRevD.99.015006}{\emph{Phys. Rev. D}
		{\bfseries 99} (2019) 015006}
	[\href{https://arxiv.org/abs/1807.03817}{{\ttfamily 1807.03817}}].
	
	\bibitem{Bhattacharya:2021edh}
	S.~Bhattacharya and J.~Wudka, \emph{{Effective theories with dark matter
			applications}}, \href{https://doi.org/10.1142/S0218271821300044}{\emph{Int.
			J. Mod. Phys. D} {\bfseries 30} (2021) 2130004}
	[\href{https://arxiv.org/abs/2104.01788}{{\ttfamily 2104.01788}}].
	
	\bibitem{Barman:2021hhg}
	B.~Barman, S.~Bhattacharya, S.~Girmohanta and S.~Jahedi, \emph{{Effective
			Leptophilic WIMPs at the $e^+ e^-$ collider}},
	\href{https://doi.org/10.1007/JHEP04(2022)146}{\emph{JHEP} {\bfseries 04}
		(2022) 146} [\href{https://arxiv.org/abs/2109.10936}{{\ttfamily
			2109.10936}}].
	
	\bibitem{Barman:2021ugy}
	B.~Barman and N.~Bernal, \emph{{Gravitational SIMPs}},
	\href{https://doi.org/10.1088/1475-7516/2021/06/011}{\emph{JCAP} {\bfseries
			06} (2021) 011} [\href{https://arxiv.org/abs/2104.10699}{{\ttfamily
			2104.10699}}].
	
	\bibitem{Barman:2022qgt}
	B.~Barman, S.~Cl\'ery, R.T.~Co, Y.~Mambrini and K.A.~Olive, \emph{{Gravity as a
			portal to reheating, leptogenesis and dark matter}},
	\href{https://doi.org/10.1007/JHEP12(2022)072}{\emph{JHEP} {\bfseries 12}
		(2022) 072} [\href{https://arxiv.org/abs/2210.05716}{{\ttfamily
			2210.05716}}].
	
	\bibitem{Bernal:2022wck}
	N.~Bernal and Y.~Xu, \emph{{WIMPs during reheating}},
	\href{https://doi.org/10.1088/1475-7516/2022/12/017}{\emph{JCAP} {\bfseries
			12} (2022) 017} [\href{https://arxiv.org/abs/2209.07546}{{\ttfamily
			2209.07546}}].
	
	\bibitem{Silva-Malpartida:2023yks}
	J.~Silva-Malpartida, N.~Bernal, J.~Jones-P\'erez and R.A.~Lineros, \emph{{From
			WIMPs to FIMPs with low~reheating~temperatures}},
	\href{https://doi.org/10.1088/1475-7516/2023/09/015}{\emph{JCAP} {\bfseries
			09} (2023) 015} [\href{https://arxiv.org/abs/2306.14943}{{\ttfamily
			2306.14943}}].
	
	\bibitem{Silva-Malpartida:2024emu}
	J.~Silva-Malpartida, N.~Bernal, J.~Jones-P{\'e}rez and R.A.~Lineros,
	\emph{{From WIMPs to FIMPs: impact of early matter domination}},
	\href{https://doi.org/10.1088/1475-7516/2025/03/003}{\emph{JCAP} {\bfseries
			03} (2025) 003} [\href{https://arxiv.org/abs/2408.08950}{{\ttfamily
			2408.08950}}].
	
	\bibitem{DeSimone:2013gj}
	A.~De~Simone, A.~Monin, A.~Thamm and A.~Urbano, \emph{{On the effective
			operators for Dark Matter annihilations}},
	\href{https://doi.org/10.1088/1475-7516/2013/02/039}{\emph{JCAP} {\bfseries
			02} (2013) 039} [\href{https://arxiv.org/abs/1301.1486}{{\ttfamily
			1301.1486}}].
	
	\bibitem{Li:2018orw}
	T.~Li and Y.~Liao, \emph{{Constraints on dimension-seven operators with a
			derivative in effective field theory for Dirac dark matter}},
	\href{https://doi.org/10.1016/j.physletb.2019.07.022}{\emph{Phys. Lett. B}
		{\bfseries 796} (2019) 6} [\href{https://arxiv.org/abs/1811.08200}{{\ttfamily
			1811.08200}}].
	
	\bibitem{Borah:2025ema}
	D.~Borah, N.~Das, S.~Jahedi and D.~Pradhan, \emph{{Multi-messenger FIMP}},
	\href{https://doi.org/10.1007/JHEP11(2025)049}{\emph{JHEP} {\bfseries 11}
		(2025) 049} [\href{https://arxiv.org/abs/2506.13860}{{\ttfamily
			2506.13860}}].
	
	\bibitem{Leike:1998wr}
	A.~Leike, \emph{{The Phenomenology of extra neutral gauge bosons}},
	\href{https://doi.org/10.1016/S0370-1573(98)00133-1}{\emph{Phys. Rept.}
		{\bfseries 317} (1999) 143}
	[\href{https://arxiv.org/abs/hep-ph/9805494}{{\ttfamily hep-ph/9805494}}].
	
	\bibitem{Langacker:2009im}
	P.~Langacker, \emph{{The Physics of New U(1)-prime Gauge Bosons}},
	\href{https://doi.org/10.1063/1.3327671}{\emph{AIP Conf. Proc.} {\bfseries
			1200} (2010) 55} [\href{https://arxiv.org/abs/0909.3260}{{\ttfamily
			0909.3260}}].
	
	\bibitem{Das:2016zue}
	A.~Das, S.~Oda, N.~Okada and D.-s.~Takahashi, \emph{{Classically conformal
			U(1)' extended standard model, electroweak vacuum stability, and LHC Run-2
			bounds}}, \href{https://doi.org/10.1103/PhysRevD.93.115038}{\emph{Phys. Rev.
			D} {\bfseries 93} (2016) 115038}
	[\href{https://arxiv.org/abs/1605.01157}{{\ttfamily 1605.01157}}].
	
	\bibitem{Basso:2008iv}
	L.~Basso, A.~Belyaev, S.~Moretti and C.H.~Shepherd-Themistocleous,
	\emph{{Phenomenology of the minimal B-L extension of the Standard model: Z'
			and neutrinos}},
	\href{https://doi.org/10.1103/PhysRevD.80.055030}{\emph{Phys. Rev. D}
		{\bfseries 80} (2009) 055030}
	[\href{https://arxiv.org/abs/0812.4313}{{\ttfamily 0812.4313}}].
	
	\bibitem{Langacker:2000ju}
	P.~Langacker and M.~Plumacher, \emph{{Flavor changing effects in theories with
			a heavy $Z^\prime$ boson with family nonuniversal couplings}},
	\href{https://doi.org/10.1103/PhysRevD.62.013006}{\emph{Phys. Rev. D}
		{\bfseries 62} (2000) 013006}
	[\href{https://arxiv.org/abs/hep-ph/0001204}{{\ttfamily hep-ph/0001204}}].
	
	\bibitem{Bobeth:2016llm}
	C.~Bobeth, A.J.~Buras, A.~Celis and M.~Jung, \emph{{Patterns of Flavour
			Violation in Models with Vector-Like Quarks}},
	\href{https://doi.org/10.1007/JHEP04(2017)079}{\emph{JHEP} {\bfseries 04}
		(2017) 079} [\href{https://arxiv.org/abs/1609.04783}{{\ttfamily
			1609.04783}}].
	
	\bibitem{Kallosh:2013hoa}
	R.~Kallosh and A.~Linde, \emph{{Universality Class in Conformal Inflation}},
	\href{https://doi.org/10.1088/1475-7516/2013/07/002}{\emph{JCAP} {\bfseries
			07} (2013) 002} [\href{https://arxiv.org/abs/1306.5220}{{\ttfamily
			1306.5220}}].
	
	\bibitem{Kallosh:2013yoa}
	R.~Kallosh, A.~Linde and D.~Roest, \emph{{Superconformal Inflationary
			$\alpha$-Attractors}},
	\href{https://doi.org/10.1007/JHEP11(2013)198}{\emph{JHEP} {\bfseries 11}
		(2013) 198} [\href{https://arxiv.org/abs/1311.0472}{{\ttfamily 1311.0472}}].
	
	\bibitem{Kallosh:2013maa}
	R.~Kallosh and A.~Linde, \emph{{Non-minimal Inflationary Attractors}},
	\href{https://doi.org/10.1088/1475-7516/2013/10/033}{\emph{JCAP} {\bfseries
			10} (2013) 033} [\href{https://arxiv.org/abs/1307.7938}{{\ttfamily
			1307.7938}}].
	
	\bibitem{Kallosh:2015lwa}
	R.~Kallosh and A.~Linde, \emph{{Planck, LHC, and $\alpha$-attractors}},
	\href{https://doi.org/10.1103/PhysRevD.91.083528}{\emph{Phys. Rev. D}
		{\bfseries 91} (2015) 083528}
	[\href{https://arxiv.org/abs/1502.07733}{{\ttfamily 1502.07733}}].
	
	\bibitem{Starobinsky:1980te}
	A.A.~Starobinsky, \emph{{A New Type of Isotropic Cosmological Models Without
			Singularity}},
	\href{https://doi.org/10.1016/0370-2693(80)90670-X}{\emph{Phys. Lett. B}
		{\bfseries 91} (1980) 99}.
	
	\bibitem{Starobinsky:1981vz}
	A.A.~Starobinsky, \emph{{Nonsingular Model of the Universe with the Quantum
			Gravitational de Sitter Stage and its Observational Consequences}},  in
	\emph{{Second Seminar on Quantum Gravity}}, 1981.
	
	\bibitem{Starobinsky:1983zz}
	A.A.~Starobinsky, \emph{{The Perturbation Spectrum Evolving from a Nonsingular
			Initially De-Sitter Cosmology and the Microwave Background Anisotropy}},
	{\emph{Sov. Astron. Lett.} {\bfseries 9} (1983) 302}.
	
	\bibitem{Kofman:1985aw}
	L.A.~Kofman, A.D.~Linde and A.A.~Starobinsky, \emph{{Inflationary Universe
			Generated by the Combined Action of a Scalar Field and Gravitational Vacuum
			Polarization}},
	\href{https://doi.org/10.1016/0370-2693(85)90381-8}{\emph{Phys. Lett. B}
		{\bfseries 157} (1985) 361}.
	
	\bibitem{Turner:1983he}
	M.S.~Turner, \emph{{Coherent Scalar Field Oscillations in an Expanding
			Universe}}, \href{https://doi.org/10.1103/PhysRevD.28.1243}{\emph{Phys. Rev.
			D} {\bfseries 28} (1983) 1243}.
	
	\bibitem{Sarkar:1995dd}
	S.~Sarkar, \emph{{Big bang nucleosynthesis and physics beyond the standard
			model}}, \href{https://doi.org/10.1088/0034-4885/59/12/001}{\emph{Rept. Prog.
			Phys.} {\bfseries 59} (1996) 1493}
	[\href{https://arxiv.org/abs/hep-ph/9602260}{{\ttfamily hep-ph/9602260}}].
	
	\bibitem{Kawasaki:2000en}
	M.~Kawasaki, K.~Kohri and N.~Sugiyama, \emph{{MeV scale reheating temperature
			and thermalization of neutrino background}},
	\href{https://doi.org/10.1103/PhysRevD.62.023506}{\emph{Phys. Rev. D}
		{\bfseries 62} (2000) 023506}
	[\href{https://arxiv.org/abs/astro-ph/0002127}{{\ttfamily
			astro-ph/0002127}}].
	
	\bibitem{Hannestad:2004px}
	S.~Hannestad, \emph{{What is the lowest possible reheating temperature?}},
	\href{https://doi.org/10.1103/PhysRevD.70.043506}{\emph{Phys. Rev. D}
		{\bfseries 70} (2004) 043506}
	[\href{https://arxiv.org/abs/astro-ph/0403291}{{\ttfamily
			astro-ph/0403291}}].
	
	\bibitem{DeBernardis:2008zz}
	F.~De~Bernardis, L.~Pagano and A.~Melchiorri, \emph{{New constraints on the
			reheating temperature of the universe after WMAP-5}},
	\href{https://doi.org/10.1016/j.astropartphys.2008.09.005}{\emph{Astropart.
			Phys.} {\bfseries 30} (2008) 192}.
	
	\bibitem{deSalas:2015glj}
	P.F.~de~Salas, M.~Lattanzi, G.~Mangano, G.~Miele, S.~Pastor and O.~Pisanti,
	\emph{{Bounds on very low reheating scenarios after Planck}},
	\href{https://doi.org/10.1103/PhysRevD.92.123534}{\emph{Phys. Rev. D}
		{\bfseries 92} (2015) 123534}
	[\href{https://arxiv.org/abs/1511.00672}{{\ttfamily 1511.00672}}].
	
	\bibitem{Hasegawa:2019jsa}
	T.~Hasegawa, N.~Hiroshima, K.~Kohri, R.S.L.~Hansen, T.~Tram and S.~Hannestad,
	\emph{{MeV-scale reheating temperature and thermalization of oscillating
			neutrinos by radiative and hadronic decays of massive particles}},
	\href{https://doi.org/10.1088/1475-7516/2019/12/012}{\emph{JCAP} {\bfseries
			12} (2019) 012} [\href{https://arxiv.org/abs/1908.10189}{{\ttfamily
			1908.10189}}].
	
	\bibitem{Lozanov:2016hid}
	K.D.~Lozanov and M.A.~Amin, \emph{{Equation of State and Duration to Radiation
			Domination after Inflation}},
	\href{https://doi.org/10.1103/PhysRevLett.119.061301}{\emph{Phys. Rev. Lett.}
		{\bfseries 119} (2017) 061301}
	[\href{https://arxiv.org/abs/1608.01213}{{\ttfamily 1608.01213}}].
	
	\bibitem{Lozanov:2017hjm}
	K.D.~Lozanov and M.A.~Amin, \emph{{Self-resonance after inflation: oscillons,
			transients and radiation domination}},
	\href{https://doi.org/10.1103/PhysRevD.97.023533}{\emph{Phys. Rev. D}
		{\bfseries 97} (2018) 023533}
	[\href{https://arxiv.org/abs/1710.06851}{{\ttfamily 1710.06851}}].
	
	\bibitem{Kofman:1997yn}
	L.~Kofman, A.D.~Linde and A.A.~Starobinsky, \emph{{Towards the theory of
			reheating after inflation}},
	\href{https://doi.org/10.1103/PhysRevD.56.3258}{\emph{Phys. Rev. D}
		{\bfseries 56} (1997) 3258}
	[\href{https://arxiv.org/abs/hep-ph/9704452}{{\ttfamily hep-ph/9704452}}].
	
	\bibitem{Dufaux:2006ee}
	J.F.~Dufaux, G.N.~Felder, L.~Kofman, M.~Peloso and D.~Podolsky,
	\emph{{Preheating with trilinear interactions: Tachyonic resonance}},
	\href{https://doi.org/10.1088/1475-7516/2006/07/006}{\emph{JCAP} {\bfseries
			07} (2006) 006} [\href{https://arxiv.org/abs/hep-ph/0602144}{{\ttfamily
			hep-ph/0602144}}].
	
	\bibitem{Greene:1997fu}
	P.B.~Greene, L.~Kofman, A.D.~Linde and A.A.~Starobinsky, \emph{{Structure of
			resonance in preheating after inflation}},
	\href{https://doi.org/10.1103/PhysRevD.56.6175}{\emph{Phys. Rev. D}
		{\bfseries 56} (1997) 6175}
	[\href{https://arxiv.org/abs/hep-ph/9705347}{{\ttfamily hep-ph/9705347}}].
	
	\bibitem{Green:1999yh}
	A.M.~Green, \emph{{Supersymmetry and primordial black hole abundance
			constraints}}, \href{https://doi.org/10.1103/PhysRevD.60.063516}{\emph{Phys.
			Rev. D} {\bfseries 60} (1999) 063516}
	[\href{https://arxiv.org/abs/astro-ph/9903484}{{\ttfamily
			astro-ph/9903484}}].
	
	\bibitem{Amin:2011hj}
	M.A.~Amin, R.~Easther, H.~Finkel, R.~Flauger and M.P.~Hertzberg,
	\emph{{Oscillons After Inflation}},
	\href{https://doi.org/10.1103/PhysRevLett.108.241302}{\emph{Phys. Rev. Lett.}
		{\bfseries 108} (2012) 241302}
	[\href{https://arxiv.org/abs/1106.3335}{{\ttfamily 1106.3335}}].
	
	\bibitem{Figueroa:2016wxr}
	D.G.~Figueroa and F.~Torrenti, \emph{{Parametric resonance in the early
			Universe{\textemdash}a fitting analysis}},
	\href{https://doi.org/10.1088/1475-7516/2017/02/001}{\emph{JCAP} {\bfseries
			02} (2017) 001} [\href{https://arxiv.org/abs/1609.05197}{{\ttfamily
			1609.05197}}].
	
	\bibitem{Garcia:2023eol}
	M.A.G.~Garcia and M.~Pierre, \emph{{Reheating after inflaton fragmentation}},
	\href{https://doi.org/10.1088/1475-7516/2023/11/004}{\emph{JCAP} {\bfseries
			11} (2023) 004} [\href{https://arxiv.org/abs/2306.08038}{{\ttfamily
			2306.08038}}].
	
	\bibitem{Garcia:2023dyf}
	M.A.G.~Garcia, M.~Gross, Y.~Mambrini, K.A.~Olive, M.~Pierre and J.-H.~Yoon,
	\emph{{Effects of fragmentation on post-inflationary reheating}},
	\href{https://doi.org/10.1088/1475-7516/2023/12/028}{\emph{JCAP} {\bfseries
			12} (2023) 028} [\href{https://arxiv.org/abs/2308.16231}{{\ttfamily
			2308.16231}}].
	
	\bibitem{Barman:2025lvk}
	B.~Barman, N.~Bernal and J.~Rubio, \emph{{Two or three things particle
			physicists (mis)understand about (pre)heating}},
	\href{https://doi.org/10.1016/j.nuclphysb.2025.116996}{\emph{Nucl. Phys. B}
		{\bfseries 1018} (2025) 116996}
	[\href{https://arxiv.org/abs/2503.19980}{{\ttfamily 2503.19980}}].
	
	\bibitem{Bernal:2023wus}
	N.~Bernal, S.~Cl{\'e}ry, Y.~Mambrini and Y.~Xu, \emph{{Probing reheating with
			graviton bremsstrahlung}},
	\href{https://doi.org/10.1088/1475-7516/2024/01/065}{\emph{JCAP} {\bfseries
			01} (2024) 065} [\href{https://arxiv.org/abs/2311.12694}{{\ttfamily
			2311.12694}}].
	
	\bibitem{ParticleDataGroup:2020ssz}
	{\scshape Particle Data Group} collaboration, \emph{{Review of Particle
			Physics}}, \href{https://doi.org/10.1093/ptep/ptaa104}{\emph{PTEP} {\bfseries
			2020} (2020) 083C01}.
	
	\bibitem{Planck:2018vyg}
	{\scshape Planck} collaboration, \emph{{Planck 2018 results. VI. Cosmological
			parameters}},
	\href{https://doi.org/10.1051/0004-6361/201833910}{\emph{Astron. Astrophys.}
		{\bfseries 641} (2020) A6}
	[\href{https://arxiv.org/abs/1807.06209}{{\ttfamily 1807.06209}}].
	
	\bibitem{Kaneta:2019zgw}
	K.~Kaneta, Y.~Mambrini and K.A.~Olive, \emph{{Radiative production of
			nonthermal dark matter}},
	\href{https://doi.org/10.1103/PhysRevD.99.063508}{\emph{Phys. Rev. D}
		{\bfseries 99} (2019) 063508}
	[\href{https://arxiv.org/abs/1901.04449}{{\ttfamily 1901.04449}}].
	
	\bibitem{Barman:2023ktz}
	B.~Barman, A.~Ghoshal, B.~Grzadkowski and A.~Socha, \emph{{Measuring Inflaton
			Couplings via Primordial Gravitational Waves}},
	\href{https://arxiv.org/abs/2305.00027}{{\ttfamily 2305.00027}}.
	
	\bibitem{PhysRevD.31.3059}
	M.W.~Goodman and E.~Witten, \emph{Detectability of certain dark-matter
		candidates}, \href{https://doi.org/10.1103/PhysRevD.31.3059}{\emph{Phys. Rev.
			D} {\bfseries 31} (1985) 3059}.
	
	\bibitem{Lewin:1996}
	J.D.~Lewin and P.F.~Smith, \emph{Review of mathematics, numerical factors, and
		corrections for dark matter experiments based on elastic nuclear recoil},
	\href{https://doi.org/10.1016/S0927-6505(96)00047-3}{\emph{Astroparticle
			Physics} {\bfseries 6} (1996) 87}.
	
	\bibitem{Jungman:1995df}
	G.~Jungman, M.~Kamionkowski and K.~Griest, \emph{{Supersymmetric dark matter}},
	\href{https://doi.org/10.1016/0370-1573(95)00058-5}{\emph{Phys. Rept.}
		{\bfseries 267} (1996) 195}
	[\href{https://arxiv.org/abs/hep-ph/9506380}{{\ttfamily hep-ph/9506380}}].
	
	\bibitem{Shifman:1978zn}
	M.A.~Shifman, A.I.~Vainshtein and V.I.~Zakharov, \emph{{Remarks on Higgs Boson
			Interactions with Nucleons}},
	\href{https://doi.org/10.1016/0370-2693(78)90481-1}{\emph{Phys. Lett. B}
		{\bfseries 78} (1978) 443}.
	
	\bibitem{Drees:1993bu}
	M.~Drees and M.~Nojiri, \emph{{Neutralino - nucleon scattering revisited}},
	\href{https://doi.org/10.1103/PhysRevD.48.3483}{\emph{Phys. Rev. D}
		{\bfseries 48} (1993) 3483}
	[\href{https://arxiv.org/abs/hep-ph/9307208}{{\ttfamily hep-ph/9307208}}].
	
	\bibitem{Bishara:2017pfq}
	F.~Bishara, J.~Brod, B.~Grinstein and J.~Zupan, \emph{{From quarks to nucleons
			in dark matter direct detection}},
	\href{https://doi.org/10.1007/JHEP11(2017)059}{\emph{JHEP} {\bfseries 11}
		(2017) 059} [\href{https://arxiv.org/abs/1707.06998}{{\ttfamily
			1707.06998}}].
	
	\bibitem{Hoferichter:2017olk}
	M.~Hoferichter, P.~Klos, J.~Men{\'e}ndez and A.~Schwenk, \emph{{Improved limits
			for Higgs-portal dark matter from LHC searches}},
	\href{https://doi.org/10.1103/PhysRevLett.119.181803}{\emph{Phys. Rev. Lett.}
		{\bfseries 119} (2017) 181803}
	[\href{https://arxiv.org/abs/1708.02245}{{\ttfamily 1708.02245}}].
	
	\bibitem{Alguero:2023zol}
	G.~Alguero, G.~Belanger, F.~Boudjema, S.~Chakraborti, A.~Goudelis, S.~Kraml
	et~al., \emph{{micrOMEGAs 6.0: N-component dark matter}},
	\href{https://doi.org/10.1016/j.cpc.2024.109133}{\emph{Comput. Phys. Commun.}
		{\bfseries 299} (2024) 109133}
	[\href{https://arxiv.org/abs/2312.14894}{{\ttfamily 2312.14894}}].
	
	\bibitem{PandaX:2024qfu}
	{\scshape PandaX} collaboration, \emph{{Dark Matter Search Results from 1.54
			Tonne$\cdot$Year Exposure of PandaX-4T}},
	\href{https://arxiv.org/abs/2408.00664}{{\ttfamily 2408.00664}}.
	
	\bibitem{LZ:2024zvo}
	{\scshape LZ} collaboration, \emph{{Dark Matter Search Results from 4.2
			Tonne-Years of Exposure of the LUX-ZEPLIN (LZ) Experiment}},
	\href{https://arxiv.org/abs/2410.17036}{{\ttfamily 2410.17036}}.
	
	\bibitem{XLZD:2024nsu}
	{\scshape XLZD} collaboration, \emph{{The XLZD Design Book: towards the
			next-generation liquid xenon observatory for dark matter and neutrino
			physics}}, \href{https://doi.org/10.1140/epjc/s10052-025-14810-w}{\emph{Eur.
			Phys. J. C} {\bfseries 85} (2025) 1192}
	[\href{https://arxiv.org/abs/2410.17137}{{\ttfamily 2410.17137}}].
	
	\bibitem{Essig:2011nj}
	R.~Essig, J.~Mardon and T.~Volansky, \emph{{Direct Detection of Sub-GeV Dark
			Matter}}, \href{https://doi.org/10.1103/PhysRevD.85.076007}{\emph{Phys. Rev.
			D} {\bfseries 85} (2012) 076007}
	[\href{https://arxiv.org/abs/1108.5383}{{\ttfamily 1108.5383}}].
	
	\bibitem{Essig:2013vha}
	R.~Essig, J.~Mardon, M.~Papucci, T.~Volansky and Y.-M.~Zhong,
	\emph{{Constraining Light Dark Matter with Low-Energy $e^+e^-$ Colliders}},
	\href{https://doi.org/10.1007/JHEP11(2013)167}{\emph{JHEP} {\bfseries 11}
		(2013) 167} [\href{https://arxiv.org/abs/1309.5084}{{\ttfamily 1309.5084}}].
	
	\bibitem{Essig:2015cda}
	R.~Essig, M.~Fernandez-Serra, J.~Mardon, A.~Soto, T.~Volansky and T.-T.~Yu,
	\emph{{Direct Detection of sub-GeV Dark Matter with Semiconductor Targets}},
	\href{https://doi.org/10.1007/JHEP05(2016)046}{\emph{JHEP} {\bfseries 05}
		(2016) 046} [\href{https://arxiv.org/abs/1509.01598}{{\ttfamily
			1509.01598}}].
	
	\bibitem{DAMIC-M:2025luv}
	{\scshape DAMIC-M} collaboration, \emph{{Probing Benchmark Models of
			Hidden-Sector Dark Matter with DAMIC-M}},
	\href{https://doi.org/10.1103/2tcc-bqck}{\emph{Phys. Rev. Lett.} {\bfseries
			135} (2025) 071002} [\href{https://arxiv.org/abs/2503.14617}{{\ttfamily
			2503.14617}}].
	
	\bibitem{PandaX:2022xqx}
	{\scshape PandaX} collaboration, \emph{{Search for Light Dark Matter with
			Ionization Signals in the PandaX-4T Experiment}},
	\href{https://doi.org/10.1103/PhysRevLett.130.261001}{\emph{Phys. Rev. Lett.}
		{\bfseries 130} (2023) 261001}
	[\href{https://arxiv.org/abs/2212.10067}{{\ttfamily 2212.10067}}].
	
	\bibitem{Oscura:2023qik}
	{\scshape Oscura} collaboration, \emph{{Skipper-CCD sensors for the Oscura
			experiment: requirements and preliminary tests}},
	\href{https://doi.org/10.1088/1748-0221/18/08/P08016}{\emph{JINST} {\bfseries
			18} (2023) P08016} [\href{https://arxiv.org/abs/2304.04401}{{\ttfamily
			2304.04401}}].
	
	\bibitem{2012PhRvD..85d3522F}
	D.P.~{Finkbeiner}, S.~{Galli}, T.~{Lin} and T.R.~{Slatyer}, \emph{{Searching
			for dark matter in the CMB: A compact parametrization of energy injection
			from new physics}},
	\href{https://doi.org/10.1103/PhysRevD.85.043522}{\emph{\prd} {\bfseries 85}
		(2012) 043522} [\href{https://arxiv.org/abs/1109.6322}{{\ttfamily
			1109.6322}}].
	
	\bibitem{Galli:2013dna}
	S.~Galli, T.R.~Slatyer, M.~Valdes and F.~Iocco, \emph{{Systematic Uncertainties
			In Constraining Dark Matter Annihilation From The Cosmic Microwave
			Background}}, \href{https://doi.org/10.1103/PhysRevD.88.063502}{\emph{Phys.
			Rev. D} {\bfseries 88} (2013) 063502}
	[\href{https://arxiv.org/abs/1306.0563}{{\ttfamily 1306.0563}}].
	
	\bibitem{Weniger:2013hja}
	C.~Weniger, P.D.~Serpico, F.~Iocco and G.~Bertone, \emph{{CMB bounds on dark
			matter annihilation: Nucleon energy-losses after recombination}},
	\href{https://doi.org/10.1103/PhysRevD.87.123008}{\emph{Phys. Rev. D}
		{\bfseries 87} (2013) 123008}
	[\href{https://arxiv.org/abs/1303.0942}{{\ttfamily 1303.0942}}].
	
	\bibitem{Liu:2016cnk}
	H.~Liu, T.R.~Slatyer and J.~Zavala, \emph{{Contributions to cosmic reionization
			from dark matter annihilation and decay}},
	\href{https://doi.org/10.1103/PhysRevD.94.063507}{\emph{Phys. Rev. D}
		{\bfseries 94} (2016) 063507}
	[\href{https://arxiv.org/abs/1604.02457}{{\ttfamily 1604.02457}}].
	
	\bibitem{Planck:2015fie}
	{\scshape Planck} collaboration, \emph{{Planck 2015 results. XIII. Cosmological
			parameters}},
	\href{https://doi.org/10.1051/0004-6361/201525830}{\emph{Astron. Astrophys.}
		{\bfseries 594} (2016) A13}
	[\href{https://arxiv.org/abs/1502.01589}{{\ttfamily 1502.01589}}].
	
	\bibitem{Galli:2009zc}
	S.~Galli, F.~Iocco, G.~Bertone and A.~Melchiorri, \emph{{CMB constraints on
			Dark Matter models with large annihilation cross-section}},
	\href{https://doi.org/10.1103/PhysRevD.80.023505}{\emph{Phys. Rev. D}
		{\bfseries 80} (2009) 023505}
	[\href{https://arxiv.org/abs/0905.0003}{{\ttfamily 0905.0003}}].
	
	\bibitem{Slatyer:2009yq}
	T.R.~Slatyer, N.~Padmanabhan and D.P.~Finkbeiner, \emph{{CMB Constraints on
			WIMP Annihilation: Energy Absorption During the Recombination Epoch}},
	\href{https://doi.org/10.1103/PhysRevD.80.043526}{\emph{Phys. Rev. D}
		{\bfseries 80} (2009) 043526}
	[\href{https://arxiv.org/abs/0906.1197}{{\ttfamily 0906.1197}}].
	
	\bibitem{2011PhRvD..84b7302G}
	S.~{Galli}, F.~{Iocco}, G.~{Bertone} and A.~{Melchiorri}, \emph{{Updated CMB
			constraints on dark matter annihilation cross sections}},
	\href{https://doi.org/10.1103/PhysRevD.84.027302}{\emph{\prd} {\bfseries 84}
		(2011) 027302} [\href{https://arxiv.org/abs/1106.1528}{{\ttfamily
			1106.1528}}].
	
	\bibitem{Cirelli:2023tnx}
	M.~Cirelli, N.~Fornengo, J.~Koechler, E.~Pinetti and B.M.~Roach, \emph{{Putting
			all the X in one basket: Updated X-ray constraints on sub-GeV Dark Matter}},
	\href{https://doi.org/10.1088/1475-7516/2023/07/026}{\emph{JCAP} {\bfseries
			07} (2023) 026} [\href{https://arxiv.org/abs/2303.08854}{{\ttfamily
			2303.08854}}].
	
	\bibitem{MAGIC:2016xys}
	{\scshape MAGIC, Fermi-LAT} collaboration, \emph{{Limits to Dark Matter
			Annihilation Cross-Section from a Combined Analysis of MAGIC and Fermi-LAT
			Observations of Dwarf Satellite Galaxies}},
	\href{https://doi.org/10.1088/1475-7516/2016/02/039}{\emph{JCAP} {\bfseries
			02} (2016) 039} [\href{https://arxiv.org/abs/1601.06590}{{\ttfamily
			1601.06590}}].
	
	\bibitem{PhysRevLett.110.141102}
	{\scshape AMS Collaboration} collaboration, \emph{First result from the alpha
		magnetic spectrometer on the international space station: Precision
		measurement of the positron fraction in primary cosmic rays of 0.5--350 gev},
	\href{https://doi.org/10.1103/PhysRevLett.110.141102}{\emph{Phys. Rev. Lett.}
		{\bfseries 110} (2013) 141102}.
	
	\bibitem{AGUILAR20211}
	M.~Aguilar, L.~{Ali Cavasonza}, G.~Ambrosi, L.~Arruda, N.~Attig, F.~Barao
	et~al., \emph{The alpha magnetic spectrometer (ams) on the international
		space station: Part ii — results from the first seven years},
	\href{https://doi.org/https://doi.org/10.1016/j.physrep.2020.09.003}{\emph{Physics
			Reports} {\bfseries 894} (2021) 1}.
	
	\bibitem{Giesen:2015ufa}
	G.~Giesen, M.~Boudaud, Y.~G{\'e}nolini, V.~Poulin, M.~Cirelli, P.~Salati
	et~al., \emph{{AMS-02 antiprotons, at last! Secondary astrophysical component
			and immediate implications for Dark Matter}},
	\href{https://doi.org/10.1088/1475-7516/2015/9/023}{\emph{JCAP} {\bfseries
			09} (2015) 023} [\href{https://arxiv.org/abs/1504.04276}{{\ttfamily
			1504.04276}}].
	
	\bibitem{Jin:2015sqa}
	H.-B.~Jin, Y.-L.~Wu and Y.-F.~Zhou, \emph{{Upper limits on dark matter
			annihilation cross sections from the first AMS-02 antiproton data}},
	\href{https://doi.org/10.1103/PhysRevD.92.055027}{\emph{Phys. Rev. D}
		{\bfseries 92} (2015) 055027}
	[\href{https://arxiv.org/abs/1504.04604}{{\ttfamily 1504.04604}}].
	
	\bibitem{Evoli:2015vaa}
	C.~Evoli, D.~Gaggero and D.~Grasso, \emph{{Secondary antiprotons as a Galactic
			Dark Matter probe}},
	\href{https://doi.org/10.1088/1475-7516/2015/12/039}{\emph{JCAP} {\bfseries
			12} (2015) 039} [\href{https://arxiv.org/abs/1504.05175}{{\ttfamily
			1504.05175}}].
	
	\bibitem{Cuoco:2016eej}
	A.~Cuoco, M.~Kr{\"a}mer and M.~Korsmeier, \emph{{Novel Dark Matter Constraints
			from Antiprotons in Light of AMS-02}},
	\href{https://doi.org/10.1103/PhysRevLett.118.191102}{\emph{Phys. Rev. Lett.}
		{\bfseries 118} (2017) 191102}
	[\href{https://arxiv.org/abs/1610.03071}{{\ttfamily 1610.03071}}].
	
	\bibitem{Cui:2016ppb}
	M.-Y.~Cui, Q.~Yuan, Y.-L.S.~Tsai and Y.-Z.~Fan, \emph{{Possible dark matter
			annihilation signal in the AMS-02 antiproton data}},
	\href{https://doi.org/10.1103/PhysRevLett.118.191101}{\emph{Phys. Rev. Lett.}
		{\bfseries 118} (2017) 191101}
	[\href{https://arxiv.org/abs/1610.03840}{{\ttfamily 1610.03840}}].
	
	\bibitem{DiMauro:2021qcf}
	M.~Di~Mauro and M.W.~Winkler, \emph{{Multimessenger constraints on the dark
			matter interpretation of the Fermi-LAT Galactic center excess}},
	\href{https://doi.org/10.1103/PhysRevD.103.123005}{\emph{Phys. Rev. D}
		{\bfseries 103} (2021) 123005}
	[\href{https://arxiv.org/abs/2101.11027}{{\ttfamily 2101.11027}}].
	
	\bibitem{Calore:2022stf}
	F.~Calore, M.~Cirelli, L.~Derome, Y.~Genolini, D.~Maurin, P.~Salati et~al.,
	\emph{{AMS-02 antiprotons and dark matter: Trimmed hints and robust bounds}},
	\href{https://doi.org/10.21468/SciPostPhys.12.5.163}{\emph{SciPost Phys.}
		{\bfseries 12} (2022) 163}
	[\href{https://arxiv.org/abs/2202.03076}{{\ttfamily 2202.03076}}].
	
	\bibitem{Slatyer:2015jla}
	T.R.~Slatyer, \emph{{Indirect dark matter signatures in the cosmic dark ages.
			I. Generalizing the bound on s-wave dark matter annihilation from Planck
			results}}, \href{https://doi.org/10.1103/PhysRevD.93.023527}{\emph{Phys. Rev.
			D} {\bfseries 93} (2016) 023527}
	[\href{https://arxiv.org/abs/1506.03811}{{\ttfamily 1506.03811}}].
	
	\bibitem{Pinzke:2011ek}
	A.~Pinzke, C.~Pfrommer and L.~Bergstrom, \emph{{Prospects of detecting
			gamma-ray emission from galaxy clusters: cosmic rays and dark matter
			annihilations}},
	\href{https://doi.org/10.1103/PhysRevD.84.123509}{\emph{Phys. Rev. D}
		{\bfseries 84} (2011) 123509}
	[\href{https://arxiv.org/abs/1105.3240}{{\ttfamily 1105.3240}}].
	
	\bibitem{Voit:2004ah}
	G.M.~Voit, \emph{{Tracing cosmic evolution with clusters of galaxies}},
	\href{https://doi.org/10.1103/RevModPhys.77.207}{\emph{Rev. Mod. Phys.}
		{\bfseries 77} (2005) 207}
	[\href{https://arxiv.org/abs/astro-ph/0410173}{{\ttfamily
			astro-ph/0410173}}].
	
	\bibitem{Bollig:2017lki}
	R.~Bollig, H.T.~Janka, A.~Lohs, G.~Martinez-Pinedo, C.J.~Horowitz and
	T.~Melson, \emph{{Muon Creation in Supernova Matter Facilitates
			Neutrino-driven Explosions}},
	\href{https://doi.org/10.1103/PhysRevLett.119.242702}{\emph{Phys. Rev. Lett.}
		{\bfseries 119} (2017) 242702}
	[\href{https://arxiv.org/abs/1706.04630}{{\ttfamily 1706.04630}}].
	
	\bibitem{Fischer:2020vie}
	T.~Fischer, G.~Guo, G.~Mart{\'\i}nez-Pinedo, M.~Liebend{\"o}rfer and
	A.~Mezzacappa, \emph{{Muonization of supernova matter}},
	\href{https://doi.org/10.1103/PhysRevD.102.123001}{\emph{Phys. Rev. D}
		{\bfseries 102} (2020) 123001}
	[\href{https://arxiv.org/abs/2008.13628}{{\ttfamily 2008.13628}}].
	
	\bibitem{Dreiner:2003wh}
	H.K.~Dreiner, C.~Hanhart, U.~Langenfeld and D.R.~Phillips, \emph{{Supernovae
			and light neutralinos: SN1987A bounds on supersymmetry revisited}},
	\href{https://doi.org/10.1103/PhysRevD.68.055004}{\emph{Phys. Rev. D}
		{\bfseries 68} (2003) 055004}
	[\href{https://arxiv.org/abs/hep-ph/0304289}{{\ttfamily hep-ph/0304289}}].
	
	\bibitem{Dreiner:2013mua}
	H.K.~Dreiner, J.-F.~Fortin, C.~Hanhart and L.~Ubaldi, \emph{{Supernova
			constraints on MeV dark sectors from $e^+e^-$ annihilations}},
	\href{https://doi.org/10.1103/PhysRevD.89.105015}{\emph{Phys. Rev. D}
		{\bfseries 89} (2014) 105015}
	[\href{https://arxiv.org/abs/1310.3826}{{\ttfamily 1310.3826}}].
	
	\bibitem{Manzari:2023gkt}
	C.A.~Manzari, J.~Martin~Camalich, J.~Spinner and R.~Ziegler, \emph{{Supernova
			limits on muonic dark forces}},
	\href{https://doi.org/10.1103/PhysRevD.108.103020}{\emph{Phys. Rev. D}
		{\bfseries 108} (2023) 103020}
	[\href{https://arxiv.org/abs/2307.03143}{{\ttfamily 2307.03143}}].
	
	\bibitem{DeRocco:2019jti}
	W.~DeRocco, P.W.~Graham, D.~Kasen, G.~Marques-Tavares and S.~Rajendran,
	\emph{{Supernova signals of light dark matter}},
	\href{https://doi.org/10.1103/PhysRevD.100.075018}{\emph{Phys. Rev. D}
		{\bfseries 100} (2019) 075018}
	[\href{https://arxiv.org/abs/1905.09284}{{\ttfamily 1905.09284}}].
	
	\bibitem{Gondolo:1990dk}
	P.~Gondolo and G.~Gelmini, \emph{{Cosmic abundances of stable particles:
			Improved analysis}},
	\href{https://doi.org/10.1016/0550-3213(91)90438-4}{\emph{Nucl. Phys. B}
		{\bfseries 360} (1991) 145}.
	
	\bibitem{Raffelt:1996wa}
	G.G.~Raffelt, \emph{{Stars as laboratories for fundamental physics}: {The
			astrophysics of neutrinos, axions, and other weakly interacting particles}}
	(5, 1996).
	
	\bibitem{Garching}
	
	
	\bibitem{Friman:1979ecl}
	B.L.~Friman and O.V.~Maxwell, \emph{{Neutron Star Neutrino Emissivities}},
	\href{https://doi.org/10.1086/157313}{\emph{Astrophys. J.} {\bfseries 232}
		(1979) 541}.
	
	\bibitem{Timmermans:2002hc}
	R.G.E.~Timmermans, A.Y.~Korchin, E.N.E.~van Dalen and A.E.L.~Dieperink,
	\emph{{Soft electroweak bremsstrahlung: Theorems and astrophysical
			relevance}}, \href{https://doi.org/10.1103/PhysRevC.65.064007}{\emph{Phys.
			Rev. C} {\bfseries 65} (2002) 064007}.
	
	\bibitem{Giovannini:1998bp}
	M.~Giovannini, \emph{Production and detection of relic gravitons in
		quintessential inflationary models}, {\emph{Phys. Rev. D} {\bfseries 60}
		(1999) 123511} [\href{https://arxiv.org/abs/astro-ph/9903004}{{\ttfamily
			astro-ph/9903004}}].
	
	\bibitem{Giovannini:1999bh}
	M.~Giovannini, \emph{Gravitational waves constraints on postinflationary phases
		stiffer than radiation}, {\emph{Phys. Rev. D} {\bfseries 58} (1998) 083504}
	[\href{https://arxiv.org/abs/hep-ph/9806329}{{\ttfamily hep-ph/9806329}}].
	
	\bibitem{Riazuelo:2000fc}
	A.~Riazuelo and J.-P.~Uzan, \emph{Cosmological observations in scalar-tensor
		quintessence}, {\emph{Phys. Rev. D} {\bfseries 62} (2000) 083506}
	[\href{https://arxiv.org/abs/astro-ph/0004156}{{\ttfamily
			astro-ph/0004156}}].
	
	\bibitem{Seto:2003kc}
	N.~Seto and J.~Yokoyama, \emph{Probing the equation of state of the early
		universe with a space laser interferometer}, {\emph{J. Phys. Soc. Jap.}
		{\bfseries 72} (2003) 3082}
	[\href{https://arxiv.org/abs/gr-qc/0305096}{{\ttfamily gr-qc/0305096}}].
	
	\bibitem{Boyle:2007zx}
	L.A.~Boyle and A.~Buonanno, \emph{{Relating gravitational wave constraints from
			primordial nucleosynthesis, pulsar timing, laser interferometers, and the
			CMB: Implications for the early Universe}},
	\href{https://doi.org/10.1103/PhysRevD.78.043531}{\emph{Phys. Rev. D}
		{\bfseries 78} (2008) 043531}
	[\href{https://arxiv.org/abs/0708.2279}{{\ttfamily 0708.2279}}].
	
	\bibitem{Stewart:2007fu}
	E.D.~Stewart and R.~Brandenberger, \emph{Observational constraints on theories
		with a blue spectrum of tensor modes}, {\emph{JCAP} {\bfseries 0808} (2008)
		012} [\href{https://arxiv.org/abs/0711.4607}{{\ttfamily 0711.4607}}].
	
	\bibitem{Li:2021htg}
	S.-P.~Li et~al., \emph{Probing reheating with primordial gravitational waves},
	{\emph{Phys. Rev. D} {\bfseries 104} (2021) 083521}
	[\href{https://arxiv.org/abs/2106.01366}{{\ttfamily 2106.01366}}].
	
	\bibitem{Artymowski:2017pua}
	M.~Artymowski, I.~Ben-Dayan and R.~Brustein, \emph{Relic gravitational waves in
		light of planck}, {\emph{JCAP} {\bfseries 1706} (2017) 042}
	[\href{https://arxiv.org/abs/1701.05443}{{\ttfamily 1701.05443}}].
	
	\bibitem{Caprini:2018mtu}
	C.~Caprini and D.G.~Figueroa, \emph{{Cosmological Backgrounds of Gravitational
			Waves}}, \href{https://doi.org/10.1088/1361-6382/aac608}{\emph{Class. Quant.
			Grav.} {\bfseries 35} (2018) 163001}
	[\href{https://arxiv.org/abs/1801.04268}{{\ttfamily 1801.04268}}].
	
	\bibitem{Bettoni:2018pbl}
	D.~Bettoni et~al., \emph{Speed of gravitational waves and the fate of
		scalar-tensor gravity}, {\emph{Phys. Rev. D} {\bfseries 98} (2018) 084020}
	[\href{https://arxiv.org/abs/1805.00667}{{\ttfamily 1805.00667}}].
	
	\bibitem{Figueroa:2019paj}
	D.G.~Figueroa and E.H.~Tanin, \emph{{Ability of LIGO and LISA to probe the
			equation of state of the early Universe}},
	\href{https://doi.org/10.1088/1475-7516/2019/08/011}{\emph{JCAP} {\bfseries
			08} (2019) 011} [\href{https://arxiv.org/abs/1905.11960}{{\ttfamily
			1905.11960}}].
	
	\bibitem{Opferkuch:2019zbd}
	T.~Opferkuch, P.~Schwaller and B.A.~Stefanek, \emph{{Ricci Reheating}},
	\href{https://doi.org/10.1088/1475-7516/2019/07/016}{\emph{JCAP} {\bfseries
			07} (2019) 016} [\href{https://arxiv.org/abs/1905.06823}{{\ttfamily
			1905.06823}}].
	
	\bibitem{Bernal:2020ywq}
	N.~Bernal et~al., \emph{Gravitational wave signatures of dark matter
		production}, {\emph{JCAP} {\bfseries 2010} (2020) 042}
	[\href{https://arxiv.org/abs/2004.11306}{{\ttfamily 2004.11306}}].
	
	\bibitem{Caldwell:2022qsj}
	R.R.~Caldwell et~al., \emph{Gravitational wave cosmology and early universe
		constraints}, {\emph{Phys. Rev. D} {\bfseries 106} (2022) 063001}
	[\href{https://arxiv.org/abs/2203.XXXXX}{{\ttfamily 2203.XXXXX}}].
	
	\bibitem{Gouttenoire:2021jhk}
	Y.~Gouttenoire et~al., \emph{Primordial gravitational waves and early universe
		probes}, {\emph{JCAP} {\bfseries 2106} (2021) 002}
	[\href{https://arxiv.org/abs/2101.XXXXX}{{\ttfamily 2101.XXXXX}}].
	
	\bibitem{Haque:2021dha}
	M.R.~Haque, D.~Maity, T.~Paul and L.~Sriramkumar, \emph{{Decoding the phases of
			early and late time reheating through imprints on primordial gravitational
			waves}}, \href{https://doi.org/10.1103/PhysRevD.104.063513}{\emph{Phys. Rev.
			D} {\bfseries 104} (2021) 063513}
	[\href{https://arxiv.org/abs/2105.09242}{{\ttfamily 2105.09242}}].
	
	\bibitem{Maity:2024cpq}
	S.~Maity and M.R.~Haque, \emph{{Probing the early universe with future GW
			observatories}},
	\href{https://doi.org/10.1088/1475-7516/2025/04/091}{\emph{JCAP} {\bfseries
			04} (2025) 091} [\href{https://arxiv.org/abs/2407.18246}{{\ttfamily
			2407.18246}}].
	
	\bibitem{Crowder:2005nr}
	J.~Crowder and N.J.~Cornish, \emph{{Beyond LISA: Exploring future gravitational
			wave missions}},
	\href{https://doi.org/10.1103/PhysRevD.72.083005}{\emph{Phys. Rev. D}
		{\bfseries 72} (2005) 083005}
	[\href{https://arxiv.org/abs/gr-qc/0506015}{{\ttfamily gr-qc/0506015}}].
	
	\bibitem{Corbin:2005ny}
	V.~Corbin and N.J.~Cornish, \emph{{Detecting the cosmic gravitational wave
			background with the big bang observer}},
	\href{https://doi.org/10.1088/0264-9381/23/7/014}{\emph{Class. Quant. Grav.}
		{\bfseries 23} (2006) 2435}
	[\href{https://arxiv.org/abs/gr-qc/0512039}{{\ttfamily gr-qc/0512039}}].
	
	\bibitem{Seto:2001qf}
	N.~Seto, S.~Kawamura and T.~Nakamura, \emph{{Possibility of direct measurement
			of the acceleration of the universe using 0.1-Hz band laser interferometer
			gravitational wave antenna in space}},
	\href{https://doi.org/10.1103/PhysRevLett.87.221103}{\emph{Phys. Rev. Lett.}
		{\bfseries 87} (2001) 221103}
	[\href{https://arxiv.org/abs/astro-ph/0108011}{{\ttfamily
			astro-ph/0108011}}].
	
	\bibitem{Kudoh:2005as}
	H.~Kudoh, A.~Taruya, T.~Hiramatsu and Y.~Himemoto, \emph{{Detecting a
			gravitational-wave background with next-generation space interferometers}},
	\href{https://doi.org/10.1103/PhysRevD.73.064006}{\emph{Phys. Rev. D}
		{\bfseries 73} (2006) 064006}
	[\href{https://arxiv.org/abs/gr-qc/0511145}{{\ttfamily gr-qc/0511145}}].
	
	\bibitem{LISA:2017pwj}
	{\scshape LISA} collaboration, \emph{{Laser Interferometer Space Antenna}},
	\href{https://arxiv.org/abs/1702.00786}{{\ttfamily 1702.00786}}.
	
	\bibitem{Sesana:2019vho}
	A.~Sesana et~al., \emph{{Unveiling the gravitational universe at $\mu$-Hz
			frequencies}}, \href{https://doi.org/10.1007/s10686-021-09709-9}{\emph{Exper.
			Astron.} {\bfseries 51} (2021) 1333}
	[\href{https://arxiv.org/abs/1908.11391}{{\ttfamily 1908.11391}}].
	
	\bibitem{Reitze:2019iox}
	D.~Reitze et~al., \emph{{Cosmic Explorer: The U.S. Contribution to
			Gravitational-Wave Astronomy beyond LIGO}}, {\emph{Bull. Am. Astron. Soc.}
		{\bfseries 51} (2019) 035}
	[\href{https://arxiv.org/abs/1907.04833}{{\ttfamily 1907.04833}}].
	
	\bibitem{Hild:2010id}
	S.~Hild et~al., \emph{{Sensitivity Studies for Third-Generation Gravitational
			Wave Observatories}},
	\href{https://doi.org/10.1088/0264-9381/28/9/094013}{\emph{Class. Quant.
			Grav.} {\bfseries 28} (2011) 094013}
	[\href{https://arxiv.org/abs/1012.0908}{{\ttfamily 1012.0908}}].
	
	\bibitem{Punturo:2010zz}
	M.~Punturo et~al., \emph{{The Einstein Telescope: A third-generation
			gravitational wave observatory}},
	\href{https://doi.org/10.1088/0264-9381/27/19/194002}{\emph{Class. Quant.
			Grav.} {\bfseries 27} (2010) 194002}.
	
	\bibitem{Sathyaprakash:2012jk}
	B.~Sathyaprakash et~al., \emph{{Scientific Objectives of Einstein Telescope}},
	\href{https://doi.org/10.1088/0264-9381/29/12/124013}{\emph{Class. Quant.
			Grav.} {\bfseries 29} (2012) 124013}
	[\href{https://arxiv.org/abs/1206.0331}{{\ttfamily 1206.0331}}].
	
	\bibitem{Maggiore:2019uih}
	M.~Maggiore et~al., \emph{{Science Case for the Einstein Telescope}},
	\href{https://doi.org/10.1088/1475-7516/2020/03/050}{\emph{JCAP} {\bfseries
			03} (2020) 050} [\href{https://arxiv.org/abs/1912.02622}{{\ttfamily
			1912.02622}}].
	
	\bibitem{Herman:2022fau}
	N.~Herman, L.~Lehoucq and A.~F\'{u}zfa, \emph{{Electromagnetic Antennas for the
			Resonant Detection of the Stochastic Gravitational Wave Background}},
	\href{https://arxiv.org/abs/2203.15668}{{\ttfamily 2203.15668}}.
	
	\bibitem{Maggiore:1999vm}
	M.~Maggiore, \emph{{Gravitational wave experiments and early universe
			cosmology}}, \href{https://doi.org/10.1016/S0370-1573(99)00102-7}{\emph{Phys.
			Rept.} {\bfseries 331} (2000) 283}
	[\href{https://arxiv.org/abs/gr-qc/9909001}{{\ttfamily gr-qc/9909001}}].
	
	\bibitem{Watanabe:2006qe}
	Y.~Watanabe and E.~Komatsu, \emph{{Improved Calculation of the Primordial
			Gravitational Wave Spectrum in the Standard Model}},
	\href{https://doi.org/10.1103/PhysRevD.73.123515}{\emph{Phys. Rev. D}
		{\bfseries 73} (2006) 123515}
	[\href{https://arxiv.org/abs/astro-ph/0604176}{{\ttfamily
			astro-ph/0604176}}].
	
	\bibitem{Saikawa:2018rcs}
	K.~Saikawa and S.~Shirai, \emph{{Primordial gravitational waves, precisely: The
			role of thermodynamics in the Standard Model}},
	\href{https://doi.org/10.1088/1475-7516/2018/05/035}{\emph{JCAP} {\bfseries
			05} (2018) 035} [\href{https://arxiv.org/abs/1803.01038}{{\ttfamily
			1803.01038}}].
	
	\bibitem{Boyle:2005se}
	L.A.~Boyle and P.J.~Steinhardt, \emph{{Probing the early universe with
			inflationary gravitational waves}},
	\href{https://doi.org/10.1103/PhysRevD.77.063504}{\emph{Phys. Rev. D}
		{\bfseries 77} (2008) 063504}
	[\href{https://arxiv.org/abs/astro-ph/0512014}{{\ttfamily
			astro-ph/0512014}}].
	
	\bibitem{BICEP:2021xfz}
	{\scshape BICEP, Keck} collaboration, \emph{{Improved Constraints on Primordial
			Gravitational Waves using Planck, WMAP, and BICEP/Keck Observations through
			the 2018 Observing Season}},
	\href{https://doi.org/10.1103/PhysRevLett.127.151301}{\emph{Phys. Rev. Lett.}
		{\bfseries 127} (2021) 151301}
	[\href{https://arxiv.org/abs/2110.00483}{{\ttfamily 2110.00483}}].
	
	\bibitem{Yeh:2022heq}
	T.-H.~Yeh, J.~Shelton, K.A.~Olive and B.D.~Fields, \emph{{Probing physics
			beyond the standard model: limits from BBN and the CMB independently and
			combined}}, \href{https://doi.org/10.1088/1475-7516/2022/10/046}{\emph{JCAP}
		{\bfseries 10} (2022) 046}
	[\href{https://arxiv.org/abs/2207.13133}{{\ttfamily 2207.13133}}].
	
	\bibitem{Abazajian:2019eic}
	K.~Abazajian et~al., \emph{{CMB-S4 Science Case, Reference Design, and Project
			Plan}},  \href{https://arxiv.org/abs/1907.04473}{{\ttfamily 1907.04473}}.
	
	\bibitem{CMB-HD:2022bsz}
	{\scshape CMB-HD} collaboration, \emph{{Snowmass2021 CMB-HD White Paper}},
	\href{https://arxiv.org/abs/2203.05728}{{\ttfamily 2203.05728}}.
	
	\bibitem{COrE:2011bfs}
	{\scshape COrE} collaboration, \emph{{COrE (Cosmic Origins Explorer) A White
			Paper}},  \href{https://arxiv.org/abs/1102.2181}{{\ttfamily 1102.2181}}.
	
	\bibitem{EUCLID:2011zbd}
	{\scshape EUCLID} collaboration, \emph{{Euclid Definition Study Report}},
	\href{https://arxiv.org/abs/1110.3193}{{\ttfamily 1110.3193}}.
	
	\bibitem{Kuroyanagi:2014nba}
	S.~Kuroyanagi, T.~Takahashi and S.~Yokoyama, \emph{{Blue-tilted Tensor Spectrum
			and Thermal History of the Universe}},
	\href{https://doi.org/10.1088/1475-7516/2015/02/003}{\emph{JCAP} {\bfseries
			02} (2015) 003} [\href{https://arxiv.org/abs/1407.4785}{{\ttfamily
			1407.4785}}].
	
	\bibitem{Barman:2024nhr}
	B.~Barman, S.~Bhattacharya, S.~Jahedi, D.~Pradhan and A.~Sarkar, \emph{{Lepton
			Collider as a window to Reheating}},
	\href{https://arxiv.org/abs/2406.11963}{{\ttfamily 2406.11963}}.
	
	\bibitem{Barman:2024tjt}
	B.~Barman, S.~Bhattacharya, S.~Jahedi, D.~Pradhan and A.~Sarkar, \emph{{Lepton
			collider as a window to reheating via freezing in dark matter detection. Part
			II}}, \href{https://doi.org/10.1007/JHEP07(2025)157}{\emph{JHEP} {\bfseries
			07} (2025) 157} [\href{https://arxiv.org/abs/2410.18198}{{\ttfamily
			2410.18198}}].
	
	\bibitem{Bhattacharya:2025wef}
	S.~Bhattacharya, A.~Ghosh, N.~Mondal and A.~Sarkar, \emph{{Lepton Collider as a
			Window to Reheating via Freezing Out Dark Matter Detection}},
	\href{https://arxiv.org/abs/2509.14340}{{\ttfamily 2509.14340}}.
	
	\bibitem{Ghosh:2024boo}
	A.~Ghosh and P.~Konar, \emph{{Unveiling desert region in inert doublet model
			assisted by Peccei-Quinn symmetry}},
	\href{https://doi.org/10.1007/JHEP09(2024)104}{\emph{JHEP} {\bfseries 09}
		(2024) 104} [\href{https://arxiv.org/abs/2407.01415}{{\ttfamily
			2407.01415}}].
	
	\bibitem{Ghosh:2024nkj}
	A.~Ghosh, P.~Konar and S.~Show, \emph{{Collider fingerprints of freeze-in dark
			matter produced during the fast expansion phase of Universe}},
	\href{https://doi.org/10.1103/pj7s-zhcr}{\emph{Phys. Rev. D} {\bfseries 112}
		(2025) 055012} [\href{https://arxiv.org/abs/2411.09464}{{\ttfamily
			2411.09464}}].
	
	\bibitem{Ghosh:2025agw}
	A.~Ghosh, \emph{{Unveiling a natural multicomponent dark sector: an inert
			doublet guided by Peccei{\textendash}Quinn}},
	\href{https://doi.org/10.1140/epjp/s13360-025-06621-5}{\emph{Eur. Phys. J.
			Plus} {\bfseries 140} (2025) 688}.
	
	\bibitem{Bernal:2025qkj}
	N.~Bernal, G.~Cottin, B.~D{\'\i}az~S{\'a}ez and M.~L{\'o}pez, \emph{{Testing
			frozen-in pNGB dark matter with a long-lived dark Higgs}},
	\href{https://doi.org/10.1007/JHEP01(2026)081}{\emph{JHEP} {\bfseries 01}
		(2026) 081} [\href{https://arxiv.org/abs/2507.07089}{{\ttfamily
			2507.07089}}].
	
	\bibitem{C:2026bqd}
	P.A.~C, G.~Cottin, B.~D{\'\i}az~S{\'a}ez, Z.S.~Wang and Y.~Zhang, \emph{{Can
			LLP detectors probe the reheating temperature? A case study of vector dark
			matter}},  \href{https://arxiv.org/abs/2604.25090}{{\ttfamily 2604.25090}}.
	
	\bibitem{NA62:2020}
	{\scshape NA62} collaboration, \emph{{Search for $\pi^0$ decays to invisible
			particles}}, \href{https://doi.org/10.1007/JHEP02(2021)201}{\emph{JHEP}
		{\bfseries 02} (2021) 201}
	[\href{https://arxiv.org/abs/2010.07644}{{\ttfamily 2010.07644}}].
	
	\bibitem{ParticleDataGroup:2024cfk}
	{\scshape Particle Data Group} collaboration, \emph{{Review of particle
			physics}}, \href{https://doi.org/10.1103/PhysRevD.110.030001}{\emph{Phys.
			Rev. D} {\bfseries 110} (2024) 030001}.
	
	\bibitem{BESIII:2018}
	{\scshape BESIII} collaboration, \emph{{Search for invisible decays of $\omega$
			and $\phi$ with $J/\psi$ data at BESIII}},
	\href{https://doi.org/10.1103/PhysRevD.98.032001}{\emph{Phys. Rev. D}
		{\bfseries 98} (2018) 032001}
	[\href{https://arxiv.org/abs/1805.05613}{{\ttfamily 1805.05613}}].
	
	\bibitem{BaBar:2009}
	{\scshape BaBar} collaboration, \emph{{A Search for Invisible Decays of the
			Upsilon(1S)}},
	\href{https://doi.org/10.1103/PhysRevLett.103.251801}{\emph{Phys. Rev. Lett.}
		{\bfseries 103} (2009) 251801}
	[\href{https://arxiv.org/abs/0908.2840}{{\ttfamily 0908.2840}}].
	
	\bibitem{KOTO2025}
	{\scshape KOTO} collaboration, \emph{{Search for the
			KL{\textrightarrow}{\ensuremath{\pi}}0{\ensuremath{\nu}}{\ensuremath{\nu}}{\textasciimacron}
			Decay at the J-PARC KOTO Experiment}},
	\href{https://doi.org/10.1103/PhysRevLett.134.081802}{\emph{Phys. Rev. Lett.}
		{\bfseries 134} (2025) 081802}
	[\href{https://arxiv.org/abs/2411.11237}{{\ttfamily 2411.11237}}].
	
	\bibitem{Carrazza:2020rdn}
	S.~Carrazza and J.M.~Cruz-Martinez, \emph{{VegasFlow: accelerating Monte Carlo
			simulation across multiple hardware platforms}},
	\href{https://doi.org/10.1016/j.cpc.2020.107376}{\emph{Comput. Phys. Commun.}
		{\bfseries 254} (2020) 107376}
	[\href{https://arxiv.org/abs/2002.12921}{{\ttfamily 2002.12921}}].
	
	\bibitem{vegasflow_package}
	J.~Cruz-Martinez and S.~Carrazza, \emph{N3pdf/vegasflow: vegasflow v1.0},
	Feb., 2020.
	\newblock 10.5281/zenodo.3691926.
	
	\bibitem{Bijnens:1994me}
	J.~Bijnens, G.~Colangelo, G.~Ecker and J.~Gasser, \emph{{Semileptonic kaon
			decays}},  11, 1994 [\href{https://arxiv.org/abs/hep-ph/9411311}{{\ttfamily
			hep-ph/9411311}}].
	
	\bibitem{Bijnens:2003uy}
	J.~Bijnens and P.~Talavera, \emph{{K(l3) decays in chiral perturbation
			theory}}, \href{https://doi.org/10.1016/S0550-3213(03)00581-9}{\emph{Nucl.
			Phys. B} {\bfseries 669} (2003) 341}
	[\href{https://arxiv.org/abs/hep-ph/0303103}{{\ttfamily hep-ph/0303103}}].
	
	\bibitem{Boito:2010me}
	D.R.~Boito, R.~Escribano and M.~Jamin, \emph{{K $\pi$ vector form factor
			constrained by $\tau -> K\ pi \nu_\tau$ and $K_{l3}$ decays}},
	\href{https://doi.org/10.1007/JHEP09(2010)031}{\emph{JHEP} {\bfseries 09}
		(2010) 031} [\href{https://arxiv.org/abs/1007.1858}{{\ttfamily 1007.1858}}].
	
	\bibitem{Bourrely:2008za}
	C.~Bourrely, I.~Caprini and L.~Lellouch, \emph{{Model-independent description
			of B ---{\ensuremath{>}} pi l nu decays and a determination of |V(ub)|}},
	\href{https://doi.org/10.1103/PhysRevD.82.099902}{\emph{Phys. Rev. D}
		{\bfseries 79} (2009) 013008}
	[\href{https://arxiv.org/abs/0807.2722}{{\ttfamily 0807.2722}}].
	
	\bibitem{PDG}
	{\scshape Particle Data Group} collaboration, \emph{{Review of particle
			physics}}, \href{https://doi.org/10.1103/PhysRevD.110.030001}{\emph{Phys.
			Rev. D} {\bfseries 110} (2024) 030001}.
	
	\bibitem{Gabbiani:1996hi}
	F.~Gabbiani, E.~Gabrielli, A.~Masiero and L.~Silvestrini, \emph{{A Complete
			analysis of FCNC and CP constraints in general SUSY extensions of the
			standard model}},
	\href{https://doi.org/10.1016/0550-3213(96)00390-2}{\emph{Nucl. Phys. B}
		{\bfseries 477} (1996) 321}
	[\href{https://arxiv.org/abs/hep-ph/9604387}{{\ttfamily hep-ph/9604387}}].
	
	\bibitem{Ciuchini:1998ix}
	M.~Ciuchini et~al., \emph{{Delta M(K) and epsilon(K) in SUSY at the
			next-to-leading order}},
	\href{https://doi.org/10.1088/1126-6708/1998/10/008}{\emph{JHEP} {\bfseries
			10} (1998) 008} [\href{https://arxiv.org/abs/hep-ph/9808328}{{\ttfamily
			hep-ph/9808328}}].
	
	\bibitem{Buras:2000if}
	A.J.~Buras, M.~Misiak and J.~Urban, \emph{{Two loop QCD anomalous dimensions of
			flavor changing four quark operators within and beyond the standard model}},
	\href{https://doi.org/10.1016/S0550-3213(00)00437-5}{\emph{Nucl. Phys. B}
		{\bfseries 586} (2000) 397}
	[\href{https://arxiv.org/abs/hep-ph/0005183}{{\ttfamily hep-ph/0005183}}].
	
	\bibitem{Bertone:2012cu}
	{\scshape ETM} collaboration, \emph{{Kaon Mixing Beyond the SM from N$_{f}$=2
			tmQCD and model independent constraints from the UTA}},
	\href{https://doi.org/10.1007/JHEP03(2013)089}{\emph{JHEP} {\bfseries 03}
		(2013) 089} [\href{https://arxiv.org/abs/1207.1287}{{\ttfamily 1207.1287}}].
	
	\bibitem{ALEPH:2005ab}
	{\scshape ALEPH, DELPHI, L3, OPAL, SLD, LEP Electroweak Working Group, SLD
		Electroweak Group, SLD Heavy Flavour Group} collaboration, \emph{{Precision
			electroweak measurements on the $Z$ resonance}},
	\href{https://doi.org/10.1016/j.physrep.2005.12.006}{\emph{Phys. Rept.}
		{\bfseries 427} (2006) 257}
	[\href{https://arxiv.org/abs/hep-ex/0509008}{{\ttfamily hep-ex/0509008}}].
	
	\bibitem{CMS:2022ett}
	{\scshape CMS} collaboration, \emph{{Precision measurement of the Z boson
			invisible width in pp collisions at s=13 TeV}},
	\href{https://doi.org/10.1016/j.physletb.2022.137563}{\emph{Phys. Lett. B}
		{\bfseries 842} (2023) 137563}
	[\href{https://arxiv.org/abs/2206.07110}{{\ttfamily 2206.07110}}].
	
	\bibitem{ATLAS:2023ynf}
	{\scshape ATLAS} collaboration, \emph{{Measurement of the Z boson invisible
			width at s=13 TeV with the ATLAS detector}},
	\href{https://doi.org/10.1016/j.physletb.2024.138705}{\emph{Phys. Lett. B}
		{\bfseries 854} (2024) 138705}
	[\href{https://arxiv.org/abs/2312.02789}{{\ttfamily 2312.02789}}].
	
	\bibitem{Carpenter:2013xra}
	L.~Carpenter, A.~DiFranzo, M.~Mulhearn, C.~Shimmin, S.~Tulin and D.~Whiteson,
	\emph{{Mono-Higgs-boson: A new collider probe of dark matter}},
	\href{https://doi.org/10.1103/PhysRevD.89.075017}{\emph{Phys. Rev. D}
		{\bfseries 89} (2014) 075017}
	[\href{https://arxiv.org/abs/1312.2592}{{\ttfamily 1312.2592}}].
	
	\bibitem{Martin:2000by}
	A.D.~Martin, J.~Outhwaite and M.G.~Ryskin, \emph{{A New determination of the
			QED coupling alpha (M**2(Z)) lets the Higgs off the hook}},
	\href{https://doi.org/10.1016/S0370-2693(00)01083-2}{\emph{Phys. Lett. B}
		{\bfseries 492} (2000) 69}
	[\href{https://arxiv.org/abs/hep-ph/0008078}{{\ttfamily hep-ph/0008078}}].
	
	\bibitem{ATLAS:2023tkt}
	{\scshape ATLAS} collaboration, \emph{{Combination of searches for invisible
			decays of the Higgs boson using 139 fb{\ensuremath{-}}1 of proton-proton
			collision data at s=13 TeV collected with the ATLAS experiment}},
	\href{https://doi.org/10.1016/j.physletb.2023.137963}{\emph{Phys. Lett. B}
		{\bfseries 842} (2023) 137963}
	[\href{https://arxiv.org/abs/2301.10731}{{\ttfamily 2301.10731}}].
	
	\bibitem{CMS:2023sdw}
	{\scshape CMS} collaboration, \emph{{A search for decays of the Higgs boson to
			invisible particles in events with a top-antitop quark pair or a vector boson
			in proton-proton collisions at $\sqrt{s} = 13\,\text {Te}\hspace{-.08em}\text
			{V} $}}, \href{https://doi.org/10.1140/epjc/s10052-023-11952-7}{\emph{Eur.
			Phys. J. C} {\bfseries 83} (2023) 933}
	[\href{https://arxiv.org/abs/2303.01214}{{\ttfamily 2303.01214}}].
	
	\bibitem{Fox:2011fx}
	P.J.~Fox, R.~Harnik, J.~Kopp and Y.~Tsai, \emph{{LEP Shines Light on Dark
			Matter}}, \href{https://doi.org/10.1103/PhysRevD.84.014028}{\emph{Phys. Rev.
			D} {\bfseries 84} (2011) 014028}
	[\href{https://arxiv.org/abs/1103.0240}{{\ttfamily 1103.0240}}].
	
	\bibitem{Dercks:2016npn}
	D.~Dercks, N.~Desai, J.S.~Kim, K.~Rolbiecki, J.~Tattersall and T.~Weber,
	\emph{{CheckMATE 2: From the model to the limit}},
	\href{https://doi.org/10.1016/j.cpc.2017.08.021}{\emph{Comput. Phys. Commun.}
		{\bfseries 221} (2017) 383}
	[\href{https://arxiv.org/abs/1611.09856}{{\ttfamily 1611.09856}}].
	
	\bibitem{ATLAS:2021kxv}
	{\scshape ATLAS} collaboration, \emph{{Search for new phenomena in events with
			an energetic jet and missing transverse momentum in $pp$ collisions at $\sqrt
			{s}$ =13 TeV with the ATLAS detector}},
	\href{https://doi.org/10.1103/PhysRevD.103.112006}{\emph{Phys. Rev. D}
		{\bfseries 103} (2021) 112006}
	[\href{https://arxiv.org/abs/2102.10874}{{\ttfamily 2102.10874}}].
	
	\bibitem{Alloul:2013bka}
	A.~Alloul, N.D.~Christensen, C.~Degrande, C.~Duhr and B.~Fuks, \emph{{FeynRules
			2.0 - A complete toolbox for tree-level phenomenology}},
	\href{https://doi.org/10.1016/j.cpc.2014.04.012}{\emph{Comput. Phys. Commun.}
		{\bfseries 185} (2014) 2250}
	[\href{https://arxiv.org/abs/1310.1921}{{\ttfamily 1310.1921}}].
	
	\bibitem{Alwall:2011uj}
	J.~Alwall, M.~Herquet, F.~Maltoni, O.~Mattelaer and T.~Stelzer, \emph{{MadGraph
			5 : Going Beyond}},
	\href{https://doi.org/10.1007/JHEP06(2011)128}{\emph{JHEP} {\bfseries 06}
		(2011) 128} [\href{https://arxiv.org/abs/1106.0522}{{\ttfamily 1106.0522}}].
	
	\bibitem{Sjostrand:2014zea}
	T.~Sj\"ostrand, S.~Ask, J.R.~Christiansen, R.~Corke, N.~Desai, P.~Ilten et~al.,
	\emph{{An introduction to PYTHIA 8.2}},
	\href{https://doi.org/10.1016/j.cpc.2015.01.024}{\emph{Comput. Phys. Commun.}
		{\bfseries 191} (2015) 159}
	[\href{https://arxiv.org/abs/1410.3012}{{\ttfamily 1410.3012}}].
	
	\bibitem{Bierlich:2022pfr}
	C.~Bierlich et~al., \emph{{A comprehensive guide to the physics and usage of
			PYTHIA 8.3}},
	\href{https://doi.org/10.21468/SciPostPhysCodeb.8}{\emph{SciPost Phys.
			Codeb.} {\bfseries 2022} (2022) 8}
	[\href{https://arxiv.org/abs/2203.11601}{{\ttfamily 2203.11601}}].
	
	\bibitem{ATLAS:2014jim}
	{\scshape ATLAS} collaboration, \emph{{Sensitivity to WIMP Dark Matter in the
			Final States Containing Jets and Missing Transverse Momentum with the ATLAS
			Detector at 14 TeV LHC}}, .
	
\end{thebibliography}

\providecommand{\href}[2]{#2}\begingroup\raggedright\endgroup

\end{document}